\documentclass[11pt]{article}
\pdfoutput=1
\usepackage{jcappub,natbib,float,caption,comment,bm,epsfig}
\usepackage{graphicx}
\usepackage[dvipsnames]{xcolor}
\usepackage[utf8]{inputenc}

\usepackage{slashed}

\newcommand{\beq}{\begin{equation}}
\newcommand{\eeq}{\end{equation}}
\newcommand{\bea}{\begin{eqnarray}}
\newcommand{\eea}{\end{eqnarray}}

\newcommand{\mpl}{m_{\mbox{\tiny{Pl}}}}

\newcommand\lsim{\mathrel{\rlap{\lower4pt\hbox{\hskip1pt$\sim$}}
        \raise1pt\hbox{$<$}}}
\newcommand\gsim{\mathrel{\rlap{\lower4pt\hbox{\hskip1pt$\sim$}}
        \raise1pt\hbox{$>$}}}

\title{\center{Merger of Dark Matter Axion Clumps and Resonant Photon Emission}}
\author[a]{Mark P. Hertzberg,}
\author[b]{Yao Li,}
\author[c,d,b]{and Enrico D. Schiappacasse}
\affiliation[a\,]{Institute of Cosmology, Dept.~of Physics and Astronomy, Tufts University, Medford 02155 MA, USA}
\affiliation[b\,]{School of Physics and Astronomy, Shanghai Jiao Tong University, Shanghai 200240, China}
\affiliation[c\,]{Department of Physics, P.O. Box 35 (YFL),
FI-40014 University of Jyväskylä, Finland}
\affiliation[d\,]{Helsinki Institute of Physics, P.O. Box 64, FIN-00014 University of Helsinki, Finland}

\emailAdd{mark.hertzberg@tufts.edu}
\emailAdd{neolee@sjtu.edu.cn}
\emailAdd{enrico.e.schiappacasse@jyu.fi}

\abstract{
A portion of light scalar dark matter, especially axions, may organize into gravitationally bound clumps (stars) and be present in large number in the galaxy today. It is therefore of utmost interest to determine if there are novel observational signatures of this scenario. Work has shown that for moderately large axion-photon couplings, such clumps can undergo parametric resonance into photons, for clumps above a critical mass $M^{\star}_c$ determined precisely by some of us in Ref.~\cite{Hertzberg:2018zte}. In order to obtain a clump above the critical mass in the galaxy today would require mergers. In this work we perform full 3-dimensional simulations of pairs of axion clumps and determine the conditions under which mergers take place through the emission of scalar waves, including analyzing head-on and non-head-on collisions, phase dependence, and relative velocities. Consistent with other work in the literature, we find that the final mass from the merger $M^{\star}_{\text{final}}\approx 0.7(M^{\star}_1+M^{\star}_2)$ is larger than each of the original clump masses (for $M^{\star}_1\sim M^{\star}_2$). Hence, it is possible for sub-critical mass clumps to merge and become super-critical and therefore undergo parametric resonance into photons. We find that mergers are expected to be kinematically allowed in the galaxy today for high Peccei-Quinn scales, which is strongly suggested by unification ideas, although the collision rate is small. While mergers can happen for axions with lower Peccei-Quinn scales due to statistical fluctuations in relative velocities, as they have a high collision rate. We estimate the collision and merger rates within the Milky Way galaxy today. We find that a merger leads to a flux of energy on earth that can be appreciable and we mention observational search strategies.
}

\begin{document}

\maketitle
\flushbottom


\section{Introduction}
\label{sec:section1}

A wide range of astrophysical observations, including large scale structure, anisotropies of the cosmic microwave background radiation, gravitational lensing, and galactic rotation curves, are well explained after including cold dark matter~\cite{Peebles:2013hla}. However, its particle physics origin remains unknown. By considering shortcomings in the Standard Model of particle physics, the QCD axion is one of the strongest dark matter candidates. Some of the primary motivations for axions come from the strong CP problem~\cite{Peccei:1977hh, Weinberg:1977ma, Wilczek:1977pj} and unification with gravity in the framework of string theory (e.g., see \cite{Svrcek:2006yi,Douglas:2006es,Arvanitaki:2009fg}). 

In recent years, the searching for axion dark matter has captured a large amount of attention from the physics community. 
This interest has lead to the development of a diverse search program including, but not limited to, haloscopes~\cite{PhysRevD.42.1297, PhysRevD.64.092003, PhysRevLett.120.151301, Zhong:2018rsr}, helioscopes~\cite{Anastassopoulos:2017ftl, Armengaud_2014},  axion-induced oscillating electric dipole moment~\cite{Graham:2011qk, Budker:2013hfa, Barbieri:2016vwg},  atomic and molecular transitions induced by axions~\cite{PhysRevLett.113.201301, 2018IJMPA..3344030F}, and  indirect axion searches~\cite{Abramowicz:2017zbp,Iwazaki:2014wka,Iwazaki:2017rtb,Caputo:2018vmy,Huang:2018lxq, Hook:2018iia, Safdi:2018oeu,Buckley:2020fmh,Arza:2020eik}.  A significant part of these searches are based on the axion-photon coupling through the dimension 5 operator $\Delta \mathcal{L} \sim g_{a\gamma \gamma} \phi\, {\bf{E}} \cdot{\bf{B}}$, where $\phi$ is the axion field,  $g_{a \gamma \gamma}$ is the axion-photon coupling constant and ${\bold{E}}$ and ${\bold{B}}$ are the electromagnetic field components, respectively. In particular, ground based experiments, such as the ADMX experiment~\cite{2010PhRvL.104d1301A} in which axions move through a large magnetic field in order to produce a cavity photon, try to detect the axion by exploiting the axion-photon coupling. Even though a detection of axions is plausible in the next few years, such phenomena have not been observed yet. 

Hence, it is very important to explore possible novel phenomena associated with this coupling in different contexts, including astrophysics. In this paper, we continue our investigation from Refs.~\cite{Schiappacasse:2017ham, Hertzberg:2018lmt,Hertzberg:2018zte} about Bose-Einstein condensates (BECs) of axion dark matter and their possible astrophysical consequences. Of particular interest to us here are the properties of gravitationally bound objects, known in the literature by various names, including ``dark matter clumps" or ``axion stars" or ``Bose stars"~\footnote{Axion stars are a particular type of boson star. For a complete review about dynamic properties of boson stars see, for example,~\cite{Liebling2012}. For novel extensions in this topic, see ~\cite{Horvat:2012aq, PhysRevD.96.084066, Choi:2019mva}}. 
These clumps are held together by the inward gravitational force and the outward pressure provided by the fact that at high occupancy the axion is accurately described by classical field theory with an associated pressure from field gradients. These clumps can have a range of masses, but they have an upper limit beyond which there is an instability due to the axion's attractive self-interactions; we will return to all these details in later sections.

Of particular interest in this work will be on merger of these clumps, and the subsequent possible phenomenon of parametric resonance of the clumps into photons in the Milky Way galaxy today. Other important work on this subject includes \cite{Tkachev:1986tr,Tkachev:1987cd,Tkachev:2014dpa,Levkov:2020txo}. 
In the following, we will first briefly explain the different scenarios at which axion clumps may be formed. Then, we shall 
describe the main features of axion clumps and how these astrophysical objects may be detected today in our proposed set up. 

The QCD axion is a pseudo-Goldstone boson associated with a spontaneously broken PQ symmetry, introduced as a solution of the strong CP problem \cite{Peccei:1977hh,Weinberg:1977ma,Wilczek:1977pj}. After the QCD phase transition, the axion acquires mass and begin to behave as cold dark matter.  In the scenario at which the PQ symmetry is broken after inflation, the axion field remain inhomogeneous from one Hubble path to the next by causality. Hence, large field fluctuations after the QCD phase transition allows the field to undergo strong mode-mode gravitational interactions and re-organize into a type of BEC of short-range order~\cite{Guth:2014hsa}; this is the ``clump" or ``axion star" \cite{Kolb:1993zz}. Dynamical relaxation can occur in the so-called kinetic regime \cite{Levkov:2018kau}, while another relevant time scale ($\Gamma_{\text{cond}}\sim 8\pi G_N m_{\phi} n_{\phi}/k^2$) occurs in the so-called condensation regime~\cite{Sikivie:2009qn,Erken:2011dz}. 
Here one normally needs to assume axion models with a domain wall number equal to the unity $N_{\text{DW}} = 1$, so that the axion overabundance coming from the decay of topological defects is avoided~\footnote{The QCD axion can be rescued for $N_{\text{DW}} > 1$ by the inclusion of a \textit{bias} term in the PQ potential~\cite{PhysRevD.23.852, Kawasaki:2014sqa}.}. When the axion mass is the order of the Hubble time, the axion begins to oscillate and roll downs to one of the $N_{\text{DW}}$ degenerate minima.  For example, in the Kim-Shifman-Vainshtein-Zakharov (KSVZ) model~\cite{PhysRevLett.43.103, SHIFMAN1980493}, N$_{\text{DW}}$ corresponds to the number of heavy quarks carrying $U(1)_{\text{PQ}}$ charge so that $N_{\text{DW}} = 1$ can be realized. By contrast,  in the Dine-Fischler-Srednicki-Zhitnitsky (DFSZ) model~\cite{Zhitnitsky:1980tq, DINE1981199}, $N_{\text{DW}}$ is single or double of the number of flavours of quark which carry $U(1)_{\text{PQ}}$ charge, e.g.  $N_{\text{DW}}= 3 \,\,\text{or}\,\, 6$. There is no clear theoretical reason to choose one model over others, although one might argue that phenomenologically the $N_{\text{DW}} = 1$ case may be more reasonable. However, if this is the case, the axion can be the cold dark matter in the Universe in the mass range  $10^{-2}\,\mbox{eV}\lesssim m_a\lesssim 10^{-4}$\,eV, which is related to $10^{9}\, \text{GeV} \lesssim F_a \lesssim   10^{11}\, \text{GeV}$ for the standard QCD axion~\cite{Kawasaki:2014sqa}~\footnote{An even slighlty the lower bound of $F>\mbox{few}\times 10^8$\,GeV comes from constraining the cooling rate originated by the production of axions in the core of the supernova SN 1987A~\cite{Raffelt:2006cw}, though for such low values of $F_a$, it is difficult to constitute all the dark matter}. 

In the scenario at which the PQ symmetry is broken before or during inflation, the axion field is driven to be highly  homogeneous on large scales, so that it is unclear if the axion may form a BEC in the late Universe in the  way we explained above. However, we recently showed for the first time in Ref.~\cite{Hertzberg:2020hsz} that axion dark matter clumps may kinetically nucleate in dark mini-halos around primordial black holes (PBHs)~\footnote{PBHs behave as a cold dark matter being stable for sufficiently large masses. Since the first detection of two merging black holes by LIGO-Virgo Collaboration, the possibility of the existence of PBHs have been strongly revitalized. For a review about PBHs and their mechanism of formation in the early Universe, see Refs.~\cite{Hawking:1971ei, Carr:1974nx, Carr:1975qj, Kawasaki:1997ju, GarciaBellido:1996qt}. For novel physics phenomena associated with PBHs, such as novel contraints in mixed dark matter scenarios with WIMPS, primordial gravitational waves, or effects on direct detection of dark matter, you may read Refs.~\cite{Boucenna:2017ghj, Nakama:2015nea, Nakama:2016enz, Hertzberg:2019exb}.}. If PBHs exists, as is a possible interpretation of the gravitational waves events detected by LIGO-Virgo collaboration~\cite{Abbott:2017vtc, Abbott:2017iws, Abbott:2016drs, TheLIGOScientific:2016htt, Abbott:2016blz, Sasaki:2016jop, Eroshenko:2016hmn, Ali-Haimoud:2017rtz, Raidal:2017mfl}, and the axion is the dominant component of dark matter, they will unavoidably acquire dark mini-halos mainly during the matter dominated era. These mini-halos satisfy the needed conditions to form kinetically axion dark matter clumps before first galaxies formation. Nucleation likely occurs in the so-called kinetic regime, where the wavelength of the axion field is much smaller than the mini-halo radius and the relaxation rate reads as $\Gamma_{\text{kin}}\sim n_{\phi} \sigma_{\text{gr}} v_{\phi} \mathcal{N}$~\cite{Levkov:2018kau, Hertzberg:2020hsz}. Here $\sigma_{\text{gr}} \propto (G_N m_{\phi}/v_{\phi}^2)^2$ is the scattering cross section due to gravitational interaction, $\mathcal{N}$ is the ocupancy number related to Bose enhancement and $v_{\phi}$ is the axion virial velocity in mini-halos.  Depending on the PBH masses, we expect the nucleation of axion clumps composed by the QCD axion or string axions. Indeed, taking a conservative $0.5\%$ in the fraction of dark matter in axion stars, we expect up to $10^{17}$ and $10^{9}$ axion clumps in the solar neighborhood  for the QCD axion and string axions, respectively. Complementary to this scenario, recently authors in Ref.~\cite{Fukunaga:2020mvq} show that clumps composed by axion-like-particles may form when the PQ symmetry is broken before or during inflation. The formation mechanism is mainly via tachyonic instability after considering a multiple cosine potential for the axion-like-particles potential coming from non-perturbative corrections.

In the scenario at which the PQ symmetry is broken before or during inflation, topological defects are not an issue due to the exponential cosmic expansion during inflation. Thus, the axion abundance is given in terms of the initial misalignment angle $\Theta_i$ as~\footnote{We are assuming that there is no dilution coming from, for example, late decays of particles beyond the standard model.}
\begin{equation}
\Omega_a h^2 \sim 0.7 \left( \frac{F_a}{10^{12} \text{GeV}} \right)^{7/6} \left( \frac{\Theta_i}{\pi} \right)^2\,.
\end{equation} 
(where the power of $7/6$ comes from estimates of the axion mass temperature dependence). 
If the initial $\Theta_i$ is taken to be $\mathcal{O}(1)$, then this shows that QCD axions must satisfy the constraint $F_a \lesssim 10^{12}\,\text{GeV}$ (or equivalently $m_a \gtrsim\, 10^{-5}\,\text{eV}$) to avoid the over-closure of the Universe~\cite{PRESKILL1983127, Abbott:1982af,Dine:1982ah}. However, there is no upper bound on the axion decay constant coming if a small initial $\Theta_i$ is considered. The presence of inflation also ensures that there is no additional problems from relic density constraints. This small $\Theta_i$ may require explanation as it appears as an additional fine-tuning. It is sometimes referred to as axion anthropic window, where $F_a \gg 10^{12}\,\text{GeV}$ and $\Theta_i \ll 1$~\cite{PhysRevLett.52.1725,Linde:1991km,Wilczek:2004cr,PhysRevD.73.023505}.  However, the axion field acquires fluctuations proportional to the Hubble parameter during inflation leading to large isocurvature density perturbations~\cite{1983PhLB..126..178A, PhysRevD.32.3178, PhysRevLett.66.5,Fox:2004kb,Hertzberg:2008wr}. Since these kind of perturbations are tightly constrained by cosmic microwave background observations, the so-called isocurvature perturbation problem arises. However, several solutions have been proposed to this problem in the literature as the proposed in Ref.~\cite{Kawasaki:2013iha}. Altogether, for the QCD axion as dark matter, the axion decay constant can conceivably span the range $10^9\, \text{GeV} \lesssim F_a \lesssim 10^{17}\,\text{GeV}$. Here the upper bound comes from black hole spins measurements~\cite{Arvanitaki:2010sy}. 

Furthermore, we note that ideas associated with unification and string theory often point to high values of $F_a$, and such values are often suggested by various kinds of axions beyond QCD and axion-like particles \cite{Arvanitaki:2009fg}. We will return to these high $F_a$ later in our work. 

We are mainly interested in spherically symmetric axion dark matter clumps which correspond to a true BEC. 
The combination of gravity, the axion attractive self-interaction, and kinetic pressure together allow for the formation of stable configurations in the non-relativistic regime. Due to the fact that this condensate is a coherently oscillating axion field, we can expect axion clumps to undergo parametric resonance of the electromagnetic field from the axion-photon coupling under suitable conditions. The output of coherent radio waves may potentially be detected on the earth. 

In conventional QCD axion models, we have $g_{a\gamma\gamma} F_a \sim\mathcal{O}(10^{-2})$. However as we showed in Ref.~\cite{Hertzberg:2018zte}, for spherically symmetric axion clumps, the necessary condition for parametric resonance is $g_{a\gamma\gamma} F_a > 0.3$ (earlier estimates include Refs.~\cite{Tkachev:1986tr,Tkachev:1987cd,Tkachev:2014dpa}). Hence for the conventional QCD axion models, resonance from spherically symmetric clumps would not be possible. 
At these values, resonance is not possible for true BEC of axion dark matter. However, we can have $g_{a\gamma\gamma} F_a \gtrsim 1$ for unconventional QCD axion models,  axion coupling with hidden sector photons~\cite{Daido:2018dmu}, or from axion-like particles, so that the resonant decay of axions may happen. Furthermore, the resonance condition is altered for clumps that carry finite angular momentum; indeed non-spherical symmetric QCD axion clumps may undergo resonant decay for sufficiently large angular momentum, as were analyzed in Ref.~\cite{Hertzberg:2018lmt}.   

For a given value of the axion-photon coupling constant $g_{a\gamma\gamma}$, there exists a critical number of particles  which allows for resonance in a clump, $N^{\star}_{c}$. Consider an axion clump with a number of particles $N_{\star}$  and a value of the axion-photon coupling $g_{a\gamma\gamma}$. Suppose that this coupling is large enough so that axion clumps may undergo resonant decay if their number of particles is larger than the critical one. After axion dark matter clump formation, one would expect a distribution for their masses. On the one hand, clumps with a number of particles greater than the critical number, i.e., $N_{\star} > N^{\star}_{c}$, will undergo resonant decay into photons. These clumps will quickly lose mass until their number of particles reaches the critical number, i.e., $N_{\star} \rightarrow N^{\star}_{c}$. On the other hand, clumps with a number of particles less than the critical number, i.e.,  $N_{\star} < N^{\star}_{c}$, may capture axion dark matter from the smooth background so that $N_{\star}$ will slowly grow towards $N^{\star}_{c}$. Thus, we expect a kind of mass pile-up at a unique value  $M^{\star}_{c} =  N^{\star}_{c} m_{\phi}$. Interestingly, this unique number depend only on fundamental constants. If this scenario is realized, we should expect to have a mass pile-up of axion stars in the Milky Way halo today. After suitable conditions, these astrophysical objects may collide and merge leading to a new BEC axion clumps with a total number of particles greater than the critical one. Thus, emission of photons via axion resonant decay could then happen and be relevant in the galaxy today. We also note that in the halo today, the effective plasma mass of photons is very small, allowing this process to potentially occur, while it would be forbidden in the early universe due to the higher plasma mass back then.

The outline of this paper is as follows: In Section 2 we briefly explain the main features of  axion field theory, axion dark matter clumps, and the parametric resonance phenomenon associated with these astrophysical objects. We summarize main results from our previous work in Refs.~\cite{Schiappacasse:2017ham, Hertzberg:2018zte}.  In Section 3  we numerically study the collision of spherically symmetric axion dark matter clumps and obtain the needed conditions for mergers to take place. In Section 4 we analyze the collision and merger rates for axion clumps in the Milky Way halo, as well as the main features of the parametric resonance phenomenon. In Section 5 we present our summary and outlook. Finally, in Appendices A and B we explicitly compute some needed results for the numerical set up performed in Section 2.

\section{Axions and Photons}
\label{sec:section2}
The general dynamics of the QCD axion has been studied and reviewed in many papers. Here we focus only on points which are relevant for this work. For a general review, see, for example, Refs.~\cite{Duffy:2009ig, Masso:2002ip, Marsh:2015xka}.

\subsection{Axion Field Theory}
The QCD axion $\phi$ is the pseudo-Nambu-Goldstone boson of the Peccei-Quinn (PQ) solution to the strong CP problem~\footnote{The fact that the axion solves the strong CP problem makes it a strong dark matter candidate. However, several solutions to this problem have been proposed in the literature. For discrete symmetry solutions, see Refs.~\cite{NELSON1984387, PhysRevLett.53.329, Carena:2019nnd}. Recently, it was proposed in Ref.~\cite{Choi:2019omm} an interesting new solution which relies on the horizontal gauge symmetry and CP invariance in a full theory.} in the Standard Model~\cite{Peccei:1977hh,Weinberg:1977ma,Wilczek:1977pj}. While the axion is massless at the classical level, non-perturbative quantum effects in QCD give rise to a potential for the $\phi$ at low temperatures. Starting from the very early universe, this potential becomes relevant at temperatures of order the confinement scale. At that time, the axion acquires a small mass and the field relaxes to the CP conserving minimum. As a dark matter candidate, cold axions are sufficiently light to be in the high occupancy regime  and, as a result, they are well described by classical field theory (after performing a suitable ensemble averaging~\cite{Hertzberg:2016tal}).

In the effective theory for axions, the Lagrangian density of the field can be written in the canonical form as~\footnote{We work in natural units $\hbar=c=1$ with a metric signature of (+ - - -).}
\begin{equation}
\mathcal{L} = \sqrt{-g} \left[ \frac{\mathcal{R}}{2 \kappa^2} + \frac{g^{\mu\nu}}{2}\nabla_{\nu}\phi \nabla_{\mu}\phi - V(\phi) \right]\,,
\end{equation}
where $g =\text{det}(g_{\mu\nu})$ is the determinant of the metric tensor, $\kappa = \sqrt{8\pi G_N}$ is the gravitational coupling and $\mathcal{R}$ is the Ricci scalar. Since we shall focus only on the non-relativistic regime for axions, we can expand the potential $V(\phi)$ around the CP conserving minimum $\phi = 0$ and keep the first two leading terms as 
\begin{equation}
V(\phi) = \frac{1}{2}m_{\phi}^2\phi^2 + \frac{\lambda}{4!} \phi^4 + \mathcal{O}(\lambda^2\phi^6/m_\phi^2)\,,
\end{equation}
where $m_{\phi}$ is the axion mass and $\lambda$ is the quartic coupling constant.  
For the standard QCD axion,  the axion mass is given by~\cite{Weinberg:1977ma}
\begin{equation}
m_{\phi} = \frac{\sqrt{m_u m_d}}{(m_u+m_d)}\frac{f_{\pi} m_{\pi}}{F_a} \approx 10^{-5}\,\text{eV} \,\left( \frac{6\times10^{11}\,\text{GeV}}{F_a} \right)\,\label{axionmassfa}
\end{equation}
where $m_u, m_d, m_{\pi}$ are the up quark, down quark and pion masses, $f_{\pi}$ is the pion decay constant and $F_a$ is the PQ symmetry breaking scale (or axion decay constant). Note that here we have taken $F_a=6\times 10^{11}$\,GeV as a typical reference value for axions in the classic window. However, as we will discuss later, higher values of $F_a$ are of considerable interest to the phenomenology; these correspond to lighter axion masses $m_a$.

The self coupling constant $\lambda$ can be parameterized in terms of the axion mass and the PQ symmetry breaking scale as
\begin{equation}
\lambda = -\gamma \frac{m_{\phi}^2}{F_a^2}\,,
\end{equation} 
which is negative for the attractive axion self-interaction and $\gamma \sim \mathcal{O}(1)$ is a parameter of order unity. As we mentioned earlier, the potential for the axion arises from QCD instantons. While the computation of this potential under the standard dilute instanton gas approximation leads to $\gamma = 1$,   a more accurate computation combining chiral perturbation theory plus lattice QCD leads to $\gamma = 1 - 3m_u m_d /(m_u + m_d)^2 \approx 0.3$~\cite{diCortona:2015ldu}.

To take the non-relativistic limit, it is useful  to express the real axion field in terms of a slowly varying complex scalar field          $\psi({\bf{x}},t)$ as follows
\begin{equation}
\phi({\bf{x}},t) = \frac{1}{\sqrt{2m_{\phi}}} \left[ \text{e}^{-im_{\phi} t}\psi({\bf{x}},t) +  \text{e}^{im_{\phi} t}\psi^{*}({\bf{x}},t) \right]\,.\label{phirel}
\end{equation}
The complex field $\psi({\bf{x}},t)$ introduces small corrections to the fundamental
frequency $\omega_0 = m_{\phi}$ in the non-relativistic regime. Inserting this expression into the above Lagrangian density, taking the non-relativistic limit in a non-expanding background (as we are interested in behavior in the galaxy today), and passing to the Hamiltonian formalism, the dynamics of the axion is determined for the
following non-relativistic Hamiltonian  
\begin{equation}
H_{\text{nr}} = H_{\text{kin}} + H_{\text{int}} + H_{\text{grav}}\,,
\label{Ham}
\end{equation}
where
\begin{align}
H_{\text{kin}}& = \frac{1}{2m_{\phi}} \int d^3x \nabla \psi^* \cdot  \nabla \psi\,,\\
H_{\text{int}} &=  \frac{\lambda}{16 m^2_{\phi}} \int d^3x\, \psi^{*2}\psi^2 \,,\\
H_{\text{grav}}& = -\frac{G_N m_{\phi}^2}{2} \int d^3x \int d^3x' \frac{\psi^*({\bf{x}})\psi^*({\bf{x}'})\psi({\bf{x}'})\psi({\bf{x}})}{|{\bf{x}}-{\bf{x}'}|}\,.
\end{align}
Here $H_{\text{kin}}$, $H_{\text{int}}$, and  $H_{\text{grav}}$ refer to the different components of the non-relativistic Hamiltonian, e.g. the kinetic, the self-interacting, and the gravitational energy, respectively. This Hamiltonian can also be derived by using many-particle quantum mechanics as shown in Ref.~\cite{Guth:2014hsa}. The Hamiltonian in Eq.~(\ref{Ham}) is invariant under the field transformation $\psi \rightarrow \psi e^{i\beta}$, where $\beta$ is a constant. This global U(1) symmetry is associated with a conserved number of particles
\begin{equation}
N = \int d^3x\, \psi^*({\bf{x}})\psi({\bf{x}})\,,
\end{equation} 
which is expected in the non-relativistic limit where particle-number violated processes are usually suppressed.
However, there can still unavoidably number changing processes through the axion-photon coupling, as we discuss later in the paper, 
where we will be interested in the resonance regime at which axions decay in pair of photons with exponential growth. The output of this phenomenon are classical electromagnetic waves.

Using the Hamilton equation, we can readily derive the equation of motion of the field in the non-relativistic regime. This, together with the Newton-Poisson equation for the (non-dynamical) Newtonian potential, $\phi_N=\phi_N(\psi,\psi^*)$, are a pair of coupled partial differential equations governing the time evolution of the system. We have,
\begin{align}
  \label{dschrodinger} i \dot{\psi} & =  - \frac{1}{2 m_{\phi}}
 \nabla^2 \psi + m_{\phi}\psi \phi_N - \frac{|\lambda|\, \psi^* \psi^2}{8 m_{\phi}^2}\,,\\
  \label{dpoission} \nabla^2 \phi_N & =  4 \pi G_N m_{\phi} | \psi |^2\,.
\end{align}
For numerical purposes, it is convenient to rescale the axion field, the Newtonian potential, and the temporal and spatial coordinates to go to the dimensionless version of these equations.  Recalling that $|\lambda| = \gamma m_{\phi}^2/F_a^2$, the suitable transformations for the variables are the following:
\begin{align}
x &= \left( \frac{\mpl \gamma^{1/2}}{m_{\phi}F_a} \right)\tilde{x}\,,\hspace{1.5 cm}t =  \left( \frac{\mpl^2 \gamma}{m_{\phi} F_a^2} \right)\tilde{t}\,, \label{xtrescale}\\
\psi& = \left( \frac{m^{1/2}_{\phi}F_a^2}{\mpl \gamma} \right)\tilde{\psi}\,,\hspace{1 cm}\phi_N = \left( \frac{F_a^2}{\mpl^2 \gamma} \right)\tilde{\phi}_N\,,
\end{align}
where a similar transformation for spatial varables ($y$, $z$) is understood. Here,  $\mpl = 1/\sqrt{G_N}$ is the Planck mass and variables with tilde accent marks refer to dimensionless quantities. Then, we can rewrite Eqs.~(\ref{dschrodinger}, \ref{dpoission}) 
as follows
\begin{align}
  \label{dschrodingernodim} i \dot{\tilde{\psi}} & =  -\frac{1}{2}
 \tilde{\nabla}^2 \tilde{\psi} + \tilde{\psi} \tilde{\phi}_N - \frac{\, \tilde{\psi}^* \tilde{\psi}^2}{8}\,,\\
  \label{dpoissionnodim} \tilde{\nabla}^2 \tilde{\phi}_N & =  4 \pi | \tilde{\psi} |^2\,.
\end{align}
Later, we shall analyze the merger of pairs of axion stars by starting from two initially separate star configurations, which independently satisfy the time independent version of Eq.~(\ref{dschrodingernodim}), and then track their self consistent non-linear evolution. 

\subsection{Axion Dark Matter Clumps}
\label{ADMC}

\begin{figure}[t!]
\includegraphics[scale=0.33]{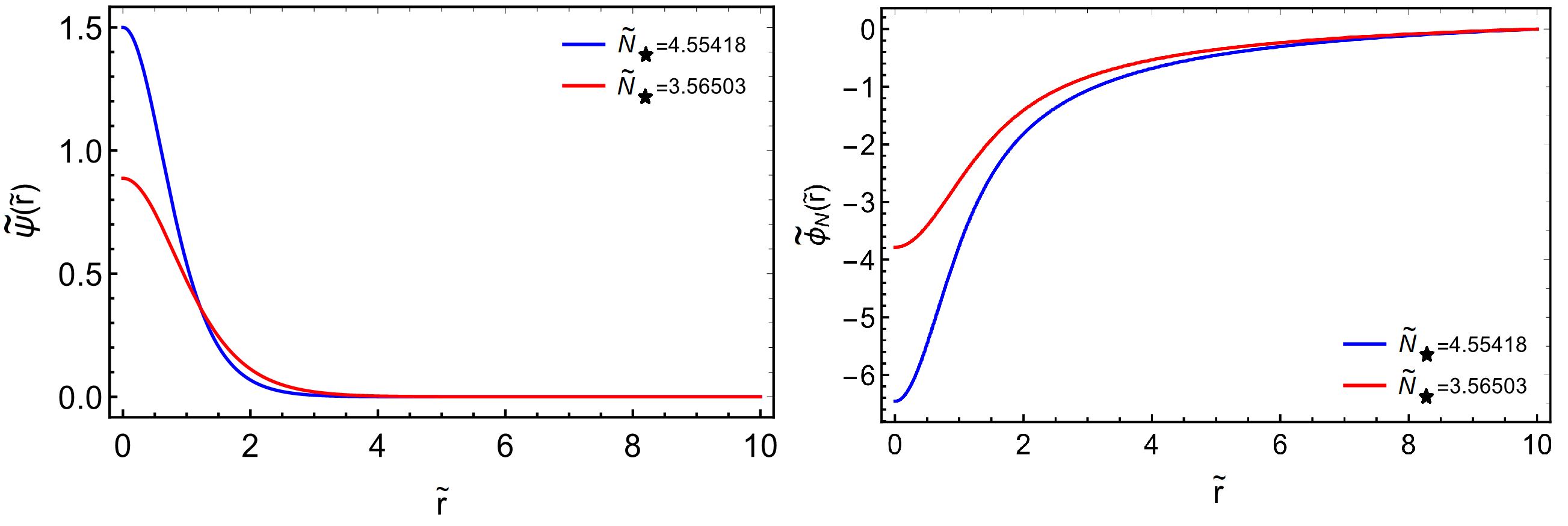}\!\!\!\!\!\!\!\!\!\!\!\!\!
\caption{Radial profiles of the axion field (right) and corresponding Newtonian potentials (left) of the stable ground state configurations with a number of particles $ \tilde{N}_{\star} = 4.55418$ (blue solid line) and $\tilde{N}_{\star} = 3.56503$ (red solid line). }
\label{IC}
\end{figure}

The axion BEC is defined by a fixed number of particles. While the true BEC is spherically symmetric, higher eigenstates
of the axion condensate includes the presence of non-zero angular momentum.  Here we will  mainly focus on the true BEC
configurations which corresponds to the state of minimum energy at fixed number of particles.   On the other hand, we will also study mergers of non head-on collisions, which can lead to some non-zero angular momentum, albeit typically small. 
We will recap the most important features of spherically symmetric axion clumps,  which were studied in detail in Refs.~\cite{Schiappacasse:2017ham,Chavanis:2011zi,Chavanis:2011zm}.
A ground state configuration can be written as a spherically symmetric stationary solution as
\begin{equation}
\psi(r,t) =  \Psi(r)e^{-i\mu t},\label{gs}
\end{equation} 
where $\mu \approx m_{\phi}$ as expected in the non-relativistic limit and $\Psi(r)$ describes the radial profile. For example, the radial profile and the corresponding Newtonian potential for a pair of ground state solutions is given in Fig.~\ref{IC}. 

For most of this paper we will obtain precise numerical results of the equations of motion, but it is useful to compare to analytical approximations, which we discuss now. 
We showed in Ref.~\cite{Schiappacasse:2017ham} that a sech and exponential-linear ansatz for the radial profile are very accurate in comparison to $\Psi(r)$ obtained numerically. 
In particular, the sech ansatz reads as
\begin{equation}
\Psi_{R}(r) = \sqrt{\frac{3 N_{\star}}{\pi^3 R^3}} \,\text{sech}(r/R)\,,\label{ansatz}
\end{equation}
where the shape of the clump is controlled by the length scale $R$ and the coefficients in front of the function ensure the normalization, i.e., $N_{\star} = \int_0^{\infty}dr\,4 \pi r^2 \Psi^2(r)$. We use the variational method to find equilibrium solutions of the axion field. Using the dimensionless variables defined previously and replacing any localized ansatz depending on a single scale into the Hamiltonian, the energy of the system after a suitable integration takes the form  
\begin{equation}
\tilde{H}(\tilde{R})= a\frac{\tilde{N}_{\star}}{\tilde{R}^2} - b\frac{\tilde{N}_{\star}^2}{\tilde{R}} - c\frac{\tilde{N}_{\star}^2}{\tilde{R}^3}\,,\label{hamil}
\end{equation} 
where coefficients $a,b,c$ depend on the localized ansatz, and, as we will see, are $\mathcal{O}(1)$ numbers. Here, the dimensionless Hamiltonian, $\tilde{H}$, the dimensionless scale length, $\tilde{R}$, and the dimensionless number of particles, $\tilde{N}_{\star}$, are related by
\begin{equation}
H = \left( \frac{F_a^3}{\mpl m_{\phi}\gamma^{3/2}} \right) \tilde{H}\,,\hspace{0.5 cm} 
R = \left( \frac{\mpl \gamma^{1/2}}{m_{\phi} F_a} \right) \tilde{R}\,,\hspace{0.5 cm}
N_{\star} = \left(\frac{\mpl F_a}{m_{\phi}^2\gamma^{1/2}} \right) \tilde{N}_{\star}\,,\hspace{0.5cm} (M_\star=m_\phi\,N_\star).\label{DimV}
\end{equation}
Now, as we will see, the typical axion stars have dimensionless quantities $\tilde{H}$, $\tilde{R}$, $\tilde{N}_\star$ that are not especially large or small (at least for the heavier stars), so the dimensionful prefactors give one a rough idea as to their values. In particular, we see that for higher $F_a$, the stars will carry a larger (negative) binding energy and be more robust to undergo mergers (we will return to this later in our work), as well being more massive (note that $m_a F_a$ is fixed in terms of the QCD scale for the QCD axion, so the radius does not change with larger $F_a$). 

\begin{figure}[t!]
\centering
\includegraphics[scale=0.18]{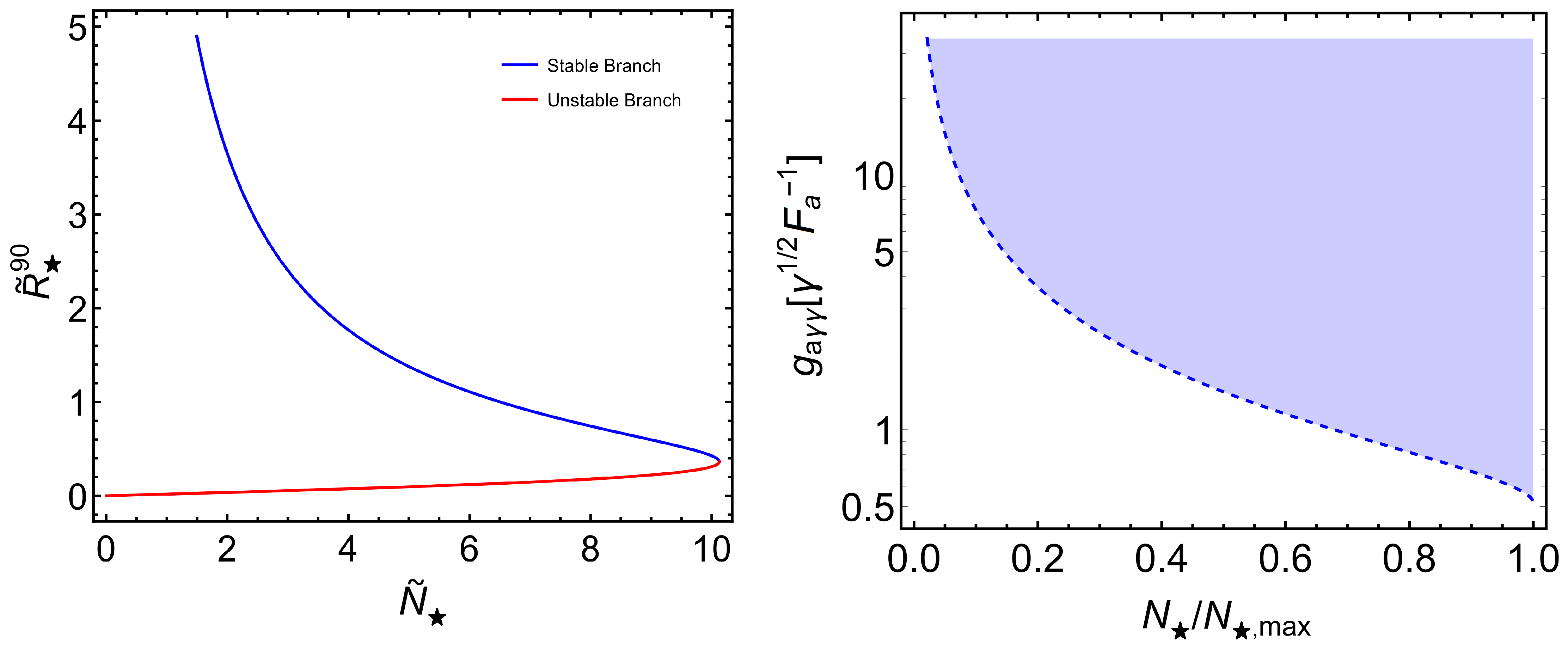}
\caption{(Left) Two branches of solutions when the axion system is treated in the non-relativistic regime for the
dimensionless radius  $\tilde{R}^{90}_{\star,\text{min}}$ (which encloses 90$\%$ of the clump mass) 
as a function of the dimensionless number of particles $\tilde{N}_{\star}$. The upper blue curve corresponds to stable solutions for spherically symmetric axion clumps,  which are the configurations of our interest. The lower red curve refers to the corresponding unstable
configurations. The sech ansatz approximation in Eq.~(\ref{ansatz}) is used  to draw both curves. (Right) Parameter space for the axion-photon coupling $g_{a\gamma \gamma} [\gamma^{1/2} F_a^{-1}]$ with respect to the
number of particles on the stable blue branch, normalized to $N_{\star,\text{max}}\, [\mpl F_a m^{-2}_{\phi}\, \gamma^{-1/2}]$. Parametric resonance of axion clumps into photons occurs in the upper right blue shaded region.}
\label{stableunstable}
\end{figure}

For the specific case of the sech ansatz, coefficients in Eq.~(\ref{hamil}) are given by
\begin{equation}
a=\frac{12 + \pi^2}{6 \pi^2}\,, \hspace{0.5cm} b= \frac{6[12 \zeta(3)-\pi^2]}{\pi^4}\,, \hspace{0.5cm} c = \frac{\pi^2-6}{8\pi^5}\label{abc} 
\end{equation}
Extremizing the Hamiltonian with respect to the variational parameter $R$ at fixed number of particles in Eq.~(\ref{hamil}),  we can map out the basic solutions of the axion-gravity-self-interacting system in the non-relativistic regime. As shown in Fig.~\ref{stableunstable} (left), there are two branches of solutions which are associated with the extrema of the Hamiltonian according to
\begin{equation}
\tilde{R}= \frac{a \pm \sqrt{a^2-3bc\tilde{N}_{\star}^2}}{b\tilde{N}_{\star}}\,.
\label{RextremeN}
\end{equation}
 For a given $N_{\star}$,  while the global maximum of $\tilde{H}(\tilde{R})$ corresponds to an unstable solution, the local minimum is a stable solution.  When gravity dominates over self-interactions (upper blue curve), BEC axion clumps are stable against perturbations and can be spatially large. By contrast, when axion self-interactions dominates over gravity (lower red curve), BEC axion clumps are unstable against perturbations and can be spatially small (for very small clumps, higher order terms of the potential $V(\phi)$ eventually become important, leading to new solutions called axitons \cite{Kolb:1993hw}, but this will not be our focus here). Since we are interested in to analyze the resonance phenomenon of photons coming from merger of axion clumps, we will focus from now exclusively on the stable branch. 
 
From Eq.~(\ref{RextremeN}), we see there is a maximum number of particles which can be in a clump, $\tilde{N}_{\star,\text{max}} \equiv a/(\sqrt{3bc})$, which is associated with a minimum length scale. At this particular point of the parameter space of stationary solutions, gravity and self-interacting forces are comparable and the stable and unstable branches converge. 

Returning to dimensionful variables, we can express the maximum allowed mass for an axion clump,  $M_{\star, \text{max}} = N_{\star, \text{max}}m_{\phi}$, and its associated radius in terms of the axion mass and the PQ scale. Since axion clumps do not have a hard surface, we define the clump radius $R_{\star}^{90}$ as  the radius at which is enclosed the 90\% of the total clump mass. Using the sech ansatz approximation, we obtain
\begin{align}
M_{\star, \text{max}} & \approx 2.4 \times 10^{19}\, \text{kg}\, \left( \frac{10^{-5}\,\text{eV}}{m_{\phi}}  \right) \left( \frac{F_a}{6\times10^{11}\,\text{GeV}}  \right) \left(  \frac{0.3}{\gamma} \right)^{1/2}\,,\label{Mmax}\\
R^{90}_{\star, \text{min}} & \approx 80\,\text{km} \left( \frac{10^{-5}\,\text{eV}}{m_{\phi}}  \right)
\left( \frac{6\times10^{11}\,\text{GeV}}{F_a}  \right) \left(  \frac{\gamma}{0.3} \right)^{1/2}\,.\label{Rmin}
\end{align}
where  $\tilde{R}^{90}_{\star, \text{min}} \approx 2.799 \tilde{R}(\tilde{N}_{\star,\text{max}})$ for the sech ansatz approximation.  

A simple manipulation allows us to express $M_{\star}$ and $R_{\star}$ of any axion clump in function of $M_{\star, \text{max}}(R^{90}_{\star, \text{min}})$  and $R^{90}_{\star, \text{min}}$, respectively, as
\begin{align}
M_{\star}(R_{\star}) & = \alpha\,M_{\star, \text{max}}(R^{90}_{\star, \text{min}})\,,\label{Malpha}\\
R_{\star}& = g(\alpha)\,R^{90}_{\star, \text{min}}\,, \label{Ralpha}
\end{align}
 where 
\begin{equation}
g(\alpha) \equiv (1 + \sqrt{1-\alpha^2})/\alpha\,\,\,\,\,\text{with}\,\,\,\, \alpha \in (0,1]. 
\end{equation}
where the dimensionless quantity $\alpha$ parameterizes the branch of stable solutions:  $\alpha\to0$ corresponds to going far up the upper blue branch to small masses and large radii. While $\alpha\to1$ corresponds to the end of the stable blue branch where the clumps are more massive and smaller radii and where it meets the unstable red branch.

The corresponding relativistic field $\phi(\bf{x},t)$  is obtained by replacing in Eq.~(\ref{phirel}) the ground state configuration $\psi(r,t)$ from Eq.~(\ref{gs}) to obtain 
\begin{equation}
\phi(r,t) =\Phi(r) \text{cos}(\omega_0 t) \,,
\end{equation} 
where the radial profile is  $\Phi(r) = \sqrt{2/m_{\phi}} \Psi(r)$ and the oscillation frequency is $\omega_0 = m_{\phi} + \mu \approx m_{\phi}$. Coherent harmonic oscillations with a frequency close to the axion mass can lead to resonance of the electromagnetic field, depending on the value of the axion-photon coupling, as we shall explain now.

\subsection{Parametric Resonance of Photons}
\label{resonancephotons}
Here we briefly review the axion-photon interaction and the parametric resonance phenomenon in axion clumps. The axion clump  resonance of photons was studied in detail by us in Ref.~\cite{Hertzberg:2018zte}, while other work includes Refs.~\cite{Tkachev:1986tr,Tkachev:1987cd,Tkachev:2014dpa,Levkov:2020txo}. 

In essentially all axion models, the axion couples to photons through the chiral anomaly, where a pair of photons is connected with the axion by a fermion loop. The interacting Lagrangian density is given by
\begin{equation}
\mathcal{L}_{a\gamma \gamma}= g_{a \gamma \gamma} \phi {\bold{E}}\cdot {\bold{B}}\,,\label{lag}
\end{equation} 
where $g_{a \gamma \gamma}$ is the axion-photon coupling constant, which has units of inverse mass, and ${\bold{E}}$ and ${\bold{B}}$ are the electromagnetic field components. The coupling to ${\bold{E}}\cdot {\bold{B}}$ is appropriate since the axion is also a pseudo-scalar and so this term is CPT invariant. 
As mentioned earlier, detection of axions in ground based experiments often relies on the axion-photon coupling. Thus, different values taken by $g_{a \gamma \gamma}$ in different theoretical realizations is of crucial importance for many experiments. In some classic models of the QCD axion, the coupling constant is written as
\begin{equation}
g_{a \gamma \gamma} = \frac{\alpha}{2 \pi F_a} \left[\frac{\bar{E}}{\bar{N}} - \frac{2(4 + m_u/m_d + m_u/m_s )}{3(1 + m_u/m_d + m_u/m_s )}  \right]\,,
\end{equation}
where $\alpha$ is the fine structure constant, $\bar{E}$ and $\bar{N}$ are the electromagnetic and color anomalies, respectively, and $m_u$, $m_d$ and $m_s$ are the usual quark masses.  The ratio between these anomalies is model dependent, but for conventional axion models 
$|g_{a \gamma \gamma}| F_a \sim \mathcal{O}(10^{-2})$~\footnote{For ease notation, we will send $g_{a\gamma \gamma} \rightarrow |g_{a\gamma  \gamma}|$ due to only its magnitude is relevant to the resonance phenomenon.}.

Since the axion dark matter clumps are coherently oscillating, the axion condensate may undergo parametric resonance of photons. During the resonance, there is an exponential growth in photon occupancy number and subsequent radio wave emission.  Due to this exponential growth, the final result is essentially classical electromagnetic waves. The electromagnetic background coming from CMB and astrophysical sources, plus the inevitable quantum fluctuations in the vacuum, ensure the presence of seed fluctuations to trigger the resonance. 

As a useful, though as we shall see overly simplistic, starting point, one may treat the axion field as a homogeneous condensate. Even though such a configuration is unstable against perturbations to collapse from gravity and attractive self-interactions, we shall see later that the resonance phenomenon in such configuration sets a relevant time scale for the growth rate associated with the resonance in localized axion clumps. For small field amplitudes, the homogeneous axion takes the form 
\begin{equation}
\phi(t) = \phi_0 \text{cos}(\omega_0 t),  
\end{equation}
where $\omega_0 \approx m_{\phi}$ and $\phi_0$ is the amplitude for oscillations. 

Consider the quantized four vector potential $\hat{A}^\mu = (\hat{A}_0, \hat{\bf{A}})$ and take the Coulomb gauge $\nabla \cdot \hat{\bf{A}} = 0$. The variation of the 
Lagrangian density of axion-photon interaction in Eq.~(\ref{lag}) with respect to the four vector potential leads to the following modified Maxwell equation for the two photon propagating degrees of freedom:
\begin{equation}
\ddot{\hat{{\bf{A}}}} -\nabla^2 \hat{{\bf{A}}} + g_{a\gamma \gamma} \nabla \times (\partial_t \phi \hat{{\bf{A}}}) = 0\,,\label{vpot}
\end{equation} 
where gradients of the axion field are neglected within the non-relativistic approximation.  Passing to Fourier space in this background homogeneous approximation, the electromagnetic modes decouple and the corresponding mode functions $s_{\bf k}$ of the vector potential satisfy the classic Mathieu equation as
\begin{equation}
\ddot{s}_{{\bf{k}}} + \omega_k^2(t) s_{{\bf{k}}} = 0\,,
\end{equation}   
where $\omega_k = k^2 - g_{a\gamma \gamma} \omega_0 k \phi_0 \text{sin}(\omega_0 t)$ is an effective frequency in the k-space. Since the frequency of the pump is periodic, i.e., $\omega_k^2(t) = \omega_k^2(t + T)$ with $T = 2\pi/\omega_0$, there is parametric resonance for modes with certain values of $k$. In the small amplitude regime, a spectrum of narrow resonant bands is observed equally spaced at $k^2 \approx (n/2)^2\omega_0^2$ for $n = 1,2,3,...$. The resonance is dominated by the first instability band having a maximum growth rate given by
\begin{equation}
\mu_{H} \approx g_{a\gamma \gamma} m_{\phi} \phi_0 /4.\label{hom}
\end{equation}
The center of the resonance band is $k = (\omega_0/2)\sqrt{1 + g_{a\gamma\gamma}^2\phi_0^2/2}$, which can be approximated as $k \approx m_{\phi}/2$ for small amplitudes. The bandwidth is 
\begin{equation}
\Delta k = k_{r,\text{edge}} - k_{l,\text{edge}} \approx g_{a\gamma\gamma} m_{\phi} \phi_0 / 2
\end{equation}
where  $k_{r,\text{edge}}$ and  $k_{l,\text{edge}}$ are the left and right hand edge of the first instability band, respectively. 

Now we turn to the important situation at hand, involving inhomogeneous spherically symmetric clumps on the stable branch. Here the equation of motion of the quantized vector potential in Eq.~(\ref{vpot}) can be expressed in terms of their mode functions $v$ and $w$ through a vector spherical harmonic decomposition involving functions ${\bf{M}}_{\text{lm}}$ and  ${\bf{N}}_{\text{lm}}$. By exploiting the spherically symmetry of the axion field, the complicated tridimensional problem is transformed into an effective 1-dimensional problem.
This decomposition reads as   
\begin{equation}
\hat{\bf{A}}({\bf{x}},t)=\int \frac{d^3k}{(2\pi)^3} \sum_{\text{lm}} 
\left[ \hat{a}(k) v_{\text{lm}}(k,t) {\bf{M}}_{\text{lm}}(k,{\bf{x}}) - \hat{b}(k) w_{\text{lm}}(k,t) {\bf{N}}_{\text{lm}}(k,{\bf{x}})  + h.c.   \right]\,.
\end{equation}
The vector spherical harmonics are defined in terms of the scalar spherical harmonics $Y_{\text{lm}}(\theta, \varphi)$ and the spherical Bessel functions $j_{\text{l}}$ as  usual as
\begin{align}
{\bf{M}}_{\text{lm}}(k,{\bf{x}}) &= \frac{i j_{\text{l}}(kr)}{\sqrt{\text{l}(\text{l}+1)}}  \left[ \frac{i \text{m}}{\sin \theta}  Y_{\text{lm}} \hat{\theta} - \frac{\partial Y_{\text{lm}}}{\partial \theta} \hat{\varphi}  \right]\,,\\
ik{\bf{N}}_{\text{lm}}(k,{\bf{x}}) &= - \nabla \times {\bf M}_{\text{lm}} (k,{\bf{x}})\,,
\end{align}
where $r=|{\bf x}|$ is radius, $\theta$ is polar angle, and $\varphi$ is azimuthal angle. Neglecting gradients of the axion field, the equation of motion of the vector potential becomes a coupled system of equations for the mode functions $v_{\text{lm}}$ and $w_{\text{lm}}$ as follows
\begin{eqnarray}
&&\int \frac{d^3k}{(2\pi)^3} \sum_{\text{lm}} \Bigg{[}  \left(  \ddot{v}_{\text{lm}} + k^2 v_{\text{lm}} - i k g_{a\gamma \gamma} \partial_t \phi w_{\text{lm}} \right) {\bf{M}}_{\text{lm}} \nonumber\\
&&\,\,\,\,\,\,\,\,\,\,\,\,\,\,\,\,\,\,\,\,\,\,\,\,\,\,\,\,\,\,\,\, - \left(  \ddot{w}_{\text{lm}} + k^2 w_{\text{lm}} + i k g_{a\gamma \gamma} \partial_t \phi v_{\text{lm}} \right) {\bf{N}}_{\text{lm}}    \Bigg{]}=0\,.
\end{eqnarray} 
Choosing one resonant channel for simplicity, considering axion field configurations which  slowly vary in space, and rewriting the
axion spatial profile by a 1-dimensional (real) Fourier transform, the system of coupled differential equations become simpler to treat numerically. For the specific channel $l=1$ and $m=0$, for example, the modes functions obey the relation $w_{10}=\pm iv_{10}(k,t)$, where 
\begin{equation}
\ddot{w}_{10}(k,t) + k^2\omega_{10}(k,t) - ig_{a\gamma \gamma}\omega_0 k \text{sin}(\omega_0 t) \int \frac{dk'}{2\pi} v_{10}(k')\tilde{\Phi}_{1d}(k-k') = 0\,. 
\end{equation}
Here $2\pi\Phi(r)=\int d\tilde{k} \cos(\tilde{k}r)\tilde{\Phi}_{1d}(\tilde{k})$ and $\omega_0$ is the fundamental frequency for the corresponding homogeneous case. 

Numerical solution of the system based on Floquet Theory in our earlier work shows that the growth rate of the resonance phenomenon is well approximated by 
\begin{equation}
\mu^{\star} \approx \left \{ \begin{matrix} \mu_{H} - \mu_{\text{esc}},\,\,\,\,\,\, \mu_{H} > \mu_{\text{esc}}
\\ \,\,\,\,\,\,\,\,\,\,\,\,0,\,\,\,\,\,\,\,\,\,\,\,\,\,\,\,\,\,\mu_{H} < \mu_{\text{esc}} \end{matrix}\right. 
\end{equation}
where $\mu_{H} \approx g_{a\gamma \gamma} m_{\phi} \phi_0 /4$ is the maximum growth rate for the case of a homogeneous condensate as shown in Eq.~(\ref{hom}) and $\mu_{esc} \approx 1/(2R_{\star})$ is the photon escape rate. Taking the sech ansatz radial profile to set the axion field amplitude as
\begin{equation}
\phi_0 = \sqrt{\frac{2}{m_{\phi}}} \Psi_0 =  \sqrt{\frac{6 N_{\star}}{m_{\phi} \pi^3 R^3}}\,, 
\end{equation}
the resonance condition takes the form
\begin{equation}
g_{a\gamma \gamma} F_a > 0.28 \left( \frac{\gamma}{0.3} \right)^{1/2} \left[\frac{g(\alpha)}{\alpha}\right]^{1/2}\,,
\label{gagammaN}
\end{equation}
where we recall from earlier that $g(\alpha) \equiv (1+\sqrt{1-\alpha^2})/\alpha$ and $\alpha \in (0,1]$. When $\alpha = 1$, the number of particles of the axion clump reaches $N_{\star, \text{max}}$ leading to $g_{a \gamma \gamma,\text{min}} = 0.28\, F_a^{-1} (\gamma/0.3)^{1/2}$. Equation~(\ref{gagammaN}) sets a general relation between the axion-photon coupling constant and the number of particles in the clump to obtain photon emission via resonance. This inequality is shown in Fig.~\ref{stableunstable} (right). For a fixed axion-photon coupling,  lighter clumps are less likely to undergo parametric resonance. While for the conventional models for the QCD axion $g_{a\gamma \gamma} F_a \sim \mathcal{O}(10^{-2})$ does not satisfy the resonance condition, axion models with atypical axion-photon coupling $g_{a\gamma \gamma} F_a \gtrsim 1$ as well as with couplings to hidden sector photons may undergo parametric resonance. In particular,  a theoretical realization of a QCD axion model in the GUT framework with hidden sector photons is performed in Ref.~\cite{Daido:2018dmu}. These authors obtain an enhancement factor for the axion-photon coupling of about 10-100 for $F_a \in [10^{10},10^{16}]$ GeV.   However, it would obviously be of particular interest to have large couplings to visible sector photons for observational purposes.

\section{Axion Stars Merger}
\label{sec:section3}

\subsection{Numerical Setup}
Now let us focus on the procedure that we shall follow to analyze the merging process of axion stars by numerical simulations. 
 We work in rectangular coordinates under periodic conditions defining the spatial and time domains as $[ \bold{\tilde{x}}_{\text{start}},  \bold{\tilde{x}}_{\text{end}}]$ and $[\tilde{t}_{\text{initial}},\tilde{t}_{\text{final}}]$, respectively. We discretize  the $\tilde{x}$-rectangular coordinate as $\tilde{x}_i = \tilde{x}_{\text{start}} + i \Delta \tilde{x}$\, for $i = 0, ..., N-1$. An analogous discretization  
apply for $\tilde{y}$ and $\tilde{z}$-coordinates. In addition, we discretize the time coordinate as $\tilde{t}_q = \tilde{t}_{\text{initial}} + q \Delta \tilde{t}$\,  for $q = 0, ..., M$. Here $\Delta \tilde{\bold{x}}$  and $\Delta \tilde{t}$ correspond to the dimensionless spatial and time step sizes.

Generally speaking, the Schr{\"o}dinger equation can be solved numerically by using a symmetric
split-step beam method~\cite{Poon:2006, Zhang:2008} according to 
\begin{eqnarray}
\label{formal_solution_schrowdinger}  \tilde{\psi} (\tilde{\bold{x}}, \tilde{t} + \Delta \tilde{t}) & = & e^{- i \int_{\tilde{t}}^{\tilde{t} + \Delta
  \tilde{t}}  \tilde{H} d \tilde{t}'} \tilde{\psi} (\tilde{\bold{x}}, \tilde{t})\,, \nonumber\\
  & \simeq & e^{- i \tilde{H} \Delta \tilde{t}} \tilde{\psi} (\tilde{\bold{x}}, \tilde{t})\,,
  \nonumber\\
  & \simeq & e^{- i \left( - \frac{\tilde{\nabla}^2}{2} + \tilde{V} \right) \Delta \tilde{t} }
  \tilde{\psi} (\tilde{\bold{x}}, \tilde{t})\,,
\end{eqnarray}
where in the second line we have taken the integrand to be approximately constant for enough small $\Delta \tilde{t}$. Here $\tilde{\nabla}^2$ is the Laplacian operator with respect to the dimensionless coordinates $\tilde{x}, \tilde{y}$, and  $\tilde{z}$ and $\tilde{H} = (-\tilde{\nabla}^2/2 + \tilde{V})$ as usual. Since we do not expect that 
$\tilde{\nabla}^2$ and $\tilde{V}$ operators commute each other, we have $\text{exp}[-i(-\tilde{\nabla}^2/2 + \tilde{V})\Delta \tilde{t}] \neq \text{exp}[i\tilde{\nabla}^2\Delta \tilde{t}/2] \text{exp}[-i\tilde{V}\Delta \tilde{t}]$. However, one can prove that
\begin{eqnarray}
\label{symmetric_split_beam}
  e^{- i \left( - \frac{\tilde{\nabla}^2}{2} + \tilde{V} \right) \Delta \tilde{t} } \tilde{\psi}(\tilde{\bold{x}},\tilde{t})& \simeq &
  e^{- i \tilde{V} \frac{\Delta \tilde{t}}{2}} e^{i \frac{\tilde{\nabla}^2}{2} \Delta
  \tilde{t}} e^{- i \tilde{V}  \frac{\Delta \tilde{t}}{2}}\tilde{\psi}(\tilde{\bold{x}},\tilde{t})\,, 
\end{eqnarray}
where the leading order term for the error is proportional to $(\Delta \tilde{t})^3$  (see Appendix \ref{AppA} for details). The action of  $\text{exp}(i\tilde{\nabla}^2 \Delta \tilde{t}/2)$ over $\text{exp}(-i\tilde{V}\Delta \tilde{t}/2) \tilde{\psi}(\tilde{\bold{x}},\tilde{t})$ can be worked out in the momentum space according to~\cite{Poon:2006}
\begin{eqnarray}
  e^{i \frac{\tilde{\nabla}^2}{2} \Delta
  \tilde{t}} e^{- i \tilde{V} \frac{\Delta \tilde{t}}{2}}\tilde{\psi}(\tilde{\bold{x}},\tilde{t}) & = &  \mathcal{F}^{- 1} \left( e^{- i \frac{\tilde{k}^2}{2} \Delta
  \tilde{t}} \mathcal{F} \left( e^{- i \tilde{V} \frac{\Delta \tilde{t}}{2}} \tilde{\psi}
  (\tilde{x}, \tilde{t})\right) \right)\,, 
\end{eqnarray}
where $\tilde{k}^2 = \tilde{k}_{\tilde{x}}^2 + \tilde{k}_{\tilde{y}}^2  + \tilde{k}_{\tilde{z}}^2 $ and $\mathcal{F} (\mathcal{F}^{-1})$ is the Fourier (inverse) transform. Thus, we can express the dimensionless axion field at the $(\tilde{t} + \Delta \tilde{t})$ time as
\begin{eqnarray}
\tilde{\psi}  (\tilde{\bold{x}}, \tilde{t} + \Delta \tilde{t}) & \simeq &   e^{- i \tilde{V} \frac{\Delta \tilde{t}}{2}}  \mathcal{F}^{- 1} \left( e^{- i \frac{\tilde{k}^2}{2} \Delta
  \tilde{t}} \mathcal{F} ( e^{- i \tilde{V} \frac{\Delta \tilde{t}}{2}} \tilde{\psi}
  (\tilde{x}, \tilde{t})) \right)\,.
\label{psisol}
\end{eqnarray}
For the particular case $\tilde{V}(\tilde{\bold{x}},\tilde{t})=\tilde{\phi}_N(\tilde{\psi}(\tilde{\bold{x}},\tilde{t})) - |\tilde{\psi}|^2(\tilde{\bold{x}},\tilde{t})/8$, Eq.~(\ref{psisol}) corresponds to the solution of (\ref{dschrodingernodim}). Here the solution for the Newtonian potential at a time $(\tilde{t} + \Delta\tilde{t})$ is obtained by solving Eq.~(\ref{dpoissionnodim}) in the momentum space.
Take $\tilde{t}$ fix and consider the three-dimensional Fourier transform of the Newtonian potential and the local number density $\tilde{n}(\tilde{\bold{x}},\tilde{t})=|\tilde{\psi}(\tilde{\bold{x}},\tilde{t})|^2$ as
 \begin{align}
\tilde{\phi}(\tilde{\bold{x}},\tilde{t}) & = \frac{1}{(2\pi)^{3/2}} \int \tilde{\phi}_{\tilde{\bold{k}}}(\tilde{t}) e^{i\,\tilde{\bold{k}}\cdot\tilde{\bold{x}}} d^3\tilde{x}\,,\\
\tilde{n}(\tilde{\bold{x}},\tilde{t}) & = \frac{1}{(2\pi)^{3/2}} \int \tilde{n}_{\tilde{\bold{k}}}(\tilde{t}) e^{i\,\tilde{\bold{k}}\cdot\tilde{\bold{x}}} d^3\tilde{x}\,,\label{nk}
\end{align}
and insert them into Eq.~(\ref{dpoissionnodim}) to obtain
\begin{align}
\int  \left( \tilde{\phi}_{\tilde{\bold{k}}}(\tilde{t}) \tilde{k}^2 + 4\pi \tilde{n}_{\tilde{\bold{k}}}(\tilde{t})  \right) e^{i~\tilde{\bold{k}}\cdot\tilde{\bold{x}}} d^3\tilde{k}&= 0\,,\\
\tilde{\phi}(\bold{\tilde{x}},\tilde{t}) = \mathcal{F}^{-1}\left( 4\pi \tilde{\nabla}^{-2}  \tilde{n}_{\tilde{\bold{k}}}(\tilde{t})  \right)& = \mathcal{F}^{-1}\left( -\frac{4\pi  \tilde{n}_{\tilde{\bold{k}}}(\tilde{t}) }{\tilde{k}^2}  \right)\,.\label{lap}
\end{align}
We use a discrete Fourier transform for Eq.~(\ref{nk}) as
\begin{equation}
\tilde{n}_{\tilde{k}_{\tilde{x}},\tilde{k}_{\tilde{y}},\tilde{k}_{\tilde{z}}} (\tilde{t})= \sum^{N-1}_{m=0} \sum^{N-1}_{l=0} \sum^{N-1}_{n=0} \tilde{n}(\tilde{x}_m,\tilde{y}_l,\tilde{z}_{\tilde{n}},\tilde{t}) e^{-i\frac{2\pi}{N}\left( \tilde{k}_{\tilde{x}}m +  \tilde{k}_{\tilde{y}}l + \tilde{k}_{\tilde{z}}n \right)}\,,
\end{equation}
and since the Newtonian potential in Eq.~(\ref{dpoissionnodim}) is real, we have  
\begin{equation}
\tilde{\phi}_{\tilde{k}_{\tilde{x}},\tilde{k}_{\tilde{y}},\tilde{k}_{\tilde{z}}} (\tilde{t}) = \tilde{\phi}_{N-\tilde{k}_{\tilde{x}},N-\tilde{k}_{\tilde{y}},N-\tilde{k}_{\tilde{z}}} (\tilde{t})\,.
\end{equation}
The Laplacian operator is taken to be
\begin{equation}
\tilde{\nabla}^{2}_{m,l,n} = \frac{e^{- \frac{i 2 \pi}{N} m} +
  e^{\frac{i 2 \pi}{N} m} - 2}{\Delta \tilde{x}^2} + \frac{e^{- \frac{i 2 \pi}{N} l} +
  e^{\frac{i 2 \pi}{N} l} - 2}{\Delta \tilde{y}^2} + \frac{e^{- \frac{i 2 \pi}{N} n} +
  e^{\frac{i 2 \pi}{N} n} - 2}{\Delta \tilde{z}^2}\,,  
\end{equation}
where we recover the rightmost expression in Eq.~(\ref{lap}) when $N \rightarrow \infty$. At that limit, we have $\tilde{\nabla}^{2}_{m,l,n} = -(2\pi m/L)^2 -  (2\pi n/L)^2  -  (2\pi n/L)^2 $ for $L \equiv \Delta \tilde{x} N$ (see Appendix \ref{AppB} for further details).

Since we operate in a finite size box, we need to absorb outgoing modes at boundaries to avoid unphysical reflection back to the central region of the grid where the merging process takes place. We implement a sponge at the boundaries through adding an imaginary potential such that
\begin{equation}
V_{\text{sponge}} = - i\frac{V_0}{2}\left[  2 + \frac{\text{tanh}(\tilde{r} - \tilde{r}_{\text{sponge}})}{\delta}-\text{tanh}(\tilde{r}_{\text{sponge}}/\delta) \right]\,,
\label{sponge}
\end{equation}
where $\tilde{r}^2 = \tilde{x}_i^2 + \tilde{y}_j^2 + \tilde{z}_k^2$ is the radius of a given point in the discrete spatial domain. Note that this potential is a smooth version of a step function with amplitude $V_0$, where $\tilde{r}_{\text{sponge}}$ and $\delta$ are the center and the width, respectively.  The imaginary potential actually behaves as a sink of outgoing particles.  In our numerical simulations, we choose $V_0=10^4$, $\tilde{r}_{\text{sponge}}=N/2$, and $\delta=0.5$.

We choose the spatial and temporal step sizes as well as the size of the cosmological box to obtain enough resolution to accurately analyze the merging process and avoid loss of particles absorbed by the spherical sponge which are still bounded to the whole system. Typical values taken by us are a cosmological box with a volume $(384)^3$ in dimensionless units, a temporal step size $\Delta \tilde{t} \simeq 0.08$, and spatial step sizes $\Delta \tilde{x} = \Delta \tilde{y} = \Delta \tilde{z} \simeq 0.08$.

\begin{figure}[ht!]
\includegraphics[scale=0.214]{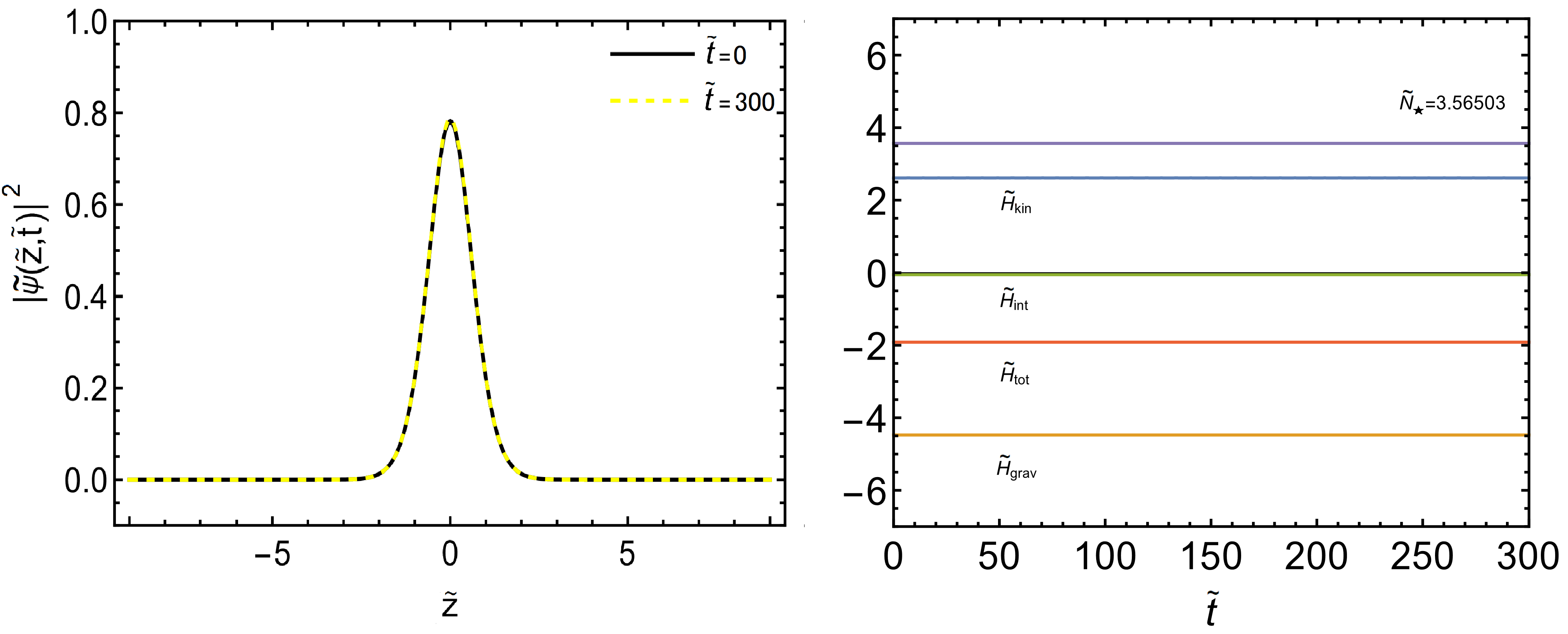}\!\!\!\!\!
\caption{(Left) Snapshots of the dimensionless local number density along the $\tilde{z}$-direction with $(\tilde{x},\tilde{y})\simeq (0,0)$ at times $\tilde{t} =0$ (bue solid line) and $\tilde{t} =300$ (yellow dashed line) for the ground state configuration with $\tilde{N}_{\star}=3.56503$. As a initial condition we use $\tilde{\psi}(\bold{\tilde{x}})_{\text{initial}}$ from (\ref{psii}) with the corresponding radial profile  shown in Fig. 1 and $\tilde{v}_{\tilde{z}} = 0$. (Right) Corresponding time evolution of the dimensionless number of particles $\tilde{N}_{\star}$ (purple line), total Hamiltonian $\tilde{H}_{\text{total}}$ (red line), $\tilde{H}_{\text{kin}}$ (blue line),  $\tilde{H}_{\text{grav}}$ (orange line), and  $\tilde{H}_{\text{int}}$ (green line) for the ground state configuration shown in the left panel. This shows that the code is faithfully preserving the expected conserved quantities in the system.}
\label{ICa}
\end{figure}

We check the numerical stability of our code by evolving in time a ground state configuration placed at the origin. Figure~\ref{ICa} (left) shows two snapshots of the dimensionless local number density  $\tilde{n}(\tilde{{\bf{x}}},\tilde{t}) = |\psi(\tilde{{\bf{x}}},\tilde{t})|^2$ along the $\tilde{z}$-direction with $(\tilde{x},\tilde{y})\simeq (0,0)$ of a stable ground state configuration with $ \tilde{N}_{\star} = 3.56503$ that we showed earlier in Fig.~\ref{IC}. 
 While the black solid line refers to the initial time, the yellow dashed line refers to the time $\tilde{t}=300$. Figure~\ref{ICa} (right) shows the corresponding evolution of the dimensionless number of particles and Hamiltonian. As expected, the shape of the square of the field norm  keeps unchanged as times goes on as well as the dimensionless number of particles and energy of the system. Indeed, the change rate of the dimensionless number of particles is negligible at all time during the simulation. In particular, $d\tilde{N}_{\star}/d\tilde{t} \sim \mathcal{O}(10^{-12})$ at  $\tilde{t}=300$. Thus our code is faithfully preserving the conserved quantities in the system accurately. 
 Note that the conservation of energy is non-trivial because the evolution operator used is no longer exactly the Hamiltonian, and furthermore, the conservation of energy and particle number is non-trivial because of the absorbing boundary condition which breaks unitarity.

In addition, we check our program code by running the time evolution of one clump using as a initial condition at $\tilde{t}=0$ 
\begin{equation}
\tilde{\psi}(\tilde{\bold{x}})_{\text{initial}} = \tilde{\psi}(\tilde{r}) e^{i\tilde{v}_{\tilde{z}}\tilde{z}}\,,\,\,\,\,\,\,\text{with}\,\,\,\,\tilde{v}_{\tilde{z}} = (\mpl \gamma^{1/2}/F_a)\,v_z\,,
\label{psii}
\end{equation}
where $\tilde{v}_{\tilde{z}}$ is the dimensionless clump velocity in the $\tilde{z}$-direction. Note that  $p_{z} z = m_{\phi}v_z z = \tilde{v}_{\tilde{z}} \tilde{z}$, where $p_z$ is the associated linear momentum.  
Here $\tilde{\psi}(\tilde{r})$ corresponds to the radial profile of a ground state (stable solution)  with $\tilde{N}_{\star} = 3.56503$ or $\tilde{N}_{\star} = 4.55418$ for the ground state configurations that we showed earlier in Fig.~\ref{IC}. Here  $\tilde{\psi}_{\text{ground-state}} =  \tilde{\psi}(\tilde{r}) \text{exp}(-i \tilde{\mu} \tilde{t})$ where $\tilde{\mu}$ is the (dimensionless) eigenfrequency as was mentioned in Sec.~\ref{sec:section2}.

\begin{figure}[ht!]
\includegraphics[scale=0.3]{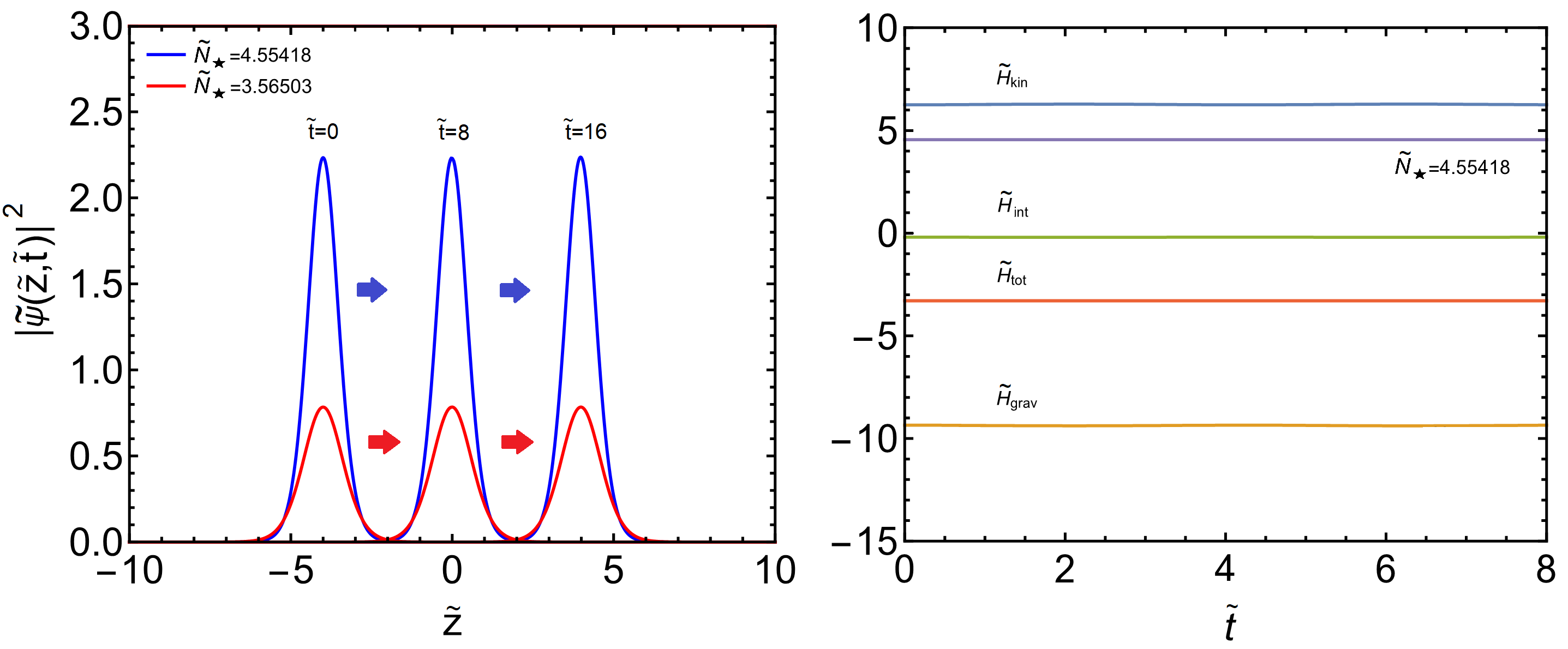}\!\!\!\!\!\!
\caption{(Left) Snapshots of the dimensionless local number density along the $\tilde{z}$-direction with $(\tilde{x},\tilde{y}) \simeq (0,0)$ at times $\tilde{t} =0, 8, 16$ for the ground state configuration with $\tilde{N}_{\star}=4.55418$ (blue solid line) and $\tilde{N}_{\star}=3.56503$ (red solid line). As a initial condition we use $\tilde{\psi}(\bold{\tilde{x}})_{\text{initial}}$ from (\ref{psii}) with the corresponding radial profile  shown in Fig. 1 and $\tilde{v}_{\tilde{z}} = 0.5$. 
(Right) Corresponding time evolution of the dimensionless number of particles $\tilde{N}$ (purple line), total Hamiltonian $\tilde{H}_{\text{total}}$ (red line), $\tilde{H}_{\text{kin}}$ (blue line),  $\tilde{H}_{\text{grav}}$ (orange line), and  $\tilde{H}_{\text{int}}$ (green line) for the ground state configuration shown in blue line in the left panel. This system is Galilean boosted compared to the type of system considered in Fig.~\ref{ICa} and so it again shows that the code is faithfully preserving the expected conserved quantities in the system.}
\label{IC2}
\end{figure}

 We take $\tilde{v}_{\tilde{z}} = 0.5$ and run the simulations.  Figure~\ref{IC2} (left) shows three snapshots of the dimensionless local number density along the $\tilde{z}$-direction with $(\tilde{x},\tilde{y})\simeq(0,0)$ at the times  $\tilde{t}=0, 8, 16$ of both configurations. Figure~\ref{IC2} (right)
shows the corresponding time evolution for the dimensionless Hamiltonian for the case $\tilde{N}_{\star}=4.55418$. As we expect, both clump solutions just travel in the $\tilde{z}$-direction without changing in shape and energy (and total number of particles).   

\subsection{Head-on collision between two ground state axion stars }
\label{head}
Here we study the head-on collision between two ground state axion stars. Since we are mainly interested in collision between clumps of the same (or similar) number of particles, we focus on the special case $\tilde{N}_{\star,1} = \tilde{N}_{\star,2}$, where $(\tilde{N}_{\star, i})^{i=1,2}$ refers to the initial number of particles of each clump.  The generalization for the initial wave function in Eq.~(\ref{psii}) for the case of two clumps traveling towards each other reads as
\begin{equation}
\tilde{\psi}(\tilde{\bold{x}})_{\text{initial}} = \tilde{\psi}(\sqrt{\tilde{x}^2+\tilde{y}^2+(\tilde{z}+\tilde{z_0})^2}) e^{i\tilde{v}_{\tilde{z}}\tilde{z}}+\tilde{\psi}(\sqrt{\tilde{x}^2+\tilde{y}^2+(\tilde{z}-\tilde{z_0})^2}) e^{-i\tilde{v}_{\tilde{z}}\tilde{z}} \,,
\label{twoc}
\end{equation}
where $2\tilde{z}_0$ is the distance between the respective center of mass (COM) of the two clumps and $\tilde{v}_{\tilde{z}}$ is the magnitude of the initial velocity of the clumps in the $\tilde{z}$-direction. Equation~(\ref{twoc}) is the sum of the wave functions of each clump. Even though we expect these wave functions hold uncorrelated complex phases between them, for now we will only consider a null phase difference (and explore the phase dependence shortly). In this scenario, for a fixed initial distance between clumps, they will merge or pass through each other depending on the  initial total energy of the system, $\tilde{H}_{\text{tot}}^{\text{initial}}$.   

\begin{figure}[t!]
\includegraphics[scale=0.29]{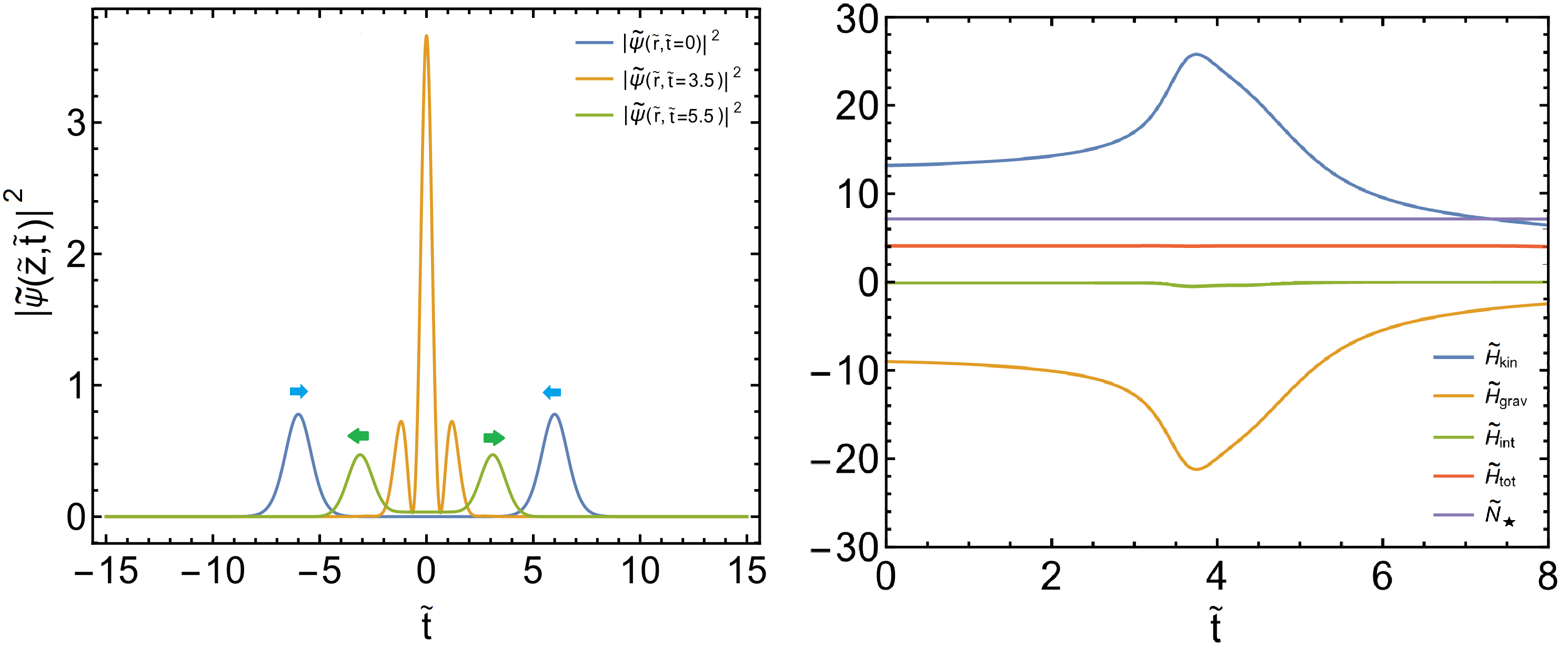}\!\!\!\!\!\!\!\!\!\!\!\!\!
\caption{Head-on collision of two clumps each with $\tilde{N}_{\star} = 3.5650$. Here $\tilde{v}_{\tilde{z}} = 1.5$ in Eq.~(\ref{twoc}). (Left) Dimensionless local number density along the $\tilde{z}$-direction with $(\tilde{x},\tilde{y})\simeq(0,0)$  at times: $\tilde{t}=0$, $\tilde{t}=3.5$, and $\tilde{t}=5.5$. (Right)  Time evolution of the dimensionless total energy (and their components) from the collision of the two clumps. The traveling clumps pass through each other without merging.}
\label{TCP}
\end{figure}

We take $\tilde{v}_{\tilde{z}}=1.5$, $\tilde{z_0} = 6$, and $\tilde{N}_{\star,1} = \tilde{N}_{\star,2}  =  3.56503$ as initial conditions in Eq.~(\ref{twoc}).  Figure \ref{TCP} (right) shows  three snapshots at different times of the dimensionless local number density.  Clumps approach and pass through each other without a final merge. Note that the  total energy of the system is constant, which shows the code is working well. And importantly, the total energy is positive, so this is expected to be an unbounded system, which is consistent with the results of Fig.~\ref{TCP} (left).
 
\begin{figure}[t!]
\centering
\includegraphics[width=\columnwidth]{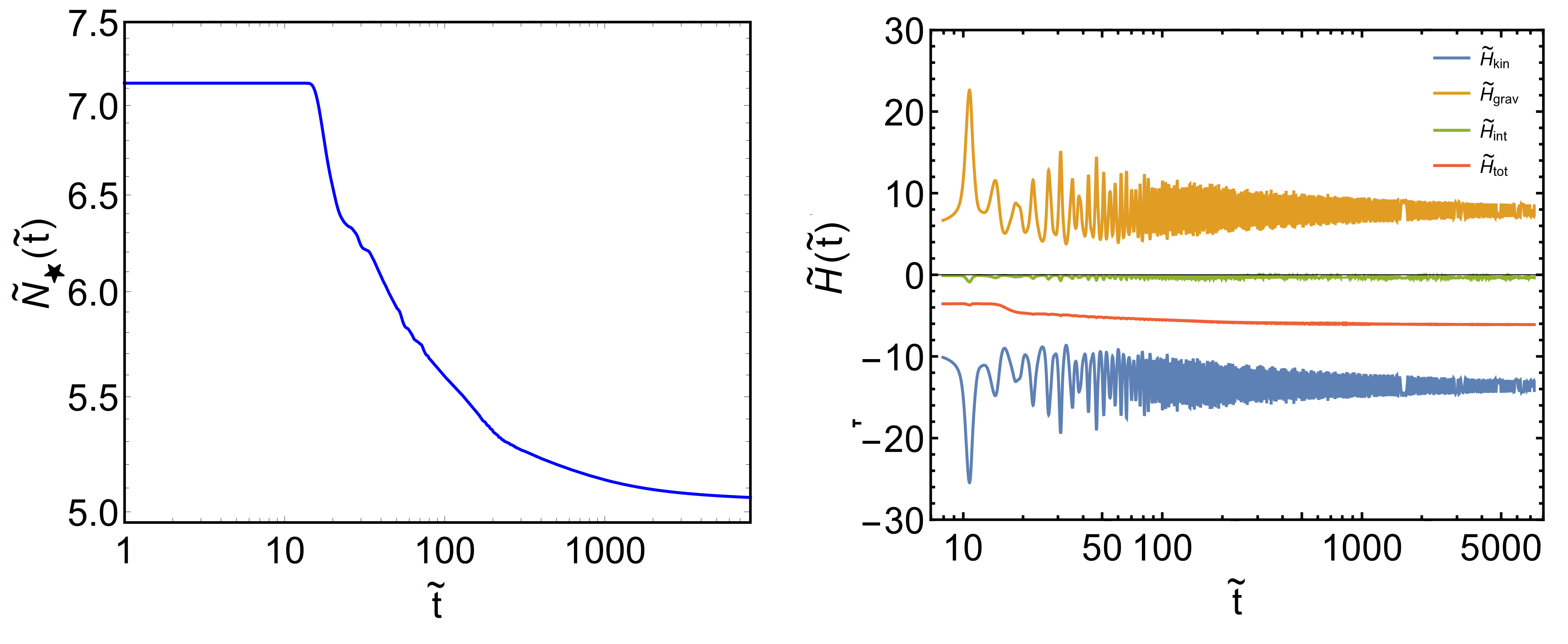}
\caption{Head-on collision of two clumps each of one with $\tilde{N}_{\star} = 3.56503$. Here $\tilde{v}_{\tilde{z}} = 0.3$ and $\tilde{z}_0 = 6$ in (\ref{twoc}). We have used a cosmological box with a volume $(384)^3$ in dimensionless units and a temporal and spatial sizes equal to $\Delta \tilde{t}=0.082$ and $(\Delta\tilde{x},\Delta\tilde{y},\Delta\tilde{z}) \simeq 0.078$, respectively. 
(Left) Evolution of the dimensionless number of particles of the system during the whole simulation. 
(Right)  Evolution of the total energy of the system and their different components ($\tilde{H}_{\text{kin}}$, $\tilde{H}_{\text{grav}}$, and $\tilde{H}_{\text{int}}$). 
The reduction in total number and total energy over time is due to emission of scalar waves that go into absorbing boundary conditions.}
\label{TCM}
\end{figure}

To analyze the merger of axion stars in their ground state configurations, we repeat the same initial conditions as before but we decrease the initial velocity as $\tilde{v}_{\tilde{z}}=0.3$. Figure \ref{TCM} (left) shows the temporal evolution of the dimensionless total number of particles of the system.  At the beginning of the merger process around $\tilde{t} ~\sim (15-25)$ a violent relaxation happens and a large number of particles escape. The rate for the loss of particles reach values about $d\tilde{N}_{\star}/d\tilde{t} \sim (10^{-2}-10^{-1})$. After that, the resultant clump keeps losing particles but with a lower rate. The particles which escape from the central clump are absorbed by the sponge when they reach the boundary. Number of particles approach to the asymptotically value $\tilde{N}_{\star} \sim 5$ at around $\tilde{t}~\sim 8 \times 10^{3}$, which is equivalent to (recall Eq.~(\ref{xtrescale}))
\begin{equation}
t \sim 2\, \text{yrs} \left( \frac{\gamma}{0.3} \right) \left( \frac{10^{-5}\,\text{eV}}{m_{\phi}} \right) \left( \frac{6\times10^{11}\,\text{GeV}}{F_a} \right)^2\,.
\end{equation}
(later we will explain that it will be interesting to also consider large $F_a\sim 10^{16}$\,eV, or so, so this time scale can be relatively short, $t\sim1$\,hour). At that time, the ejecting rate for particles is quite small: $d\tilde{N}_{\star}/d\tilde{t}\sim \mathcal{O}(10^{-7})$. 
Figure \ref{TCM} (right) shows the temporal evolution of the total energy of the system and its different components. The negative initial energy (red line) indicates a bounded system and so one can expect the merger to occur \cite{Guzman:2018evm}. All components of the energy show an oscillatory behavior which tend to stabilize as the resultant clump approaches to its ground state configuration. Note that the total energy of the system (red line) decreases as a result of the ejection of particles during the merger process and their subsequent destruction after hitting the sponge at boundaries.  

\begin{figure}[t!]
\centering
\includegraphics[width=\columnwidth]{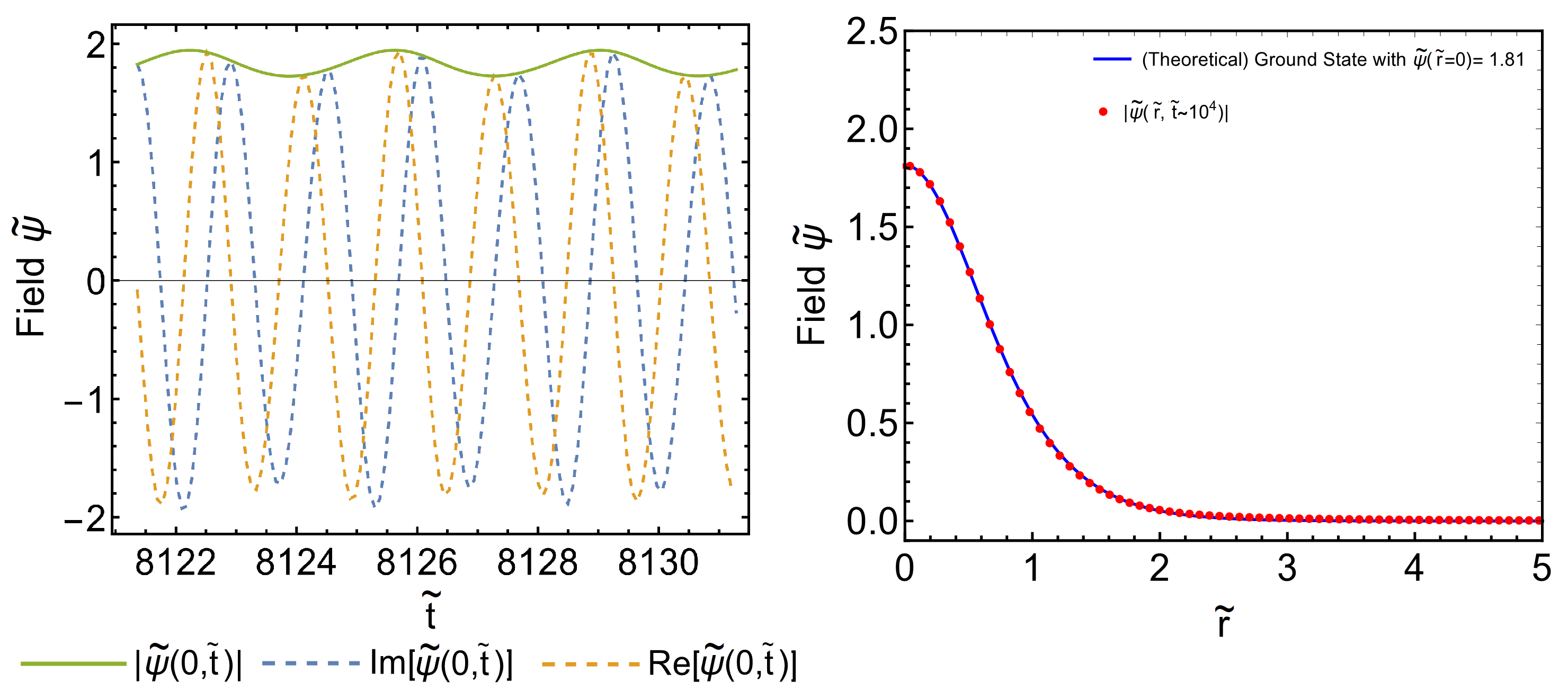}
\caption{(Left) Value of the field at late times after the merger of  two (stable) ground state clumps as detailed in Fig.~\ref{TCM}. (Left) The field value at the origin $(\tilde{x},\tilde{y},\tilde{z})= (0,0,0)$; green solid curve is absolute value of field, dashed blue curve is imaginary part of field, and dashed orange curve is real part of field. (Right) Red points are absolute value of the field in the $\tilde{z}$-direction with $(\tilde{x},\tilde{y})= (0,0)$ of the resultant clump at $\tilde{t} = 8121$. Blue solid curve is theoretical (stable) ground state configuration obtained by solving Eqs.~(\ref{dschrodingernodim}, \ref{dpoissionnodim}) with $\tilde{\Psi}(\tilde{r}=0)=1.814$ in Eq.~(\ref{gs}).}
\label{match}
\end{figure}

We stop our simulation at $\tilde{t} \sim 8 \times 10^3$,  when the final clump seems to stabilize. At lates times the absolute value of the field at the center of mass of the clump oscillates around a central value of $\tilde{\psi}(\tilde{r}=0) \simeq 1.8$; this is shown in Fig.~\ref{match} (left). To confirm that the resultant clump is close to a ground state configuration (stable solution), we compare this solution with the theoretical solution obtained by solving the pair of Eqs.~(\ref{dschrodingernodim}, \ref{dpoissionnodim}) using a stationary solution $\tilde{\psi}(\tilde{r},\tilde{t}) = \tilde{\Psi}(\tilde{r})e^{-i\tilde{\mu} \tilde{t}}$ with a central value for the field $\tilde{\Psi}(\tilde{r}=0)=1.814$ in Eq.~(\ref{gs}). Fig.~\ref{match} (right) shows the radial profile of the resultant clump (red points) at $\tilde{t}=8121$ and the theoretical numerical solution for a ground state with the same central value for the field (blue line). We see that both solutions agree extremely well. In addition, Table 1 shows values for the energy components of the theoretical ground state BEC and the corresponding values for the resultant clump at the end of the simulation. The percentage relative error for the total energy and its components are quite small, indicating that the resultant clump is close to a ground state axion clump, and we expect that it will approach closer to this solution over time due to slow scalar wave emission.

A very important quantity for the phenomenology that we will discuss in the next section is the ratio between the final number of particles of the resultant clump and the initial number of particles. Numerically, we find it to be
\begin{equation}
N^{\star}_{\text{final}} \simeq 0.7(N_{\star,1} + N_{\star,2})
\end{equation}
where $\tilde{N}_{\star, 1}=\tilde{N}_{\star, 2}=3.56503$. Note that since $M_\star=m_\phi\,N_\star$, this relation also applies to the final mass; as reported in the abstract. 
In words, the final clump is approximately formed by $70\%$ of the total initial number of particles of the original colliding clumps, whereas the remaining $30\%$ is radiated off by scalar wave emission. This is consistent with the work of Ref.~\cite{Schwabe:2016rze}. 

\begin{table}[t]
\begin{center}
\begin{tabular}{ l | l | l | l | l}
   & $\tilde{H}_{\text{kin}}$ & $\tilde{H}_{\text{grav}}$ & $\tilde{H}_{\text{int}}$ & $\tilde{H}_{\text{total}}$ \\
\hline
\hline
\vspace{0.04 cm}
\text{Stable Ground State  ($\tilde{\Psi}(\tilde{r}=0)$ = 1.814})  & 7.5356 & -14.1402 & -0.3103 & -6.9149\\
  $|\tilde{\psi}(\tilde{r},\tilde{t}\simeq8\times10^3)|$ & 7.4980 & -13.8390 & -0.3085 & -6.6495 \\
Percentage Relative Error ($\xi$)   & 0.5$\%$  & 2.1$\%$ & 0.6$\%$ &  3.8$\%$\\
\end{tabular}
\caption{Total energy (and their components) of a theoretical stable ground state configuration with $\tilde{\Psi}(\tilde{r})=1.814$ and the resultant clump at $\tilde{t}\simeq 8\times 10^3$ generated by merging two (stable) ground state clumps with $\tilde{N}=3.56503$ and $\tilde{v}_{\tilde{z}} = 0.3$.}
\end{center}
\label{table:info}
\end{table}

Since we expect a dependence on the initial total energy of the system on the final clump mass, we performed several runs with different initial energies. The final clump tends to accumulate more particles from the progenitor's clumps as the magnitude of the total initial energy increases, however this tendency was found to be somewhat weak. 
So in a rather robust way, all merger processes follow essentially the same pattern: after the coalescence, the system tends to settle down to the ground state configuration by releasing an excess of particles. Similar results were reported in Ref.~\cite{Schwabe:2016rze} through numerical simulations of the merger of solitonic cores in the context of ultra-light axion dark matter halos. However, we note that in that work, the axion self-interaction was irrelevant, while we are exploring clumps whose self-interaction is relatively important (albeit marginally sub-leading to gravity). Related work in the early universe includes Ref.~\cite{Amin:2019ums}.

\subsection{Non-head-on collision between two ground state axion stars}

So far we have only discussed head-on collisions between the axion stars. The general case for collisions can be studied with 
a non-zero impact parameter $\tilde{b}$. For the initial profile considered in Eq.~(\ref{twoc}), a finite impact parameter means the vector which joins the respective center of mass of the two clumps has a non-zero component along the $(\tilde{x},\tilde{y})$-direction. We have run several simulations with different impact parameters to verify that the primary conclusions obtained for the case
of head-on collisions remains essentially the same. If the total initial energy of the system is negative, the two clumps merge leading to a resulting clump, unless the impact parameter is quite large and the clumps completely miss each other. 

\begin{figure}[ht!]
\centering
\includegraphics[scale=0.18]{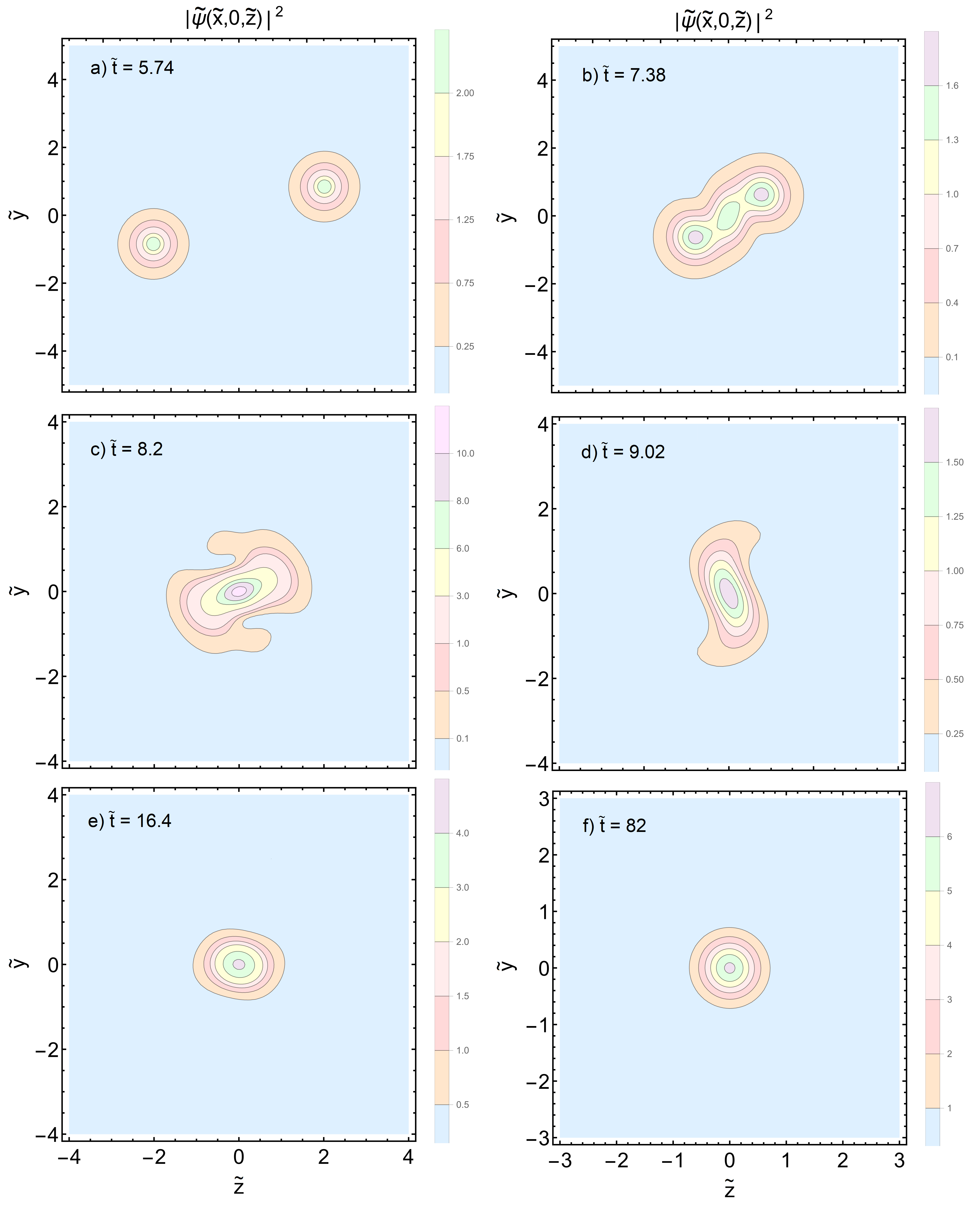}
\caption{Non-head-on collision between two clumps that are originally in their ground state configurations; both with a number of particles $\tilde{N}_{\star} = 4.55418$. The center of mass of the two clumps are initially separated  by a distance equal to $\sqrt{(2\tilde{z}_0)^2 + (2\tilde{x}_0)^2} = \sqrt{12^2+2^2}$ and have an initial velocity in the $\tilde{z}$-direction equal to $\tilde{v}_{\tilde{z}}=0.5$.}
\label{impact}
\end{figure}

Figure~\ref{impact} shows countour levels of the local number density $|\tilde{\psi}(\tilde{x},0,\tilde{z})|^2$ at different times  for the non-head-on collision of two identical ground state configurations with number of particles $\tilde{N}=4.55418, \tilde{v}_{\tilde{z}}=0.5, \tilde{z}_0 = 6$, and an  impact parameter $\tilde{b}=2$ along the $\tilde{x}$-direction. We see that as clumps approach they begin to interact with each other in a rather complicated way. The coalescence process takes a longer time in comparison to the head-on collision case. We ran simulations with different impact parameters to conclude that as the impact parameter increases, clumps take a longer time to merge after undergoing an inspiral motion. However, after they finally merge the process at which the resultant clump begins to settle down to the ground state configuration by releasing the excess of particles occurs in a similar way to those for the head-on collision case. 

\subsection{Parameter Space for Merger}
As we already mentioned, if the initial total energy of the system composed by two colliding clumps is negative, i.e., $H_{\text{tot}}^{\text{initial}} < 0$, clumps will merge leading to a resulting clump. If the initial separation between clumps is large compared to the size of each of the two clumps radii, we can estimate the total energy of the system as follows: the sum of the individual energies of each configuration, i.e.,  $(H_{\text{nr},i})^{i=1,2} = (H_{\text{kin},i} + H_{\text{grav},i}+ H_{\text{int},i})^{i=1,2}$, plus the additional kinetic  ($(H^{\text{cm}}_{\text{kin},i})^{i=1,2}$) and gravitational energy ($H^{\star-\star}_{\text{grav}}$) parts associated with the kinetic energy of  the center of mass of each clump and the gravitational attraction between clumps treated as point masses in this regime (for other work see Ref.~\cite{Cotner:2016aaq}). In detail, these new extra pieces for the clumps energy reads as
\begin{align}
H^{\text{cm}}_{\text{kin},i} = \frac{M_{\star,i} v_{\text{cm},i}^2}{2} = \left( \frac{F_a^3}{\mpl m_{\phi} \gamma^{3/2}} \right) \frac{\tilde{N}_{\star,i} \tilde{v}_{\text{cm},i}^2}{2}\,,\label{kincm}\\
H_{\text{grav}}^{\star-\star} = -\frac{G_N M_{\star,1}M_{\star,2}}{d} = - \left( \frac{F_a^3}{\mpl m_{\phi} \gamma^{3/2}} \right)\frac{\tilde{N}_{\star, 1}\tilde{N}_{\star, 2}}{\tilde{d}}\,,\label{gravstarstar}
\end{align} 
where $d$ is the initial distance between the center of mass of each clump and $(v_{\text{cm},i})^{i=1,2}$ is the center of mass velocity of each clump. The corresponding dimensionless energies associated with Eqs.~(\ref{kincm}, \ref{gravstarstar}) can be readily defined as $\tilde{H}_{\text{kin},i}^{\text{cm}} =\tilde{N}_{\star,i} \tilde{v}_{\text{cm},i}^2/2$ and $\tilde{H}_{\text{grav}}^{\star-\star} = -\tilde{N}_{\star, 1} \tilde{N}_{\star, 2} /\tilde{d}$. Since we are mainly interested in the case of collisions between two clumps with $\tilde{N}_1 \simeq \tilde{N}_2$, the initial energy of the system is just given by
\begin{equation}
H_{\text{tot}}^{\text{initial}} \simeq 2 H_{\text{kin}} + 2 H_{\text{grav}} + 2 H_{\text{int}} + 2 H_{\text{kin}}^{\text{cm}}+ H_{\text{grav}}^{\star-\star}\,.\label{Einitial}
\end{equation}
By simplicity, let us assume the head-on collision of two (stable) ground state axion stars with $(\tilde{N}_{\star,1}, \tilde{N}_{\star,2})\simeq \tilde{N}_{\star}$. Both clumps are traveling towards each other in the $\tilde{z}$-direction with equal but opposite velocities. As was explained in Sec.~\ref{ADMC}, the sech ansatz for the radial profile of an axion clump is a good approximation for the ground state configuration. Thus, going to the dimensionless variables, we can reexpress Eq.~(\ref{Einitial}) in terms of this approximation as 
\begin{equation}
\tilde{H} \simeq \frac{2 a b^2 \tilde{N}_{\star}^3}{(a+\sqrt{a^2-3bc\tilde{N}_{\star}^2})^2}-\frac{2 b^2\tilde{N}_{\star}^3}{a+\sqrt{a^2-3bc\tilde{N}_{\star}^2}}-\frac{2 b^3c\tilde{N}_{\star}^5}{(a+\sqrt{a^2-3bc\tilde{N}_{\star}^2})^3} +\tilde{N}_{\star} \tilde{v}^2_{\tilde{z}}-\frac{\tilde{N}_{\star}^2}{2 \tilde{z}_0} \,,
\label{Ha}
\end{equation}
where the dimensionless clump velocity in the $\tilde{z}$-direction is defined through
\begin{equation}
v_z = \left(\frac{F_a}{\mpl \gamma^{1/2}}\right) \tilde{v}_{\tilde{z}} \,,
\end{equation}
coefficients $(a,b,c)$ are listed in Eq.~(\ref{abc}) and $2\tilde{z}_0$ is the distance between the center of mass of each clump. This scaling suggests that for velocities on the order of the ratio of PQ to Planck scale, then mergers are reasonable; we will return to this point shortly. 
We will consider collision between clumps with a number of particles such that $\tilde{N}_{\star} \lesssim 0.7 \tilde{N}_{\star,\text{max}}$ in order to avoid that the resulting clump overpasses the maximum mass for an stable configuration  leading to a collapse and explosion in relativistic axions~\cite{Levkov:2016rkk}.~\footnote{We have set the fraction for $N_{\star}/N_{\star, \text{max}}$ no larger than 0.7 considering the typical mass of the resulting clump. However, if the clump somehow overpasses $M_{\text{max}}$, numerical computations in Ref.~\cite{Levkov:2016rkk} shows a final remant $M_{\star} < M_{\star,\text{max}}$ after multiple cycles of collapse and explosion.}

\begin{figure}[t!]
\centering
\includegraphics[width=\columnwidth]{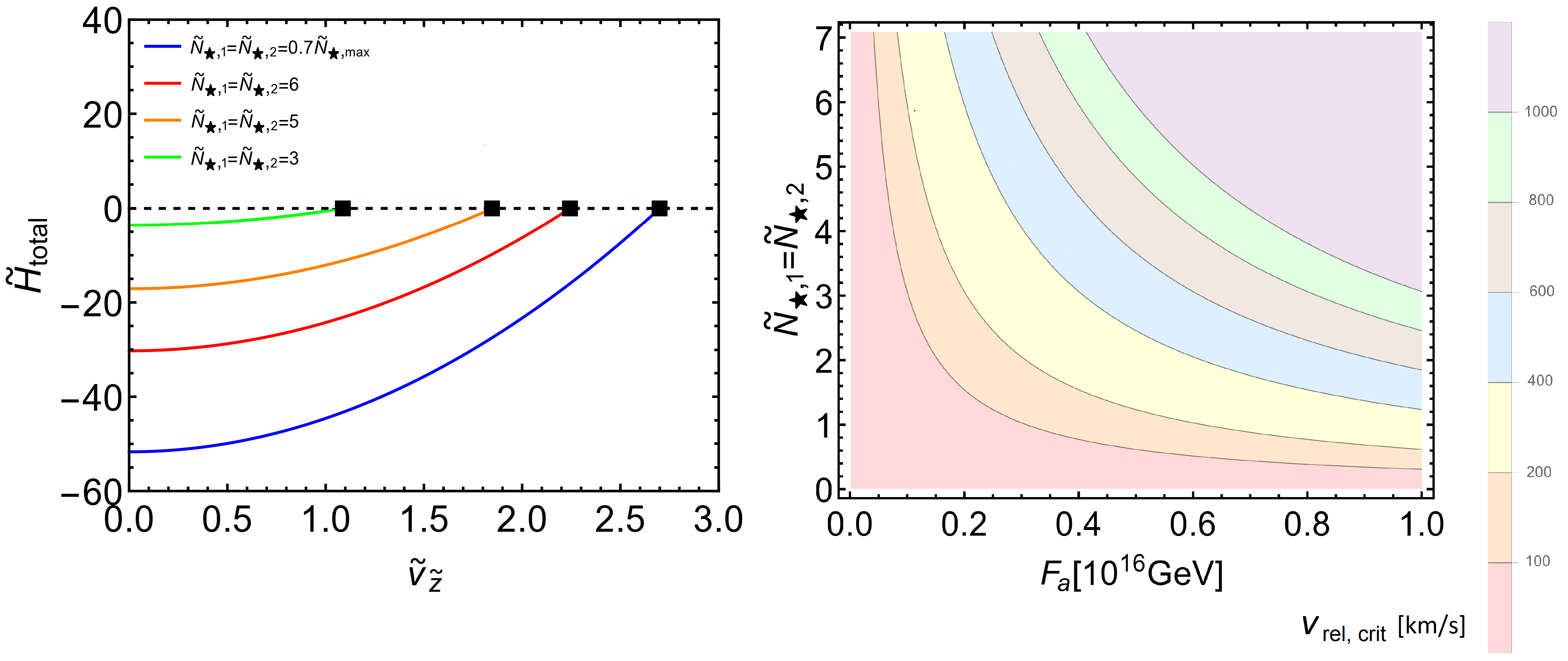}
\caption{(Left) Initial total (dimensionless) energy of the system, $\tilde{H}_{\text{total}}$, with respect to the magnitude of the maximum initial velocity of both two clumps, $\tilde{v}_{\tilde{z},\text{max}}$, to lead to a merger after a head-on collision. A sech ansatz is used to approximate the clumps radial profile. Results are shown for different initial total number of particles. In particular, green, orange, red and blue line refer to the cases $\tilde{N}_{\star,1}=\tilde{N}_{\star,2} = (3, 5, 6, 0.7\tilde{N}_{\star,\text{max}})$, respectively. The initial distance between the center of mass of the clumps is set to be $2\times8 R$, where $R$ is the clump lenght scale in Eq.~(\ref{ansatz}). (Right) Contour-level of the clumps critical relative velocity $v_{\text{rel, crit}}$ [km/s] in the parameter space $(\tilde{N}_{\star,1}=\tilde{N}_{\star,2}, F_a)$, e.g. the initial number of particles of each clump and the PQ symmetry breaking scale, respectively. The critical relative velocity for clumps is calculated using $\tilde{v}_{\tilde{z},\text{crit}} = \tilde{v}_{\tilde{z},\text{crit}}(\tilde{N}_{\star,1}=\tilde{N}_{\star,2})$ from the plot on the left and Eq.~\ref{vrelcrit}.  }
\label{ps}
\end{figure}

Figure~\ref{ps} (left) shows the evolution of the initial total (dimensionless) energy of the system for two identical clumps, Eq.~(\ref{Ha}), with respect to the magnitude of the clumps velocity $\tilde{v}_{\tilde{z}}$. Results are shown for different initial total number of particles. In detail, green, orange, red, and blue lines refer to the cases $\tilde{N}_{\star,1}=\tilde{N}_{\star,2}=(3, 5, 6, 0.7 \tilde{N}_{\star,\text{max}})$, respectively. In Eq.~(\ref{Ha}),  we have set the initial distance between clumps to be $2\tilde{z}_0 = 2 \times 8 R$, where   $R$ is the length scale of each clump as shown in Eq.~(\ref{ansatz}) . 
Note that this distance is about 11 times larger than the geometrical mean of the length scales, e.g. $\sqrt{2}R$. At that distance, numerical calculations show the Newtonian Hamiltonian in Eq.~(\ref{Ha}) is a reasonable approximation for the initial total energy of the system. We define the critical initial velocity of clumps, $\tilde{v}_{\tilde{z},\text{crit}}$, as the velocity at which the initial total energy of the system vanishes. Above this velocity the system is no longer bounded and the head-on collision will not lead to a final merger. We have marked in Fig.~\ref{ps} (left) with a square the critical velocity for different initial number of particles. The larger the initial number of particles, the larger the critical velocity because $\tilde{H}^{\text{cm}}_{\text{kin},i}$ depends only on the number of particles to the first power.  In Fig.~\ref{ps} (right) we show the contour-level of the critical relative velocity of the clumps in the parameter space $(\tilde{N}_{\star},F_a)$. This velocity is calculated using $\tilde{v}_{\tilde{z},\text{crit}} = \tilde{v}_{\tilde{z},\text{crit}}(\tilde{N}_{\star})$ from  Fig.~\ref{ps} (left) and the following transformation 
\begin{equation}
v_{\text{rel, crit}} \simeq 2 \times 448 \,\text{km/s}\, \left(\frac{F_a}{10^{16}\,\text{GeV}}\right)\,\tilde{v}_{\tilde{z},\text{crit}}(\tilde{N}_{\star}).\label{vrelcrit} 
\end{equation}
We find the following empirical relationship between the critical velocity of each clump and the number of particles in the clump:
\begin{equation}
\tilde{v}_{\tilde{z},\text{crit}}(\tilde{N}_{\star})\approx 0.4\,\tilde{N}_{\star}\label{vrelcrit2}
\end{equation}
with corresponding relative velocity $\tilde{v}_{\tilde{z},\text{rel,crit}}=2\tilde{v}_{\tilde{z},\text{crit}}$. 
Since we expect clumps today have relative velocities $\sim \mathcal{O}(10^2)\, \text{km/s}$ in the galactic halo, our results show that in order for a typical pair of clumps to readily merge, one needs an axion decay constant of
\begin{equation}
F_a \gtrsim 10^{15}\, \text{GeV}\label{Fbound}
\end{equation} 
On the other hand for $F_a\ll 10^{15}$\,GeV, mergers are still possible for situations in which the relative velocity is accidentally small. Since the distribution of velocities ${\bf v}_{\text{rel}}$ has zero mean (but large variance), this can happen occasionally, and will be estimated in the next Section. Also note that lower $F_a$ allows for more numerous clumps (since clump mass is $\propto F_a/m_a$, as seen in Eqs.~(\ref{Mmax},\,\ref{Malpha})) 
and so the collision rate will be high for lower $F_a$.

\subsection{Interference effects during the axion stars merger}

Since clumps at the initial time hold uncorrelated phases, we should include a relative complex phase $\delta $ in Eq.~(\ref{twoc}) to completely characterize the initial total wave function. Thus, we expect an interference pattern at the superposition time. 

\begin{figure}[t!]
\hspace{-0.1cm}\includegraphics[scale=0.205]{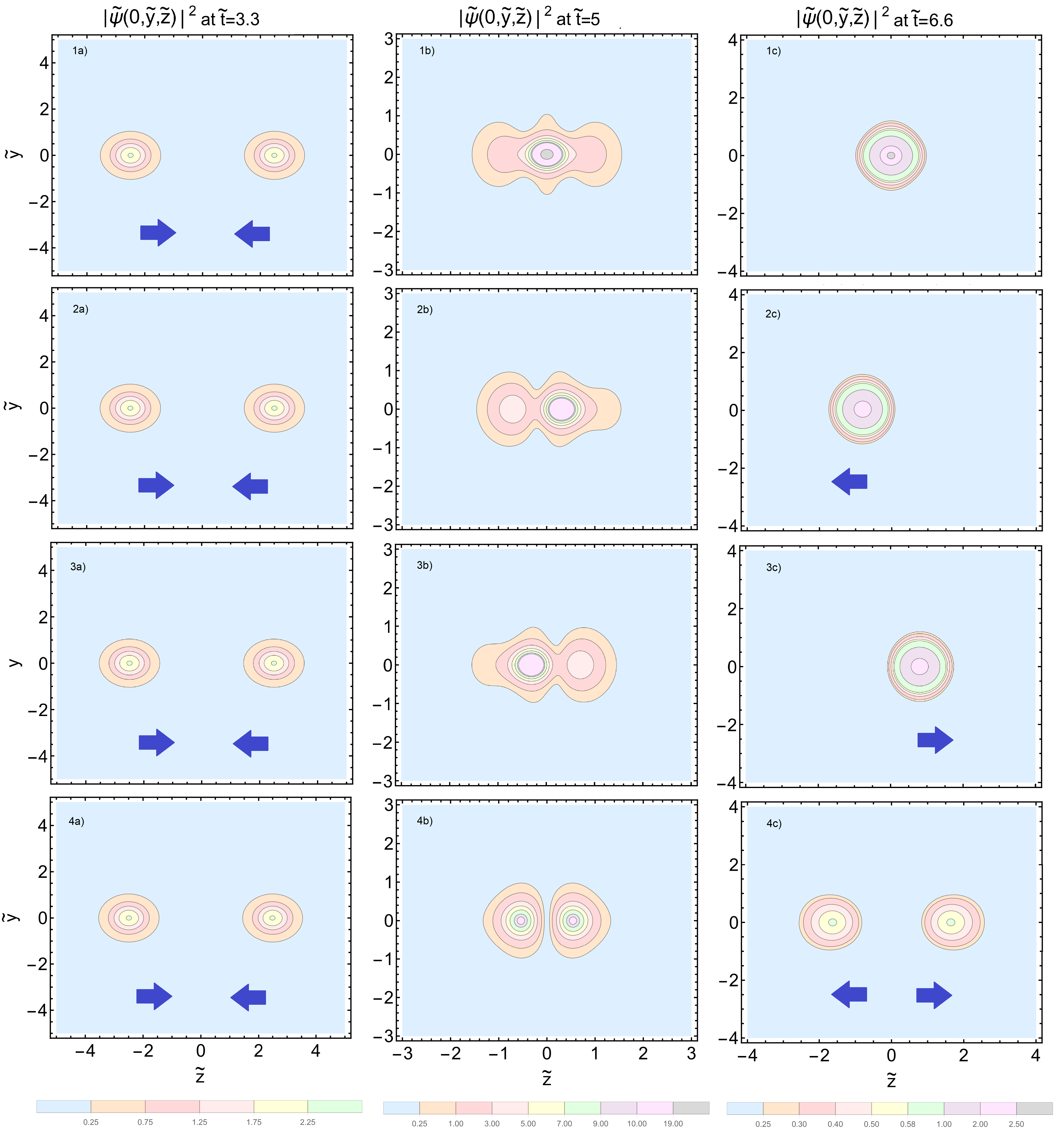}
\caption{Head-on collision between two clumps in their ground state configurations both with a number of particles $\tilde{N}_{\star} = 4.55418$ but different phase differences: $\delta = 0, \pi/2,$ $3\pi/2$ and $\pi$ for the rows (1,2,3,4), respectively. For all cases, the center of mass of the two clumps are initially separated  by a distance equal to $2\tilde{z}_0 = 12$ and have an initial velocity $\tilde{v}_{\tilde{z}}=1$.}
\label{phase}
\end{figure}

Figure~\ref{phase} shows contour levels of the local number density $|\tilde{\psi}(x \simeq 0,\tilde{y},\tilde{z})|^2$ at different times for the head-on collision of two identical ground state configurations with number of particles $\tilde{N}_{\star} = 4.55418$, $\tilde{v}_{\tilde{z}}=1$, $\tilde{z}_0 = 6$, and negative total initial energy.  The rows (1,2,3,4) refer to values of the phase difference $\delta = (0, \pi/2, 3\pi/2, \pi)$, respectively.  

As we expect, the head-on collision of the two clumps for the case of correlated phases lead to the formation of a new clump after merger (Fig.~\ref{phase}-1c). By contrast, for the case of phase opposition, destructive interference leads to a void region between clumps preventing them for merging (Fig.~\ref{phase}-4d) and acting like an effective repulsive force (Fig.~\ref{phase}-4c). We can understand this phenomenon considering the interference pattern at the time of maximum interaction $\tilde{t}_{\text{int}}$ formed by two traveling (identical) ground state configurations as follows 
\begin{align}
\tilde{\psi}({\bf{\tilde{x}}},\tilde{t}) &= \tilde{\psi}(\sqrt{\tilde{x}^2+\tilde{y}^2+(\tilde{z}+\tilde{z}_0)^2})\text{e}^{i(\tilde{v}_{\tilde{z}}\tilde{z}+\tilde{\mu} \tilde{t})} +  \tilde{\psi}(\sqrt{\tilde{x}^2+\tilde{y}^2+(\tilde{z}-\tilde{z}_0)^2})\text{e}^{-i(\tilde{v}_{\tilde{z}}\tilde{z}-\tilde{\mu} \tilde{t} + \delta)}\,,\nonumber\\
|\tilde{\psi}({\bf{\tilde{x}}},\tilde{t}_{\text{int}})|^2 &= 2|\tilde{\psi}\sqrt{\tilde{x}^2+\tilde{y}^2+\tilde{z}^2}|^2 \left[1+ \text{cos}(2\tilde{v}_{\tilde{z}}\tilde{z}+\delta)\right]\,.\label{interf}
\end{align}
The maximum interaction between clumps occurs when $\tilde{z}_{0}=0$. When $\delta = \pi$, the destructive interference leads to $|\tilde{\psi}(\tilde{x},\tilde{y},0),\tilde{t}_{\text{int}})| = 0$. This void in the plane $(\tilde{x},\tilde{y},0)$ can be seen as an effective repulsive force which acts on each clump preventing them for merger. An analog effects was reported in Ref.~\cite{Paredes:2015wga} in the context of solitonic galactic cores made of ultralight axions, where the axion self-interaction was neglected. For the case of any phase difference different from $\pi$, clumps merge. For completeness, we have added in Fig.~\ref{phase} particular cases $\delta = (\pi/2, 3\pi/2)$. For $\delta = \pi/2$, the destructive interference at $\tilde{t}_{\text{int}}$ occurs in the plane $(\tilde{x},\tilde{y},\tilde{v}_{\tilde{z}}\tilde{z}=\pi/4)$. Thus, the new clump formed by the merger feels a repulsive force from the right which pushes it to the $-\tilde{z}$-direction (Fig.~\ref{phase}-2c). Obviously, the same situation occurs for $\delta = 3\pi/2$, but now the resultant clump feels a repulsive force from its left (Fig.~\ref{phase}-3c). We note that despite appearances, linear momentum is in fact conserved in these processes. That is because the scalar wave emitted carries off the equal and opposite momentum. We have explicitly checked that the subsequent scalar wave cooling occurs in a somewhat similar way for any value of $\delta$ different from $\pi$.

\section{Astrophysical Signature via Resonant Photon Emission }

\subsection{Collision and Merger Rate for Axion Stars}
\label{rate}
We expect that axion stars form a fraction of the dark matter $f^{\text{DM}}_{\star}$ in the Milky Way halo. The collision rate per halo per year between two axion stars, with masses $(M_{\star,1}, M_{\star,2}) \simeq M_{\star}$, is given by an integral over the halo
\begin{equation}
\Gamma_{\star-\star} = 4 \pi \int_{0}^{R_{\text{halo}}} \frac{r^2}{2} \left( \frac{\rho_{\text{halo}}(r)f_{\star}^{\text{DM}}}{M_{\star}}\right)^2 \langle \sigma_{\text{eff}}(v_{\text{rel}}) v_{\text{rel}} \rangle\,dr\,\label{gammacol} 
\end{equation}
where $\rho_{\text{halo}}(r)$ is the density profile associated with the dark matter distribution within galactic halos, 
$ \langle\ldots \rangle$ is the average over the axion star relative velocity  $v_{\text{rel}}$ distribution in halos and 
$R_{\text{halo}}$ is a characteristic radius such as $R_{200}$\footnote{This is the radius at which the enclosed mass $M_{200}$ has a mean overdensity 200 times greater than the critical density.}.  The $(1/2)$ in Eq.~(\ref{gammacol}) is a symmetry factor to avoid a double counting coming from the fact collisions occur between the same kind of astrophysical objects, and we are assuming spherical symmetry of the halo for simplicity. The effective cross section of the collision,  $\sigma_{\text{eff}}$, corresponds to the usual geometric cross section enhanced  by the gravitational focus as
\begin{equation}
\sigma_{\text{eff}}(v_{\text{rel}}) = \pi (R_{\star}+R_{\star})^2 \left( 1 + \frac{v^2_{\star,\text{esc}}}{v^2_{\text{rel}}} \right)  = 4 \pi R_{\star}^2 \left(  1 + \frac{2 G_N M_{\star}}{R_{\star}v^2_{\text{rel}}} \right)\,\label{sigmav}
\end{equation} 
where $v^{2}_{\star,\text{esc}} = 2G_N (M_{\star}+M_{\star})/(R_{\star}+R_{\star})$ is the mutual escape speed between the axion stars. Considering that $v_{\text{rel}}$ is the order of the dark matter velocity in the halo, $v$, we  have
\begin{equation}
  \langle \sigma_{\text{eff}}(v) v \rangle =4 \pi \int_{0}^{v_{\text{esc}}} p(v) \sigma_{\text{eff}}(v) v^3 dv\,, 
\end{equation}
where 
\begin{equation}
p(v) = p_0\, \text{exp}[-v^2/v_0^2]
\end{equation} 
is a Gaussian velocity probability distribution in the Galactic frame depending on two parameters: a characteristic escape velocity $v_{\text{esc}}$ and $v_0$. The value of $v_0$ is usually taken as the circular velocity at the Solar position. The value of the normalization constant is obtained from $4\pi \int_0^{v_{\text{esc}}} v^2 p(v)dv = 1$ as
\begin{equation}
  p_0 = \frac{1}{(\pi v_0^2)^{3/2}} \left( \text{Erf}[v_{\text{esc}}/v_{0}] - \frac{2v_{\text{esc}} e^{-v_{\text{esc}}^2/v_0^2}}{\sqrt{\pi}v_0} \right)^{-1}\,.
\end{equation}
which is quite close to $1/(\pi v_0^2)^{3/2}$ for most cases of interest ($v_{\text{esc}}\gtrsim v_0$).

Evaluating Eq.~(\ref{sigmav}) requires an expression for the axion star mass and radius.  Before calculating $\Gamma_{\star-\star}$ in Eq.~(\ref{gammacol}), we  explicitly calculate the gravitational enhancement as
\begin{equation}
 \frac{2G_N M_{\star}}{R_{\star}v^2}  \sim  10^{-7} \left[\frac{7.46}{(g(\alpha)/\alpha)|_{\alpha=0.5}}\right] \left( \frac{F_a}{6 \times 10^{11}\,\text{GeV}} \right)^2   \left( \frac{0.3}{\gamma} \right)  \left( \frac{220 \,\text{km}\, s^{-1}}{v} \right)^2\,. \label{sigmavcalculated}
\end{equation}
We notice that even though the gravitational focusing is negligible for the classical QCD axion-window,  $10^{9}\,\text{GeV} \lesssim F_a \lesssim 10^{11}\,\text{GeV}$, we see that the gravitational attraction between axion stars when they pass near each other becomes at least as big as the geometric cross section for $F_a \gtrsim 10^{15}\,\text{GeV}$; this is indeed the same value we obtained for efficient mergers, which is not a coincidence because there is considerable overlap in the physics here. 

To obtain a first estimate of the collision rate for axion stars, we calculate Eq.~(\ref{gammacol}) for the case at which the dark matter density profile in galactic halos is homogeneous.  
Suppose that the whole dark matter resides on Milky Way-like halos with characteristic mass $M_{200} \sim 10^{12}\,M_{\odot}$ and uniform density $\bar{\rho}_{\text{halo}} \sim 200 \rho^0_m$ such that $M_{200}=(4\pi/3)\bar{\rho}_{\text{halo}}R_{200}^3$. Thus, the collision rate between axion stars per year and galaxy for the homogeneous case, $\Gamma^{\text{hom}}_{\star-\star}$, is given by         
\begin{align}
\Gamma^{\text{hom}}_{\star-\star} \,\sim\, & 3 \left[\frac{(g(\alpha)/\alpha)|_{\alpha=0.5}}{7.46}\right]^{2}   
\left( \frac{f_{\star}^{\text{DM}}}{0.01} \right)^2   \left( \frac{\gamma}{0.3} \right)^2
\left( \frac{6 \times 10^{11}\,\text{GeV}}{F_a} \right)^4  \times \nonumber \\
& \,\,\,\left[ 1 + 10^{-7} \left[\frac{7.46}{(g(\alpha)/\alpha)|_{\alpha=0.5}}\right]  \left( \frac{F_a}{6 \times 10^{11}\, \text{GeV}} \right)^2 \left( \frac{0.3}{\gamma} \right)    \right]\,\frac{\text{collision}}{\text{yr} \times \text{galaxy}}\,,\label{GammahomEst}
\end{align}
where we have taken $v_0 = 220\, \text{km s}^{-1}$ and $v_{\text{esc}}=544\,\text{km s}^{-1}$ from typical values of the Standard Halo Model~\cite{PhysRevD.33.3495, Evans:2018bqy}. Note that the collision rate does not depend directly on the axion mass. However, for the case of the QCD axion, this mass is linked to the decay constant via Eq.~(\ref{axionmassfa}).

A more accurate estimate of the collision between axion stars follows by considering a radial profile for the dark matter distribution within Milky Way-like halos. Since a fraction of the whole dark matter is in axion stars, it is plausible to consider that they distribute according to a typical halo dark matter profile such that a Navarro-Frenk-White profile ($\rho^{\text{NFW}}_{\text{halo}}$)~\cite{Navarro:1995iw} or Burkert profile ($\rho^{\text{B}}_{\text{halo}}$)~\cite{1995ApJ...447L..25B}. While an Einasto dark matter profile was assumed in Ref.~\cite{BAI2018187} for the study of radio signals generated by collisions between axion and neutron stars, a NFW profile was assumed in Ref.~\cite{Eby:2017xaw} to analyze collisions between axion stars and astrophysical objects. Both considered halo profiles parameterized as follows  
\begin{align}
\rho^{\text{NFW}}_{\text{halo}}(r) & = \rho_s \left(\frac{r}{r_s}\right)^{-1}\left[ 1+ \left( \frac{r}{r_s} \right) \right]^{-2} \,,\\
\rho^{\text{B}}_{\text{halo}}(r)& =  \rho_s \left(1+\frac{r}{r_s}\right)^{-1}\left[ 1+ \left( \frac{r}{r_s} \right)^2 \right]^{-1}   \,,
\end{align}
where $\rho_s$ and $r_s$ are the scale density and the scale radius, respectively. While $r_s$ in the NFW profile is the radius at which $d\text{log}\rho^{\text{NFW}}_{\text{halo}} / d\text{log}r = -2$, in the Burkert profile $r_s$ is the radius of the region of approximately constant density. The corresponding best fit Milky Way halo parameters associated with these specific halo mass models are summarized in Table \ref{table:nonlin}.  

\begin{table}[t]
\caption{Main dark matter halo parameters for three different Milky Way mass models (best-fit models).} 
\centering 
\begin{tabular}{c c c c c c c c c c c} 
\hline\hline 
Profile & $R_{200}$ & $M_{200}$& $r_s$ & $\rho_s$ & $\rho_{\odot}$ & $R_{\odot}$ & $v_0$ & $v_{\text{esc}}$ \\ 
 &  [kpc] &  [$M_{\odot}$] &  [kpc] &  [GeV/cm$^{3}$] &   [GeV/cm$^{3}$] & [kpc]  & [km/s] & [km/s] \\ [0.5ex]
\hline  
 NFW~\cite{2011MNRAS.414.2446M} & 237 &  1.43$\times 10^{12}$ & 20.2 & 0.32 & 0.395 & 8.29 & 239 & 622  \\ 
B~\cite{Nesti:2013uwa} & 291 &  1.11$\times 10^{12}$ &  9.26  & 1.57 & 0.487 & 7.94 & 241 & 576  \\ 
NFW~\cite{Nesti:2013uwa} & 319 & 1.53 $\times 10^{12}$ & 16.1  &  0.53 & 0.471 & 8.08 & 244 & 613 \\ 
\hline 
\end{tabular}
\label{table:nonlin} 
\end{table}

\begin{figure}[t!]
\centering
\includegraphics[width=\columnwidth]{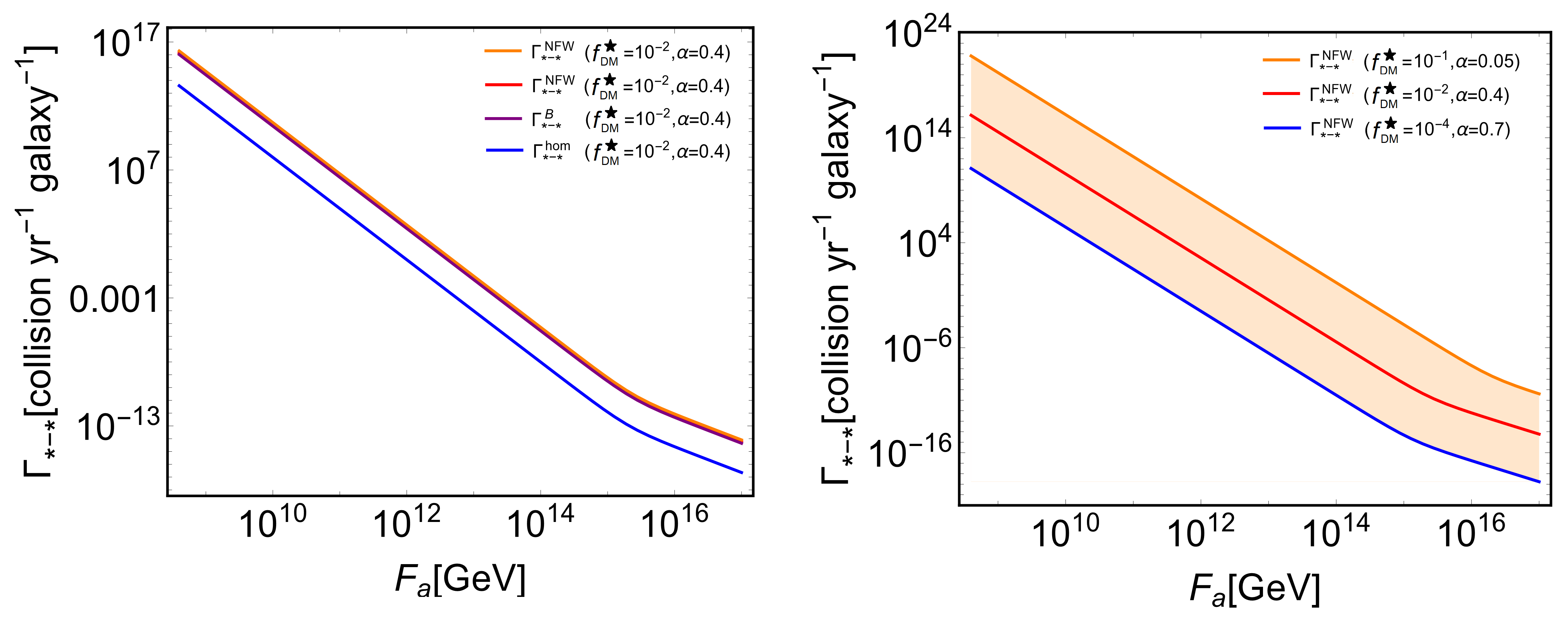}
\caption{(Left) Collision rate for close encounters between axion dark matter clumps versus the PQ scale using a homogeneous density for the Milky Way halo (blue line) and three different mass profile for the halo. We have used for all cases  $N_{\star, 1} = N_{\star, 2} = 0.4 N_{\text{max}}$ and $f_{\star}^{\text{DM}}=0.01$. We have used halo parameters (best-fit models) obtained in Refs.~\cite{2011MNRAS.414.2446M} (NFW profile), ~\cite{Nesti:2013uwa} (Burkert profile), and~\cite{Nesti:2013uwa} (NFW profile) in red, purple, and orange lines, respectively.  The purple line corresponds to the homogeneous case at which $\bar{\rho}_{\text{halo}} \sim 200 \rho^0_m$. (Right) Collision rate for close encounters between axion dark matter clumps versus the PQ scale for the parameter space $(0.05, 10^{-4}) \leq (\alpha, f_{\star}^{\text{DM}}) \leq (0.7, 10^{-1}) $ (orange shaded region), where 
$N_{\star, 1} = N_{\star, 2} = \alpha N_{\text{max}}$. In particular, blue, red and orange lines correspond to values $(\alpha, f_{\star}^{\text{DM}}) = (0.7, 10^{-4}), (0.4, 10^{-2}), (0.05, 10^{-1})$, respectively. The collision rate for all cases is calculated by using a NFW profile for the dark matter halo~\cite{2011MNRAS.414.2446M}.}
\label{CollisionRate}
\end{figure}

As shown in Fig.~\ref{CollisionRate} (left), the collision rate for the three possible parameters are approximately the same. In comparison to the homogeneous case, the gravitational focusing enhancement of the collision rate is approximately unaltered because the values of $v_0$ and $v_{\text{esc}}$ are all quite similar. However, including a mass profile for the halo significantly enhances the geometrical cross section and, as a result, the total collision rate increases by two orders of magnitude, i.e.,. $\Gamma^{\text{NFW/B}}_{\star-\star}/\Gamma^{\text{hom}}_{\star-\star} = \mathcal{O}(10^2)$. Apart from the strong dependence of the collision rate on the decay constant, this rate depends also on the number of particles of the colliding clumps and the fraction of dark matter in axion clumps, i.e.,  $\alpha N_{\star,\text{max}}$ and $f_{\star}^{\text{DM}}$. Figure~\ref{CollisionRate} (right) shows the expected range of values for the collision rate (orange shaded region) in the parameter space $(0.005,10^{-4}) \leq (\alpha,f_{\star}^{\text{DM}}) \leq (0.7,10^{-1})$. The larger $f_{\star}^{\text{DM}}$ (or the smaller $\alpha$), the larger the collision rate. 

\begin{figure}[t!]
\centering
\includegraphics[width=8cm]{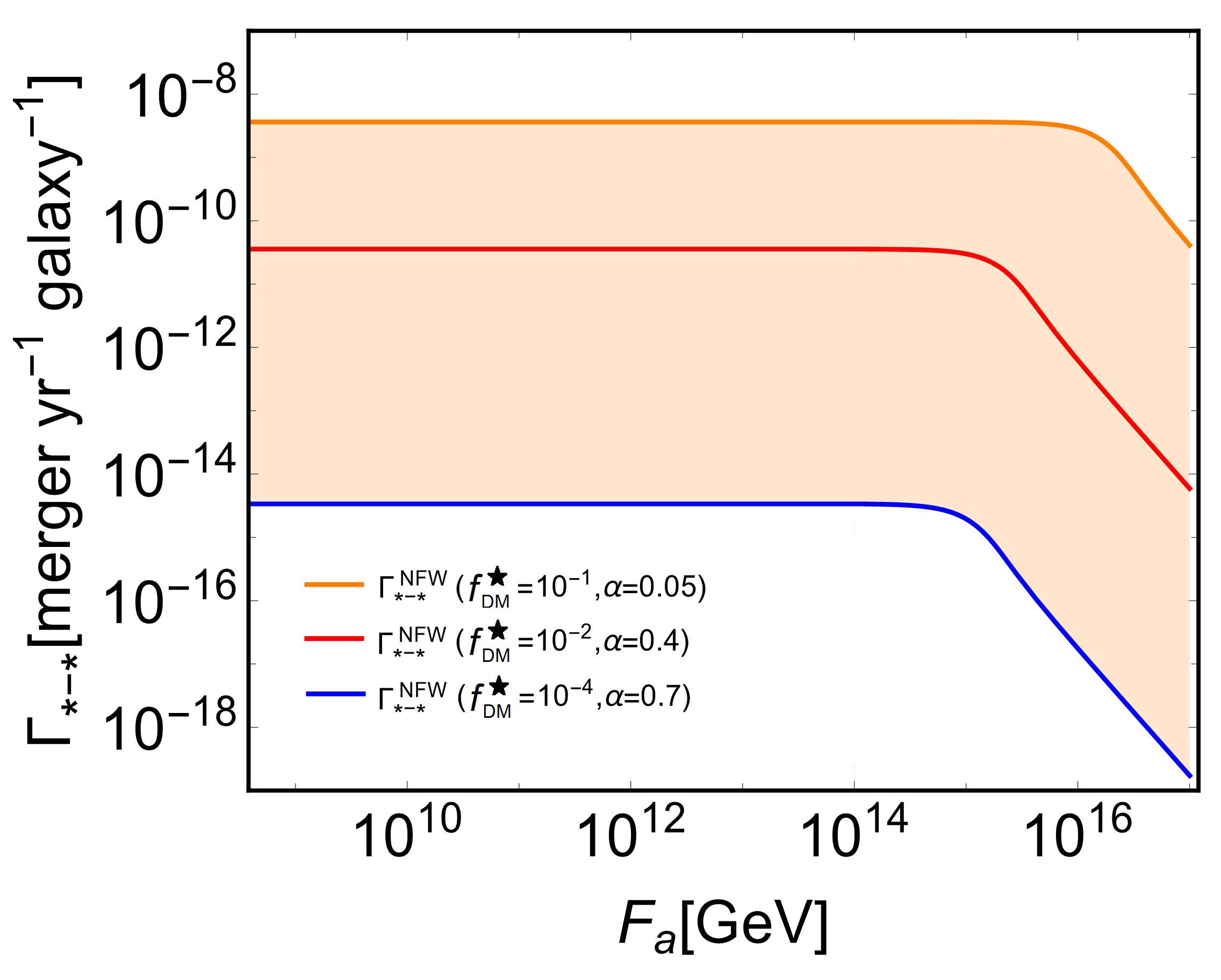}
\caption{Merger rate from pair-wise interactions from a single collision between axion dark matter clumps versus the PQ scale  to the decay constant for the parameter space $(0.05, 10^{-4}) \leq (\alpha, f_{\star}^{\text{DM}}) \leq (0.7, 10^{-1}) $  (orange shaded region), 
 where  $N_{\star, 1} = N_{\star, 2} = \alpha N_{\text{max}}$. As in Fig.~\ref{CollisionRate}, the blue, red and orange lines correspond to values $(\alpha, f_{\star}^{\text{DM}}) = (0.7, 10^{-4}), (0.4, 10^{-2}), (0.05, 10^{-1})$, respectively. The collision rate for all cases is calculated by using a NFW profile for the dark matter halo~\cite{2011MNRAS.414.2446M}, and making the simplified assumption that mergers arise from 2-body pair-wise interactions only. For small $F_a$ this provides a conservative lower bound on the actual merger rate, which can be enhanced due to 3-body processes, etc. }
\label{MergeRate}
\end{figure}

The above gives an estimate for the collision rate. However, as we discussed in Section \ref{sec:section3}, not all collisions lead to mergers. In fact the typical speeds of stars in the galaxy indicate that they will typically carry too much energy for a 2-body merger to take place when the PQ scale is small; recall Fig.~\ref{ps}. So in order to estimate the fraction that immediately lead to mergers from a single collision, we can return to the velocity distribution above, and instead of cutting off the integral at the escape speed of the galaxy $v_{esc}$, we cut off the integral at the critical velocity for merger $v_{\text{rel, crit}}$ from Eqs.~(\ref{vrelcrit},\,\ref{vrelcrit2}) (or $v_{esc}$ if it is smaller). This leads to Fig.~\ref{MergeRate}. 
Note that now for small $F_a$, instead of the rate rising as $1/F_a^4$, as indicated in Eq.~(\ref{GammahomEst}), we instead have a constant rate. This can be understood as follows: For small $F_a$, we have to take into account that $v_{\text{rel, crit}}\propto F_a$, which becomes statistically disfavored. The probability that the relative velocity is this small, scales as $v_{\text{rel, crit}}^3\propto F_a^3$. In addition the scattering rate is suppressed by another power of $v_{\text{rel, crit}}\propto F_a$, since for slow movers they take longer to reach one another. Together these 4 powers of $F_a$ in the merger rate cancel the $1/F_a^4$ of the collision rate giving a flat rate. We see that the rates seen in Fig.~\ref{MergeRate} are quite small. We note, however, that this plot indicates only a conservative lower bound on the merger rate, as it is only based on pair-wise 2-body interactions, while 3-body interactions and multiple encounter can enhance the rate, especially considering that the rate of collisions can in fact be quite high (at least for lower $F_a$). On the other hand, it is possible that the increased collisions could destabilize the clumps, since they the low $F_a$ clumps have rather lower binding energies.

Recall that for a fixed axion-photon coupling constant, clumps need to more than a critical  number of particles to undergo resonant decay into photons, as shown in Eq.~(\ref{gagammaN}) and plotted in Fig.~\ref{stableunstable} (right). Thus, considering the empirical rule for merger obtained in Sec.~\ref{head}, i.e., $\tilde{N}^{\star}_{\text{final}} \simeq 0.7 (\tilde{N}_{\star,1} + \tilde{N}_{\star,2})$, two colliding clumps both with number of particles $\tilde{N}_{\star,1}=\tilde{N}_{\star,1} =\alpha N_{\star,\text{max}}$  would lead to a new clump as $N^{\star}_{\text{final}} \simeq (1.4\times \alpha)N_{\star,\text{max}}$. This new clump can be associated with a critical axion-photon coupling constant for resonant decay through Eq.~(\ref{gagammaN}).  For values $0.05 \leq \alpha \leq 0.7$ shown in Fig.~\ref{CollisionRate}(right), the minimum value for $g_{a\gamma\gamma}$ to have resonance reads as
\begin{equation}
        0.28  \left( \frac{\gamma}{0.3} \right)^{1/2}  \lesssim g_{a\gamma\gamma,\text{min}} F_a  \lesssim 5.7 \left( \frac{\gamma}{0.3} \right)^{1/2}\,,
 \end{equation}
  where the lower bound is obtained for $\alpha \simeq 0.7$.

\subsection{Photon Emission}
After axion star formation sometime in the earlier history of the universe, we expect a distribution for number of particles as shown in the blue curve of  Fig.~\ref{stableunstable} (left).  For a fixed axion-photon coupling constant, there is a critical number of particles $N_c$ which allows for resonance. Suppose that the axion-photon coupling $g_{a\gamma\gamma}^{\star}$ lies in the blue shaded region of Fig.~\ref{stableunstable}, so that $g_{a\gamma\gamma}^{\star} \geq g_{a\gamma\gamma, \text{min}}$ and $N_c^{\star} \equiv N_c$. All clumps which have a number of particles such that $N_{\star} > N_c^{\star}$, they will undergo parametric resonance into photons. These clumps will quickly lose energy by electromagnetic radiation and decrease their number of particles (or, equivalently, their masses) until $N_{\star} \rightarrow N_c^{\star}$. After that point, the resonance phenomenon would shut-off. For example, for $g_{a\gamma\gamma}^{\star} = 2 [\gamma^{1/2} F_a^{-1}]$, we have $N_c^{\star} \simeq 3.6 [\mpl F_a m_{\phi}^{-2}\gamma^{-1/2}]$ as shown in Fig.~\ref{stableunstable} (right).  On the other hand, all clumps which initially 
after formation had a particle number less than the critical  $N_c^{\star}$ for resonance would tend to capture axion dark matter
from the background and to move down through the blue curve~\footnote{The total energy of a stable BEC axion clump decreases as the number of particles increases reaching its lower values at $\tilde{H}(\tilde{N}_{\star,\text{max}}) \simeq -a^2 b^{1/2}/(9 \sqrt{3}c^{3/2})$, where $(a,b,c)$ are given by Eqs.~(\ref{abc}).} in Fig.~\ref{stableunstable} (right) until $N \rightarrow N^{\star}_c$.   Thus, one may expect that in the Milky Way halo today a pile-up of axion dark matter clumps at a unique value of particle number or mass, which for spherically symmetric clumps, is given in terms of fundamental constants. 

The idea then is that some fraction of clumps today in the Milky Way can collide with each other leading to the formation of a new resultant clump. Conditions for an effective merger and collision rate were discussed in Sec.~\ref{sec:section3},\ref{rate} and we shall comeback to this later. Under suitable conditions, we found that clumps can merge and produce a new clump according to the relation $N^{\star}_{\text{final}} \sim 0.7 (N_{\star,1}+N_{\star,2}) \simeq  1.4 N_{\star}$, where we have 
used $N_{\star}\simeq (N_{\star,1},N_{\star,2})$. Since $N_{\star} \gtrsim N_{c}$, the new clump will undergo parametric resonance as soon as it settles down to a ground state configuration. The energy released by the electromagnetic radiation during resonance, $E_{\star,\gamma}$, can be estimated as
\begin{equation}
E_{\star,\gamma}  = \left[0.7(\tilde{N}_{\star,1}+\tilde{N}_{\star,2}) - \tilde{N}^{\star}_{c}\right] \frac{\mpl F_a}{m_{\phi} \gamma^{1/2}}\,\simeq 1.4 (\alpha - 0.71 \alpha_c) M_{\star,\text{max}}\,,\label{Erelease}
\end{equation} 
where $N_{\star} = \alpha N_{\star, \text{max}}$, $N_c^{\star} = \alpha_c N_{\star, \text{max}}$. Note that  $M_{\star,\text{max}}$ can be rewritten from Eq.~(\ref{Mmax}) as
\begin{equation}
M_{\star,\text{max}} \sim 1.4 \times 10^{46}\,\text{GeV} \left( \frac{10^{-5}\,\text{eV}}{m_{\phi}} \right) \left( \frac{F_a}{6 \times 10^{11}\,\text{GeV}} \right) \left( \frac{0.3}{\gamma} \right)^{1/2}\,.
\end{equation}
The release of this energy will occur very quickly due to it being a resonant process. When the axion-photon coupling is large enough for a fixed number of particles, the Bose-Einstein statistics allows for exponential growth of the photon occupancy number leading to a final output of classical electromagnetic waves. The time scale for this exponential growth $\tau$ can be estimated from the growth rate. As explained in Section \ref{resonancephotons}, the growth rate $\mu^{\star}$ is well approximated by  the difference between the maximum growth rate for the case of an homogeneous condensate and the photon escape rate. Thus, $\tau = 1/\mu^{\star} \approx 1/\mu_{H}$ so that the time scale reads as
\begin{equation}
\tau \lesssim 2 \times 10^{-4}\, \text{s}\,\, g(\alpha)  \left( \frac{\gamma}{0.3}  \right)^{1/2} \left( \frac{6 \times 10^{11}\,\text{GeV}}{F_a} \right) \left( \frac{10^{-5}\,\text{eV}}{m_a} \right)\,,\label{tau}
\end{equation}  
where we have used $g_{a\gamma\gamma, \text{min}}F_a = 0.28\, (\gamma/0.3)^{1/2} [g(\alpha)/\alpha]^{1/2}$ as the minimum axion-coupling constant to satisfy the resonant condition in Eq.~(\ref{gagammaN}). This is valid once the clump mass is somewhat larger than the critical mass and we have reached the ground state. In practice the approach to the ground state may be somewhat slow, so this may be somewhat of an over-estimate of the true growth rate. This deserves further investigation in future work.

The electromagnetic radiation output corresponds to a narrow line near the resonant wavelength of $\lambda_{\text{EM}} \approx 2\pi / k \approx 4 \pi/m_{\phi}$, which can be expressed as  
\begin{equation}
\lambda_{\text{EM}} \approx 0.25\, (m_{\phi}/{10^{-5}\text{eV}})^{-1}\,\mbox{meters}\,. 
\end{equation}
The bandwidth can be estimated from the width of the first instability band for the homogeneous case, $\Delta k \approx g_{a\gamma \gamma} m_{\phi} \phi_0/2$. Using the sech ansatz radial profile to set the axion field amplitude, the frequency line of emission is
\begin{equation}
\nu_{\text{EM}} \approx 1.2\, \text{GHz}\left(\frac{m_{\phi}}{10^{-5}\,\text{eV}}\right) \pm {1.6\,\text{kHz}\over g(\alpha)}  \left( \frac{F_a}{6 \times 10^{11}\,\text{GeV}} \right) \left( \frac{m_{\phi}}{10^{-5}\,\text{eV}} \right) \left(0.3\over\gamma\right)^{\!1/2}\,\label{bandwidth}
\end{equation} 
where the bandwidth in the frequency is related to the growth-time scale as $\Delta \nu_{\text{EM}} = 1/(\tau \pi)$.

The energy density flux of the signal on the Earth, the energy per unit area per unit time, from a resonance event at distance $D$ in the Milky Way halo is given by
\begin{equation}
S =  \frac{P}{4 \pi D^2}  = \frac{\Delta E/\Delta t}{4 \pi D^2} 
\end{equation}
When the axion merging is still taking place, the resonant emission would be suppressed. However, once the resonance can take place, one can expect exponential growth in photons may be possible. The energy output comes from Eq.~(\ref{Erelease}), so we can estimate $\Delta E\sim E_{\star,\gamma}$, while the characteristic time we can estimate as the growth rate $\Delta t\sim\tau$. For the distance to merger $D$, we may consider $D\sim 50$\,kpc as a typical distance in the galaxy. Altogether this gives an estimate of the energy density flux as
\begin{equation}
S \sim 5\times 10^{-3}\,\mbox{W/m}^2\left(\alpha-0.71\alpha_c\over g(\alpha)\right)\left(F_a\over 6 \times 10^{11}\,\mbox{GeV}\right)^{\!2}\left(50\,\mbox{kpc}\over D\right)^{\!2}\left(0.3\over\gamma\right)
\end{equation}
(again the true value should be lower for a slow merger).  
For comparison, we can compare this to the energy flux from the sun of $S_{sun}=1370$\,W/m$^2$. Hence for large $F_a$ mergers, which in fact are the ones most robust as discussed earlier, the energy flux on earth is appreciable. We note that relevant wavelengths are radio waves, and that for high $F_a$ axions, one is probing deep into long wavelengths, which may be difficult to achieve; but possible with future telescopes.

For astronomical observations, it is important to note the received flux per bandwidth, or the spectral flux density, $S_B=S/\Delta B$, where 
$\Delta B\sim\Delta\nu_{EM}=1/(\tau\pi)$, as mentioned above. This can be quite appreciable, since the signal is anticipated to be highly monochromatic, as indicated in Eq.~(\ref{bandwidth}). We obtain
\begin{equation}
S_B\sim 3\times 10^{-6}\,\mbox{W/m}^2/\mbox{Hz}\,
(\alpha-0.71\alpha_c)\left(F_a\over 6 \times 10^{11}\,\mbox{GeV}\right)\left( \frac{10^{-5}\,\text{eV}}{m_{\phi}} \right)\left(50\,\mbox{kpc}\over D\right)^{\!2}\left(0.3\over\gamma\right)^{\!1/2}\label{SB}
\end{equation}
Note that for high $F_a$ we also have low $m_{\phi}$, so this can be quite large (although this will be reduced for slow mergers). 

\subsection{Detectability}

The signal frequency of the resonant phenomenon is related to the axion mass as shown in Eq.~(\ref{bandwidth}). 
As we explained in Sec. 3  (see Fig.~\ref{ps}), collisions between isolated pair of axion clumps which typically lead to merger require large values of the axion decay constant. However, mergers for moderate values of $F_a$ are still viable since the collision rate is much more larger in that regime.  

For the QCD axion and an axion decay constant in the range $10^{10}\,\text{GeV} \lesssim F_a \lesssim  10^{13}\,\text{GeV}$, the central frequency of emission 
ranges as $70\,\text{MHz} \lesssim \nu_{\text{EM}} \lesssim 70\,\text{GHz}$. This range of frequencies is covered by current (and prospective) radio telescopes. For example, while the Arecibo Observatory with a 305-meter single aperture telescope covers frequencies from 300 MHz to 10 GHz~\cite{Giovanelli:2005ee}, the Five hundred meter Aperture Spherical radio Telescope (FAST) from 70 MHz to 3 GHz~\cite{2011IJMPD..20..989N}, the Karl G. Jansky Very Large Array (JVLA) from 1 GHz to 50 GHz~\cite{Gk}, and the Green Bank telescope (GBT) from 290 MHz to 115.3 GHz~\cite{GBT}. The Square Kilometre Array (SKA) phase 1 to be constructed during the period 2018-2023 can cover frequencies from 50 MHz to 350 MHz (SKA-low frequencies) and from 350 MHz to 14 GHz (SKA-mild frequencies)~\cite{SKA}.

In the regime of $F_a \gtrsim 10^{15}\,\text{GeV}$, mergers of axion clumps are  more robust, but the resonant axion decay leads to low frequency photon emission which would require radio telescopes from space to be detected. Detection of frequencies less than $\sim30\,\text{MHz}$ is highly limited on the Earth due to the absortion and scattering produced by the ionosphere. The plasma frequency of this layer is about 15 MHz (10 MHz) on the day (night) side of the Earth near sunspot maximum (minimum). Thus, the layer is opaque to all lower frequencies. 
The detection of electromagnetic waves from the galaxy  in the window $\sim(30\,\text{kHz} - 30\,\text{MHz})$~\footnote{The lower bound is set just about the local plasma frequency of the interplanetary medium.} requires a space or Lunar-based radio telescope. The Orbiting Low Frequency Antennas for Radio Astronomy Mission (OLFAR) is an ambitious plan of building a swarm of hundreds to thousands of satellites to analyze frequencies below 30 MHz. The phase 4 of the mission is planned to be carried out at 2030, which corresponds to the development and deployment of orbiting nano satellites on the far side of the moon. This mission and others similar will offer in the future a chance for detection of low frequency photons emitted by axion clumps at high $F_a$ regimes~\cite{BENTUM2020856}.

Apart from the range of frequencies involved in the phenomenon of study, the sensitivity of radio telescopes is a crucial factor to be considered.   The smallest spectral flux density that a radio telescope can detect depends on the observation time, $t_{\text{obs}}$, the bandwidth of the signal, $\Delta B$, and the ratio of the effective collecting area of the telescope, $A_{\text{eff}}$, to the system temperature, $T_{\text{sys}}$, as follows~\footnote{In unit of Jansky (Jy), which is the usual unit of measurement in radio astronomy, we have $S_{B,\text{min}} \approx 0.09\,\text{Jy}\, (1\, \text{MHz}/\Delta B)^{1/2}(1\, \text{ms}/t_{\text{obs}})^{1/2} [10^3 \text{(m}^2\text{/K) / (A}_{\text{eff}}/\text{T}_{\text{sys}})]$.}~\cite{Condon16}
\begin{equation}
S_{B,\text{min}} \approx 9\times 10^{-28}\, \text{W/m}^2\text{/Hz} \left( \frac{1\, \text{MHz}}{\Delta B} \right)^{1/2} \left( \frac{1\,\text{ms}}{t_{\text{obs}}} \right)^{1/2} \left(  \frac{10^3\,\text{m}^2\text{/K}}{A_{\text{eff}}/T_{\text{sys}}} \right)\,.
\end{equation}
This critical spectral flux needs to be compared with the spectral flux density obtained in Eq.~(\ref{SB}). For instance, for the SKA (Phase 1)~\cite{Bacon:2018dui}, the effective area to system temperature is $A_{\text{eff}}/T_{\text{sys}} = 2.7\times 10^3 \,\text{m}^2/\text{K}$. Thus, we obtain $S_{B,\text{min}} \sim 3 \times 10^{-28}\,  (1\, \text{MHz}/\Delta B)^{1/2}\,(1\, \text{ms}/t_{\text{obs}})^{1/2}W/m^2/Hz$. In the frequency range detectable for SKA, we see that $S_B \gg S_{B,\text{min}}$ for emissions in the galaxy (even when taking into account the reduced flux for a slow merger).

\section{Discussion}

In this work we have explored a novel way to possibly detect axion dark matter, by computing axion star mergers and the possible resulting resonance into photons. As our earlier work showed there is a critical mass for clumps beyond which they can undergo parametric resonance into photons, depending on the axion-photon coupling. However, it was unclear if sub-critical mass clumps could ever go beyond criticality to achieve the condition for resonance in the late universe. This work shows that this is indeed possible under the right conditions; we discuss the rates shortly. We find that when mergers take place, the resultant clump has a mass moderately larger than the original clump masses, despite the loss of mass in scalar wave emission. This means that sub-critical mass clumps can become super-critical and undergo photon emission for moderate to large axion-photon couplings. 
We note that the plasma mass in the halo of the galaxy $m_\text{plasma}^2\sim (6\times10^{-12}\,\mbox{eV})^2n_e/(0.03\mbox{cm}^{-3})$ will can make the resonance kinematically forbidden for sufficiently small axion masses. However, this is only a problem for rather extreme values of parameters. 

We found that the collision rate can be appreciable, although it depends strongly on the PQ scale $F_a$. For smaller $F_a$, the number of clumps and their cross sections are large, so the collisions are very frequent in the galaxy today (see Fig.~\ref{CollisionRate}). However, such clumps have a small binding energy and so such collisions typically don't lead to mergers, at least not between a single isolated pair (see Fig.~\ref{ps}). However,  this can happen via statistical flukes in the galaxy due to the Maxwellian distribution of relative velocities (see Fig.~\ref{MergeRate}). We emphasize that this provides only a lower bound on the merger rate, as they can be enhanced by 3-body interactions, which are not taken into account here. They could be destabilized too. We leave this for future work. 

On the other hand, for a typical collision to usually lead to mergers requires larger values of the PQ scale of $F_a\gtrsim 10^{15}$\,eV; although in this case the rate of collisions is much smaller. 
The possible signal arising from larger values of $F_a$ (from clumps that are more robust against disruption) would be very interesting, although it seems to be a very rare event. In fact it is ordinarily rather difficult to probe these high $F_a$ regimes since couplings to matter are then further suppressed since it is through dimension 5 operators. As explained in the introduction, large values of the axion decay constant are achievable in the scenario in which the PQ symmetry is broken before or during inflation, when fine tuning over the misalignment angle is allowed. 
In many string compactifications, the axion decay constant is in the range $10^{15}\,\text{GeV}-10^{18}\,\text{GeV}$, though there are exceptions \cite{Svrcek:2006yi}. Apart from the presence of the QCD axion, which can be plausibly emdedded in this framework, string models predict the presence of many axion-like-particles with masses distributed logarithmically \cite{Arvanitaki:2009fg}. 

Furthermore, kinetic nucleation of QCD and string axion stars in mini-halos around PBHs~\cite{Hertzberg:2020hsz} and axion-like-particles clumps formed by tachyonic instability~\cite{Fukunaga:2020mvq}, etc, are processes that can take place for general values of the PQ symmetry scale, so they may provide other ways to achieve enhanced merger rates. On the other hand, lower $F_a$ is the more traditional window on the QCD axion, and it has the feature that the collision rate is much larger; so it strongly suggests simulations that include 3-body interactions, etc, for a more complete understanding of the merger rate.

Higher $F_a$ models corresponds to smaller axion masses $m_a$ and since this sets the characteristic frequency of the radio wave emission, one would therefore need telescopes with sensitivity to rather low frequencies. 
While for more moderate values of $F_a$ and more moderate axion masses, the mergers are statistically rare, though they are still possible due to the fact that their collisions are much more frequent. Their characteristic time scale for relaxation towards the ground state is rather long and therefore the resonance may be a slower, gradual process, leading to a lower flux of photons on the earth. These issues all deserve further exploration.

\section*{Acknowledgments}
M. P. H. is supported in part by National Science Foundation grant PHY-1720332. E. D. S. thanks to Yu Ling Chang for
computational support at the early stage of this project. We thank Kimmo Kainulainen for useful discussions about
non-head-on collisions between axion clumps and we thank Igor Tkachev for discussion.
\appendix

\section{Proof of Evolution Approximation 
}\label{AppA}
\label{AppendA}
  First, consider the expression
\begin{equation}
e^{(A + B) h + F (h)}  =  e^{B h / 2} e^{A h} e^{B h / 2}\,,
\label{appa1}
\end{equation}
where $F(0)=0$. We take a derivative with respect to $h$ of~(\ref{appa1}) to obtain 
\begin{align}
(A + B + F' (h)) e^{(A + B) h + F (h)} & =  \frac{B}{2} e^{B
    h / 2} e^{A h} e^{B h / 2} + e^{B h / 2} A e^{A h} e^{B h / 2} + e^{B h /
    2} e^{A h} \frac{B}{2} e^{B h / 2}\,,\label{appa2}\\
(A + B + F' (h)) e^{B h / 2} e^{A h} e^{B h / 2} & = 
    \frac{B}{2} e^{B h / 2} e^{A h} e^{B h / 2} + e^{B h / 2} A e^{A h} e^{B h
    / 2} + e^{B h / 2} e^{A h} \frac{B}{2} e^{B h / 2}\,,\label{appa3}\\
 F' (h) & =  - \frac{B}{2} + e^{B h / 2} A e^{- B h / 2} - A +
    e^{B h / 2} e^{A h} \frac{B}{2} e^{- A h} e^{- B h / 2}\,,\label{appa4}
\end{align}
where we have used ~(\ref{appa1}) to obtain (\ref{appa3}). From (\ref{appa4}) we obtain
$F''(h)$ as follows
\begin{align}
 F'' (h) &  =  \frac{B}{2} e^{B h / 2} A e^{- B h / 2} - e^{B h
    / 2} A \frac{B}{2} e^{- B h / 2} + \frac{B}{2} e^{B h / 2} e^{A h}
    \frac{B}{2} e^{- A h} e^{- B h / 2}
      +\nonumber\\
& e^{B h / 2} A e^{A h} \frac{B}{2} e^{- A h} e^{- B h / 2} - e^{B h
    / 2} e^{A h} \frac{B}{2} A e^{- A h} e^{- B h / 2} 
   - e^{B h / 2} e^{A h}
    \frac{B}{2} e^{- A h} \frac{B}{2} e^{- B h / 2}\,,\nonumber\\
  & =  e^{B h / 2} \left[ \frac{B}{2}, A \right] e^{- B h / 2} + e^{B h /
    2} \left[ \frac{B}{2}, e^{A h} \frac{B}{2} e^{- A h} \right] e^{- B h / 2}\,,
\end{align}
where we have used $\frac{d}{\text{dx}} (e^{\text{Ax}} \text{Be}^{-
  \text{Ax}}) = e^{\text{Ax}} [A, B] e^{- \text{Ax}}$. Similarly, we obtain $F'''(h)$ as
\begin{align}
 F''' (h) & =  e^{B h / 2} \left[ \frac{B}{2}, \left[
    \frac{B}{2}, A \right] \right] e^{- B h / 2} + e^{B h / 2} \left[
    \frac{B}{2}, \left[ \frac{B}{2}, e^{A h} \frac{B}{2} e^{- A h} \right]
    \right] e^{- B h / 2} +\nonumber\\
    &   e^{B h / 2} \left[ \frac{B}{2}, e^{A h} \left[ A, \frac{B}{2}
    \right] e^{- A h} \right] e^{- B h / 2} + e^{B h / 2} \left[ \frac{B}{2},
    e^{A h} \left[ A, \frac{B}{2} \right] e^{- A h} \right] e^{- B h / 2} +\nonumber\\
    &   e^{B h / 2} e^{A h} \left[ A, \left[ A, \frac{B}{2} \right] \right]
    e^{- A h} e^{- B h / 2}\,.
\end{align} 
We then Taylor expand the $F(h)$ function around
$h = 0 $.  Noting that $F(0) = 0$ and the first two derivatives of $F(h)$ evaluated at $h=0$ are zero, i.e.,
  \begin{align}
    F' (0) & =  - \frac{B}{2} + A - A + \frac{B}{2} = 0\,,\\
    F'' (0) & =  \frac{B A}{2} - \frac{A B}{2} + \frac{B^2}{4} + A
    \frac{B}{2} - \frac{B}{2} A - \frac{B^2}{4} = 0\,,
\end{align}
we obtain $F(h) = F'''(h)h^3/6! +\mathcal{O}(h^4)$, where
\begin{equation}
 F''' (0)  =  \left[ \frac{B}{2}, \left[ \frac{B}{2}, A \right] \right] +
    \left[ \frac{B}{2}, \left[ A, \frac{B}{2} \right] \right] + \left[
    \frac{B}{2}, \left[ A, \frac{B}{2} \right] \right] + \left[ A, \left[ A,
    \frac{B}{2} \right] \right]
     =  \left[ A + \frac{B}{2}, \left[ A, \frac{B}{2} \right] \right]\,.
\end{equation}
 
\section{Discrete Fourier Transform (DFT)}\label{AppB}
\label{AppendB}
\quad For an arbitrary one-dimensional array $\phi_a$, the DFT $\tilde{\phi}_l$  is defined to be
\begin{eqnarray}
  \widetilde{\phi_l} & = & \sum_{a = 0}^{N - 1} \phi_a e^{- \frac{i 2 \pi}{N}
  a l} \,.
\label{DFT}
\end{eqnarray}
If  $\phi_a$'s are real, then $\widetilde{\phi}_k = \widetilde{\phi}_{N-k}$. So,
evaluating half of $\{ \widetilde{\phi}_k \}$ is enough, which is a fast
discrete Fourier transform (FFT).
Let us consider the centered finite difference scheme for the second derivative of $\phi_i$ as follows
\begin{equation}
\phi''_i \simeq \frac{\phi_{i+1} - 2\phi_i + \phi_{i-1}}{ (\Delta x)^2}\,,
\label{2d}
\end{equation}  
where the error is $\mathcal{O}((\Delta x) ^2)$. Using Eq.~(\ref{DFT}), the DFT of Eq.~(\ref{2d}) reads as
\begin{align}
\tilde{\phi}_l'' & \simeq \frac{1}{(\Delta x)^2} \left[  \sum_{a = 0}^{N - 1} \phi_{a+1} e^{- \frac{i 2 \pi}{N}
  a l} -2  \sum_{a = 0}^{N - 1} \phi_a e^{- \frac{i 2 \pi}{N}
  a l} +  \sum_{a = 0}^{N - 1} \phi_{a-1} e^{- \frac{i 2 \pi}{N}
  a l}   \right]\,,\\
& \simeq \frac{1}{(\Delta x)^2} \left[  \sum_{a = 1}^{N } \phi_{a} e^{- \frac{i 2 \pi}{N}
  (a-1) l} -2  \sum_{a = 0}^{N - 1} \phi_a e^{- \frac{i 2 \pi}{N}
  a l} +  \sum_{a = -1}^{N - 2} \phi_{a} e^{- \frac{i 2 \pi}{N}
  (a+1) l}   \right]\,,\\
& \simeq \frac{ (e^{ \frac{i 2 \pi l}{N}}
  -2 +e^{- \frac{i 2 \pi l }{N}}
   )  }{(\Delta x)^2} \left[  \sum_{a = 0}^{N-1 } \phi_{a} e^{- \frac{i 2 \pi}{N}
  a l}\right]\,,\label{bc}\\
& \simeq \frac{ (e^{ \frac{i 2 \pi l}{N}}
  -2 +e^{- \frac{i 2 \pi l }{N}}
   ) }{(\Delta x) ^2} \tilde{\phi}_l\,,
\end{align}
where we have assumed periodic boundary conditions to obtain Eq.~(\ref{bc}).
Thus, the Laplace operator in momentum space is expressed as 
\begin{eqnarray}
(\nabla^2)_{a} = \frac{e^{- \frac{i 2 \pi}{N} a} +
  e^{\frac{i 2 \pi}{N} a} - 2}{\Delta x^2} \,.\\
\end{eqnarray} 
Generalization of this formula from one to three dimensions is given by
\begin{equation}
  \label{laplace_in_k_space} (\nabla^2)_{a, b, c}  =  \frac{e^{- \frac{i 2 \pi}{N} a} +
  e^{\frac{i 2 \pi}{N} a} - 2}{\Delta x^2} + \frac{e^{- \frac{i 2 \pi}{N} b} +
  e^{\frac{i 2 \pi}{N} b} - 2}{\Delta y^2} + \frac{e^{- \frac{i 2 \pi}{N} c} +
  e^{\frac{i 2 \pi}{N} c} - 2}{\Delta z^2} \,.
\end{equation}

\bibliographystyle{JHEP}
\bibliography{AxionMergerHLS}

\providecommand{\href}[2]{#2}\begingroup\raggedright\begin{thebibliography}{100}

\bibitem{Hertzberg:2018zte}
M.~P. Hertzberg and E.~D. Schiappacasse, \emph{{Dark Matter Axion Clump
  Resonance of Photons}},
  \href{https://doi.org/10.1088/1475-7516/2018/11/004}{\emph{JCAP} {\bfseries
  1811} (2018) 004}, [\href{https://arxiv.org/abs/1805.00430}{{\ttfamily
  1805.00430}}].

\bibitem{Peebles:2013hla}
P.~Peebles, \emph{{Dark Matter}},
  \href{https://doi.org/10.1073/pnas.1308786111}{\emph{Proc. Nat. Acad. Sci.}
  {\bfseries 112} (2015) 2246},
  [\href{https://arxiv.org/abs/1305.6859}{{\ttfamily 1305.6859}}].

\bibitem{Peccei:1977hh}
R.~D. Peccei and H.~R. Quinn, \emph{{CP Conservation in the Presence of
  Instantons}}, \href{https://doi.org/10.1103/PhysRevLett.38.1440}{\emph{Phys.
  Rev. Lett.} {\bfseries 38} (1977) 1440--1443}.

\bibitem{Weinberg:1977ma}
S.~Weinberg, \emph{{A New Light Boson?}},
  \href{https://doi.org/10.1103/PhysRevLett.40.223}{\emph{Phys. Rev. Lett.}
  {\bfseries 40} (1978) 223--226}.

\bibitem{Wilczek:1977pj}
F.~Wilczek, \emph{{Problem of Strong $P$ and $T$ Invariance in the Presence of
  Instantons}}, \href{https://doi.org/10.1103/PhysRevLett.40.279}{\emph{Phys.
  Rev. Lett.} {\bfseries 40} (1978) 279--282}.

\bibitem{Svrcek:2006yi}
P.~Svrcek and E.~Witten, \emph{{Axions In String Theory}},
  \href{https://doi.org/10.1088/1126-6708/2006/06/051}{\emph{JHEP} {\bfseries
  06} (2006) 051}, [\href{https://arxiv.org/abs/hep-th/0605206}{{\ttfamily
  hep-th/0605206}}].

\bibitem{Douglas:2006es}
M.~R. Douglas and S.~Kachru, \emph{{Flux compactification}},
  \href{https://doi.org/10.1103/RevModPhys.79.733}{\emph{Rev. Mod. Phys.}
  {\bfseries 79} (2007) 733--796},
  [\href{https://arxiv.org/abs/hep-th/0610102}{{\ttfamily hep-th/0610102}}].

\bibitem{Arvanitaki:2009fg}
A.~Arvanitaki, S.~Dimopoulos, S.~Dubovsky, N.~Kaloper and J.~March-Russell,
  \emph{{String Axiverse}},
  \href{https://doi.org/10.1103/PhysRevD.81.123530}{\emph{Phys. Rev. D}
  {\bfseries 81} (2010) 123530},
  [\href{https://arxiv.org/abs/0905.4720}{{\ttfamily 0905.4720}}].

\bibitem{PhysRevD.42.1297}
C.~Hagmann, P.~Sikivie, N.~S. Sullivan and D.~B. Tanner, \emph{Results from a
  search for cosmic axions},
  \href{https://doi.org/10.1103/PhysRevD.42.1297}{\emph{Phys. Rev. D}
  {\bfseries 42} (Aug, 1990) 1297--1300}.

\bibitem{PhysRevD.64.092003}
S.~Asztalos, E.~Daw, H.~Peng, L.~J. Rosenberg, C.~Hagmann, D.~Kinion et~al.,
  \emph{Large-scale microwave cavity search for dark-matter axions},
  \href{https://doi.org/10.1103/PhysRevD.64.092003}{\emph{Phys. Rev. D}
  {\bfseries 64} (Oct, 2001) 092003}.

\bibitem{PhysRevLett.120.151301}
{\scshape ADMX Collaboration} collaboration, N.~Du, N.~Force, R.~Khatiwada,
  E.~Lentz, R.~Ottens, L.~J. Rosenberg et~al., \emph{Search for invisible axion
  dark matter with the axion dark matter experiment},
  \href{https://doi.org/10.1103/PhysRevLett.120.151301}{\emph{Phys. Rev. Lett.}
  {\bfseries 120} (Apr, 2018) 151301}.

\bibitem{Zhong:2018rsr}
{\scshape HAYSTAC} collaboration, L.~Zhong et~al., \emph{{Results from phase 1
  of the HAYSTAC microwave cavity axion experiment}},
  \href{https://doi.org/10.1103/PhysRevD.97.092001}{\emph{Phys. Rev. D}
  {\bfseries 97} (2018) 092001},
  [\href{https://arxiv.org/abs/1803.03690}{{\ttfamily 1803.03690}}].

\bibitem{Anastassopoulos:2017ftl}
{\scshape CAST} collaboration, V.~Anastassopoulos et~al., \emph{{New CAST Limit
  on the Axion-Photon Interaction}},
  \href{https://doi.org/10.1038/nphys4109}{\emph{Nature Phys.} {\bfseries 13}
  (2017) 584--590}, [\href{https://arxiv.org/abs/1705.02290}{{\ttfamily
  1705.02290}}].

\bibitem{Armengaud_2014}
E.~Armengaud, F.~T. Avignone, M.~Betz, P.~Brax, P.~Brun, G.~Cantatore et~al.,
  \emph{Conceptual design of the international axion observatory ({IAXO})},
  \href{https://doi.org/10.1088/1748-0221/9/05/t05002}{\emph{Journal of
  Instrumentation} {\bfseries 9} (may, 2014) T05002--T05002}.

\bibitem{Graham:2011qk}
P.~W. Graham and S.~Rajendran, \emph{{Axion Dark Matter Detection with Cold
  Molecules}}, \href{https://doi.org/10.1103/PhysRevD.84.055013}{\emph{Phys.
  Rev. D} {\bfseries 84} (2011) 055013},
  [\href{https://arxiv.org/abs/1101.2691}{{\ttfamily 1101.2691}}].

\bibitem{Budker:2013hfa}
D.~Budker, P.~W. Graham, M.~Ledbetter, S.~Rajendran and A.~Sushkov,
  \emph{{Proposal for a Cosmic Axion Spin Precession Experiment (CASPEr)}},
  \href{https://doi.org/10.1103/PhysRevX.4.021030}{\emph{Phys. Rev. X}
  {\bfseries 4} (2014) 021030},
  [\href{https://arxiv.org/abs/1306.6089}{{\ttfamily 1306.6089}}].

\bibitem{Barbieri:2016vwg}
R.~Barbieri, C.~Braggio, G.~Carugno, C.~Gallo, A.~Lombardi, A.~Ortolan et~al.,
  \emph{{Searching for galactic axions through magnetized media: the QUAX
  proposal}}, \href{https://doi.org/10.1016/j.dark.2017.01.003}{\emph{Phys.
  Dark Univ.} {\bfseries 15} (2017) 135--141},
  [\href{https://arxiv.org/abs/1606.02201}{{\ttfamily 1606.02201}}].

\bibitem{PhysRevLett.113.201301}
P.~Sikivie, \emph{Axion dark matter detection using atomic transitions},
  \href{https://doi.org/10.1103/PhysRevLett.113.201301}{\emph{Phys. Rev. Lett.}
  {\bfseries 113} (Nov, 2014) 201301}.

\bibitem{2018IJMPA..3344030F}
V.~V. {Flambaum}, H.~B. {Tran Tan}, I.~B. {Samsonov}, Y.~V. {Stadnik} and
  D.~{Budker}, \emph{{Resonant detection and production of axions with atoms}},
  \href{https://doi.org/10.1142/S0217751X1844030X}{\emph{International Journal
  of Modern Physics A} {\bfseries 33} (Nov., 2018) 1844030}.

\bibitem{Abramowicz:2017zbp}
M.~A. Abramowicz, M.~Bejger and M.~Wielgus, \emph{{Collisions of neutron stars
  with primordial black holes as fast radio bursts engines}},
  \href{https://doi.org/10.3847/1538-4357/aae64a}{\emph{Astrophys. J.}
  {\bfseries 868} (2018) 17},
  [\href{https://arxiv.org/abs/1704.05931}{{\ttfamily 1704.05931}}].

\bibitem{Iwazaki:2014wka}
A.~Iwazaki, \emph{{Axion stars and fast radio bursts}},
  \href{https://doi.org/10.1103/PhysRevD.91.023008}{\emph{Phys. Rev. D}
  {\bfseries 91} (2015) 023008},
  [\href{https://arxiv.org/abs/1410.4323}{{\ttfamily 1410.4323}}].

\bibitem{Iwazaki:2017rtb}
A.~Iwazaki, \emph{{Axion Stars and Repeating Fast Radio Bursts with Finite
  Bandwidths}},  \href{https://arxiv.org/abs/1707.04827}{{\ttfamily
  1707.04827}}.

\bibitem{Caputo:2018vmy}
A.~Caputo, M.~Regis, M.~Taoso and S.~J. Witte, \emph{{Detecting the Stimulated
  Decay of Axions at RadioFrequencies}},
  \href{https://doi.org/10.1088/1475-7516/2019/03/027}{\emph{JCAP} {\bfseries
  03} (2019) 027}, [\href{https://arxiv.org/abs/1811.08436}{{\ttfamily
  1811.08436}}].

\bibitem{Huang:2018lxq}
F.~P. Huang, K.~Kadota, T.~Sekiguchi and H.~Tashiro, \emph{{Radio telescope
  search for the resonant conversion of cold dark matter axions from the
  magnetized astrophysical sources}},
  \href{https://doi.org/10.1103/PhysRevD.97.123001}{\emph{Phys. Rev. D}
  {\bfseries 97} (2018) 123001},
  [\href{https://arxiv.org/abs/1803.08230}{{\ttfamily 1803.08230}}].

\bibitem{Hook:2018iia}
A.~Hook, Y.~Kahn, B.~R. Safdi and Z.~Sun, \emph{{Radio Signals from Axion Dark
  Matter Conversion in Neutron Star Magnetospheres}},
  \href{https://doi.org/10.1103/PhysRevLett.121.241102}{\emph{Phys. Rev. Lett.}
  {\bfseries 121} (2018) 241102},
  [\href{https://arxiv.org/abs/1804.03145}{{\ttfamily 1804.03145}}].

\bibitem{Safdi:2018oeu}
B.~R. Safdi, Z.~Sun and A.~Y. Chen, \emph{{Detecting Axion Dark Matter with
  Radio Lines from Neutron Star Populations}},
  \href{https://doi.org/10.1103/PhysRevD.99.123021}{\emph{Phys. Rev. D}
  {\bfseries 99} (2019) 123021},
  [\href{https://arxiv.org/abs/1811.01020}{{\ttfamily 1811.01020}}].

\bibitem{Buckley:2020fmh}
J.~H. Buckley, P.~B. Dev, F.~Ferrer and F.~P. Huang, \emph{{Fast radio bursts
  from axion stars moving through pulsar magnetospheres}},
  \href{https://arxiv.org/abs/2004.06486}{{\ttfamily 2004.06486}}.

\bibitem{Arza:2020eik}
A.~Arza, T.~Schwetz and E.~Todarello, \emph{{How to suppress exponential growth
  -- on the parametric resonance of photons in an axion background}},
  \href{https://arxiv.org/abs/2004.01669}{{\ttfamily 2004.01669}}.

\bibitem{2010PhRvL.104d1301A}
S.~J. {Asztalos}, G.~{Carosi}, C.~{Hagmann}, D.~{Kinion}, K.~{van Bibber},
  M.~{Hotz} et~al., \emph{{SQUID-Based Microwave Cavity Search for Dark-Matter
  Axions}}, \href{https://doi.org/10.1103/PhysRevLett.104.041301}{\emph{prl}
  {\bfseries 104} (Jan., 2010) 041301},
  [\href{https://arxiv.org/abs/0910.5914}{{\ttfamily 0910.5914}}].

\bibitem{Schiappacasse:2017ham}
E.~D. Schiappacasse and M.~P. Hertzberg, \emph{{Analysis of Dark Matter Axion
  Clumps with Spherical Symmetry}},
  \href{https://doi.org/10.1088/1475-7516/2018/03/E01,
  10.1088/1475-7516/2018/01/037}{\emph{JCAP} {\bfseries 1801} (2018) 037},
  [\href{https://arxiv.org/abs/1710.04729}{{\ttfamily 1710.04729}}].

\bibitem{Hertzberg:2018lmt}
M.~P. Hertzberg and E.~D. Schiappacasse, \emph{{Scalar dark matter clumps with
  angular momentum}},
  \href{https://doi.org/10.1088/1475-7516/2018/08/028}{\emph{JCAP} {\bfseries
  08} (2018) 028}, [\href{https://arxiv.org/abs/1804.07255}{{\ttfamily
  1804.07255}}].

\bibitem{Liebling2012}
S.~L. Liebling and C.~Palenzuela, \emph{Dynamical boson stars},
  \href{https://doi.org/10.12942/lrr-2012-6}{\emph{Living Reviews in
  Relativity} {\bfseries 15} (May, 2012) 6}.

\bibitem{Horvat:2012aq}
D.~Horvat and A.~Marunovi\'c, \emph{{Dark energy-like stars from nonminimally
  coupled scalar field}},
  \href{https://doi.org/10.1088/0264-9381/30/14/145006}{\emph{Class. Quant.
  Grav.} {\bfseries 30} (2013) 145006},
  [\href{https://arxiv.org/abs/1212.3781}{{\ttfamily 1212.3781}}].

\bibitem{PhysRevD.96.084066}
L.~G. Collodel, B.~Kleihaus and J.~Kunz, \emph{Excited boson stars},
  \href{https://doi.org/10.1103/PhysRevD.96.084066}{\emph{Phys. Rev. D}
  {\bfseries 96} (Oct, 2017) 084066}.

\bibitem{Choi:2019mva}
G.~Choi, H.-J. He and E.~D. Schiappacasse, \emph{{Probing Dynamics of Boson
  Stars by Fast Radio Bursts and Gravitational Wave Detection}},
  \href{https://doi.org/10.1088/1475-7516/2019/10/043}{\emph{JCAP} {\bfseries
  10} (2019) 043}, [\href{https://arxiv.org/abs/1906.02094}{{\ttfamily
  1906.02094}}].

\bibitem{Tkachev:1986tr}
I.~Tkachev, \emph{{Coherent scalar field oscillations forming compact
  astrophysical objects}}, {\emph{Sov. Astron. Lett.} {\bfseries 12} (1986)
  305--308}.

\bibitem{Tkachev:1987cd}
I.~Tkachev, \emph{{An Axionic Laser in the Center of a Galaxy?}},
  \href{https://doi.org/10.1016/0370-2693(87)91318-9}{\emph{Phys. Lett. B}
  {\bfseries 191} (1987) 41--45}.

\bibitem{Tkachev:2014dpa}
I.~Tkachev, \emph{{Fast Radio Bursts and Axion Miniclusters}},
  \href{https://doi.org/10.1134/S0021364015010154}{\emph{JETP Lett.} {\bfseries
  101} (2015) 1--6}, [\href{https://arxiv.org/abs/1411.3900}{{\ttfamily
  1411.3900}}].

\bibitem{Levkov:2020txo}
D.~Levkov, A.~Panin and I.~Tkachev, \emph{{Radio-emission of axion stars}},
  \href{https://arxiv.org/abs/2004.05179}{{\ttfamily 2004.05179}}.

\bibitem{Guth:2014hsa}
A.~H. Guth, M.~P. Hertzberg and C.~Prescod-Weinstein, \emph{{Do Dark Matter
  Axions Form a Condensate with Long-Range Correlation?}},
  \href{https://doi.org/10.1103/PhysRevD.92.103513}{\emph{Phys. Rev.}
  {\bfseries D92} (2015) 103513},
  [\href{https://arxiv.org/abs/1412.5930}{{\ttfamily 1412.5930}}].

\bibitem{Kolb:1993zz}
E.~W. Kolb and I.~I. Tkachev, \emph{{Axion miniclusters and Bose stars}},
  \href{https://doi.org/10.1103/PhysRevLett.71.3051}{\emph{Phys. Rev. Lett.}
  {\bfseries 71} (1993) 3051--3054},
  [\href{https://arxiv.org/abs/hep-ph/9303313}{{\ttfamily hep-ph/9303313}}].

\bibitem{Levkov:2018kau}
D.~Levkov, A.~Panin and I.~Tkachev, \emph{{Gravitational Bose-Einstein
  condensation in the kinetic regime}},
  \href{https://doi.org/10.1103/PhysRevLett.121.151301}{\emph{Phys. Rev. Lett.}
  {\bfseries 121} (2018) 151301},
  [\href{https://arxiv.org/abs/1804.05857}{{\ttfamily 1804.05857}}].

\bibitem{Sikivie:2009qn}
P.~Sikivie and Q.~Yang, \emph{{Bose-Einstein Condensation of Dark Matter
  Axions}}, \href{https://doi.org/10.1103/PhysRevLett.103.111301}{\emph{Phys.
  Rev. Lett.} {\bfseries 103} (2009) 111301},
  [\href{https://arxiv.org/abs/0901.1106}{{\ttfamily 0901.1106}}].

\bibitem{Erken:2011dz}
O.~Erken, P.~Sikivie, H.~Tam and Q.~Yang, \emph{{Cosmic axion thermalization}},
  \href{https://doi.org/10.1103/PhysRevD.85.063520}{\emph{Phys. Rev.}
  {\bfseries D85} (2012) 063520},
  [\href{https://arxiv.org/abs/1111.1157}{{\ttfamily 1111.1157}}].

\bibitem{PhysRevD.23.852}
A.~Vilenkin, \emph{Gravitational field of vacuum domain walls and strings},
  \href{https://doi.org/10.1103/PhysRevD.23.852}{\emph{Phys. Rev. D} {\bfseries
  23} (Feb, 1981) 852--857}.

\bibitem{Kawasaki:2014sqa}
M.~Kawasaki, K.~Saikawa and T.~Sekiguchi, \emph{{Axion dark matter from
  topological defects}},
  \href{https://doi.org/10.1103/PhysRevD.91.065014}{\emph{Phys. Rev. D}
  {\bfseries 91} (2015) 065014},
  [\href{https://arxiv.org/abs/1412.0789}{{\ttfamily 1412.0789}}].

\bibitem{PhysRevLett.43.103}
J.~E. Kim, \emph{Weak-interaction singlet and strong $\mathrm{CP}$ invariance},
  \href{https://doi.org/10.1103/PhysRevLett.43.103}{\emph{Phys. Rev. Lett.}
  {\bfseries 43} (Jul, 1979) 103--107}.

\bibitem{SHIFMAN1980493}
M.~Shifman, A.~Vainshtein and V.~Zakharov, \emph{Can confinement ensure natural
  cp invariance of strong interactions?},
  \href{https://doi.org/https://doi.org/10.1016/0550-3213(80)90209-6}{\emph{Nuclear
  Physics B} {\bfseries 166} (1980) 493 -- 506}.

\bibitem{Zhitnitsky:1980tq}
A.~Zhitnitsky, \emph{{On Possible Suppression of the Axion Hadron Interactions.
  (In Russian)}}, {\emph{Sov. J. Nucl. Phys.} {\bfseries 31} (1980) 260}.

\bibitem{DINE1981199}
M.~Dine, W.~Fischler and M.~Srednicki, \emph{A simple solution to the strong cp
  problem with a harmless axion},
  \href{https://doi.org/https://doi.org/10.1016/0370-2693(81)90590-6}{\emph{Physics
  Letters B} {\bfseries 104} (1981) 199 -- 202}.

\bibitem{Raffelt:2006cw}
G.~G. Raffelt, \emph{{Astrophysical axion bounds}},
  \href{https://doi.org/10.1007/978-3-540-73518-2\_3}{\emph{Lect. Notes Phys.}
  {\bfseries 741} (2008) 51--71},
  [\href{https://arxiv.org/abs/hep-ph/0611350}{{\ttfamily hep-ph/0611350}}].

\bibitem{Hertzberg:2020hsz}
M.~P. Hertzberg, E.~D. Schiappacasse and T.~T. Yanagida, \emph{{Axion Stars
  Nucleation in Dark Mini-Halos around Primordial Black Holes}},
  \href{https://arxiv.org/abs/2001.07476}{{\ttfamily 2001.07476}}.

\bibitem{Hawking:1971ei}
S.~Hawking, \emph{{Gravitationally collapsed objects of very low mass}},
  {\emph{Mon. Not. Roy. Astron. Soc.} {\bfseries 152} (1971) 75}.

\bibitem{Carr:1974nx}
B.~J. Carr and S.~W. Hawking, \emph{{Black holes in the early Universe}},
  {\emph{Mon. Not. Roy. Astron. Soc.} {\bfseries 168} (1974) 399}.

\bibitem{Carr:1975qj}
B.~J. Carr, \emph{{The Primordial black hole mass spectrum}},
  \href{https://doi.org/10.1086/153853}{\emph{Astrophys. J.} {\bfseries 201}
  (1975) 1--19}.

\bibitem{Kawasaki:1997ju}
M.~Kawasaki, N.~Sugiyama and T.~Yanagida, \emph{{Primordial black hole
  formation in a double inflation model in supergravity}},
  \href{https://doi.org/10.1103/PhysRevD.57.6050}{\emph{Phys. Rev.} {\bfseries
  D57} (1998) 6050--6056},
  [\href{https://arxiv.org/abs/hep-ph/9710259}{{\ttfamily hep-ph/9710259}}].

\bibitem{GarciaBellido:1996qt}
J.~Garcia-Bellido, A.~D. Linde and D.~Wands, \emph{{Density perturbations and
  black hole formation in hybrid inflation}},
  \href{https://doi.org/10.1103/PhysRevD.54.6040}{\emph{Phys. Rev.} {\bfseries
  D54} (1996) 6040--6058},
  [\href{https://arxiv.org/abs/astro-ph/9605094}{{\ttfamily
  astro-ph/9605094}}].

\bibitem{Boucenna:2017ghj}
S.~M. Boucenna, F.~Kuhnel, T.~Ohlsson and L.~Visinelli, \emph{{Novel
  Constraints on Mixed Dark-Matter Scenarios of Primordial Black Holes and
  WIMPs}}, \href{https://doi.org/10.1088/1475-7516/2018/07/003}{\emph{JCAP}
  {\bfseries 07} (2018) 003},
  [\href{https://arxiv.org/abs/1712.06383}{{\ttfamily 1712.06383}}].

\bibitem{Nakama:2015nea}
T.~Nakama and T.~Suyama, \emph{{Primordial black holes as a novel probe of
  primordial gravitational waves}},
  \href{https://doi.org/10.1103/PhysRevD.92.121304}{\emph{Phys. Rev. D}
  {\bfseries 92} (2015) 121304},
  [\href{https://arxiv.org/abs/1506.05228}{{\ttfamily 1506.05228}}].

\bibitem{Nakama:2016enz}
T.~Nakama and T.~Suyama, \emph{{Primordial black holes as a novel probe of
  primordial gravitational waves. II: Detailed analysis}},
  \href{https://doi.org/10.1103/PhysRevD.94.043507}{\emph{Phys. Rev. D}
  {\bfseries 94} (2016) 043507},
  [\href{https://arxiv.org/abs/1605.04482}{{\ttfamily 1605.04482}}].

\bibitem{Hertzberg:2019exb}
M.~P. Hertzberg, E.~D. Schiappacasse and T.~T. Yanagida, \emph{{Warning about
  Dark Matter Direct Detection in the Presence of LIGO-Motivated Primordial
  Black Holes}},  \href{https://arxiv.org/abs/1910.10575}{{\ttfamily
  1910.10575}}.

\bibitem{Abbott:2017vtc}
{\scshape LIGO Scientific, VIRGO} collaboration, B.~P. Abbott et~al.,
  \emph{{GW170104: Observation of a 50-Solar-Mass Binary Black Hole Coalescence
  at Redshift 0.2}}, \href{https://doi.org/10.1103/PhysRevLett.118.221101,
  10.1103/PhysRevLett.121.129901}{\emph{Phys. Rev. Lett.} {\bfseries 118}
  (2017) 221101}, [\href{https://arxiv.org/abs/1706.01812}{{\ttfamily
  1706.01812}}].

\bibitem{Abbott:2017iws}
{\scshape LIGO Scientific, Virgo} collaboration, B.~P. Abbott et~al.,
  \emph{{Search for intermediate mass black hole binaries in the first
  observing run of Advanced LIGO}},
  \href{https://doi.org/10.1103/PhysRevD.96.022001}{\emph{Phys. Rev.}
  {\bfseries D96} (2017) 022001},
  [\href{https://arxiv.org/abs/1704.04628}{{\ttfamily 1704.04628}}].

\bibitem{Abbott:2016drs}
{\scshape LIGO Scientific, Virgo} collaboration, B.~P. Abbott et~al.,
  \emph{{Supplement: The Rate of Binary Black Hole Mergers Inferred from
  Advanced LIGO Observations Surrounding GW150914}},
  \href{https://doi.org/10.3847/0067-0049/227/2/14}{\emph{Astrophys. J. Suppl.}
  {\bfseries 227} (2016) 14},
  [\href{https://arxiv.org/abs/1606.03939}{{\ttfamily 1606.03939}}].

\bibitem{TheLIGOScientific:2016htt}
{\scshape LIGO Scientific, Virgo} collaboration, B.~P. Abbott et~al.,
  \emph{{Astrophysical Implications of the Binary Black-Hole Merger GW150914}},
  \href{https://doi.org/10.3847/2041-8205/818/2/L22}{\emph{Astrophys. J.}
  {\bfseries 818} (2016) L22},
  [\href{https://arxiv.org/abs/1602.03846}{{\ttfamily 1602.03846}}].

\bibitem{Abbott:2016blz}
{\scshape LIGO Scientific, Virgo} collaboration, B.~P. Abbott et~al.,
  \emph{{Observation of Gravitational Waves from a Binary Black Hole Merger}},
  \href{https://doi.org/10.1103/PhysRevLett.116.061102}{\emph{Phys. Rev. Lett.}
  {\bfseries 116} (2016) 061102},
  [\href{https://arxiv.org/abs/1602.03837}{{\ttfamily 1602.03837}}].

\bibitem{Sasaki:2016jop}
M.~Sasaki, T.~Suyama, T.~Tanaka and S.~Yokoyama, \emph{{Primordial Black Hole
  Scenario for the Gravitational Wave Event GW150914}},
  \href{https://doi.org/10.1103/PhysRevLett.117.061101}{\emph{Phys. Rev. Lett.}
  {\bfseries 117} (2016) 061101},
  [\href{https://arxiv.org/abs/1603.08338}{{\ttfamily 1603.08338}}].

\bibitem{Eroshenko:2016hmn}
Y.~N. Eroshenko, \emph{{Gravitational waves from primordial black holes
  collisions in binary systems}},
  \href{https://doi.org/10.1088/1742-6596/1051/1/012010}{\emph{J. Phys. Conf.
  Ser.} {\bfseries 1051} (2018) 012010},
  [\href{https://arxiv.org/abs/1604.04932}{{\ttfamily 1604.04932}}].

\bibitem{Ali-Haimoud:2017rtz}
Y.~Ali-Haïmoud, E.~D. Kovetz and M.~Kamionkowski, \emph{{Merger rate of
  primordial black-hole binaries}},
  \href{https://doi.org/10.1103/PhysRevD.96.123523}{\emph{Phys. Rev.}
  {\bfseries D96} (2017) 123523},
  [\href{https://arxiv.org/abs/1709.06576}{{\ttfamily 1709.06576}}].

\bibitem{Raidal:2017mfl}
M.~Raidal, V.~Vaskonen and H.~Veermäe, \emph{{Gravitational Waves from
  Primordial Black Hole Mergers}},
  \href{https://doi.org/10.1088/1475-7516/2017/09/037}{\emph{JCAP} {\bfseries
  1709} (2017) 037}, [\href{https://arxiv.org/abs/1707.01480}{{\ttfamily
  1707.01480}}].

\bibitem{Fukunaga:2020mvq}
H.~Fukunaga, N.~Kitajima and Y.~Urakawa, \emph{{Can axion clumps be formed in a
  pre-inflationary scenario?}},
  \href{https://arxiv.org/abs/2004.08929}{{\ttfamily 2004.08929}}.

\bibitem{PRESKILL1983127}
J.~Preskill, M.~B. Wise and F.~Wilczek, \emph{Cosmology of the invisible
  axion},
  \href{https://doi.org/https://doi.org/10.1016/0370-2693(83)90637-8}{\emph{Physics
  Letters B} {\bfseries 120} (1983) 127 -- 132}.

\bibitem{Abbott:1982af}
L.~Abbott and P.~Sikivie, \emph{{A Cosmological Bound on the Invisible Axion}},
  \href{https://doi.org/10.1016/0370-2693(83)90638-X}{\emph{Phys.\ Lett.\ B}
  {\bfseries 120} (1983) 133--136}.

\bibitem{Dine:1982ah}
M.~Dine and W.~Fischler, \emph{{The Not So Harmless Axion}},
  \href{https://doi.org/10.1016/0370-2693(83)90639-1}{\emph{Phys. Lett. B}
  {\bfseries 120} (1983) 137--141}.

\bibitem{PhysRevLett.52.1725}
S.-Y. Pi, \emph{Inflation without tears: A realistic cosmological model},
  \href{https://doi.org/10.1103/PhysRevLett.52.1725}{\emph{Phys. Rev. Lett.}
  {\bfseries 52} (May, 1984) 1725--1728}.

\bibitem{Linde:1991km}
A.~D. Linde, \emph{{Axions in inflationary cosmology}},
  \href{https://doi.org/10.1016/0370-2693(91)90130-I}{\emph{Phys. Lett. B}
  {\bfseries 259} (1991) 38--47}.

\bibitem{Wilczek:2004cr}
F.~Wilczek, \emph{{A Model of anthropic reasoning, addressing the dark to
  ordinary matter coincidence}},
  \href{https://arxiv.org/abs/hep-ph/0408167}{{\ttfamily hep-ph/0408167}}.

\bibitem{PhysRevD.73.023505}
M.~Tegmark, A.~Aguirre, M.~J. Rees and F.~Wilczek, \emph{Dimensionless
  constants, cosmology, and other dark matters},
  \href{https://doi.org/10.1103/PhysRevD.73.023505}{\emph{Phys. Rev. D}
  {\bfseries 73} (Jan, 2006) 023505}.

\bibitem{1983PhLB..126..178A}
M.~{Axenides}, R.~{Brandenberger} and M.~{Turner}, \emph{{Development of axion
  perturbations in an axion dominated universe}},
  \href{https://doi.org/10.1016/0370-2693(83)90586-5}{\emph{Physics Letters B}
  {\bfseries 126} (June, 1983) 178--182}.

\bibitem{PhysRevD.32.3178}
D.~Seckel and M.~S. Turner, \emph{``isothermal'' density perturbations in an
  axion-dominated inflationary universe},
  \href{https://doi.org/10.1103/PhysRevD.32.3178}{\emph{Phys. Rev. D}
  {\bfseries 32} (Dec, 1985) 3178--3183}.

\bibitem{PhysRevLett.66.5}
M.~S. Turner and F.~Wilczek, \emph{Inflationary axion cosmology},
  \href{https://doi.org/10.1103/PhysRevLett.66.5}{\emph{Phys. Rev. Lett.}
  {\bfseries 66} (Jan, 1991) 5--8}.

\bibitem{Fox:2004kb}
P.~Fox, A.~Pierce and S.~D. Thomas, \emph{{Probing a QCD string axion with
  precision cosmological measurements}},
  \href{https://arxiv.org/abs/hep-th/0409059}{{\ttfamily hep-th/0409059}}.

\bibitem{Hertzberg:2008wr}
M.~P. Hertzberg, M.~Tegmark and F.~Wilczek, \emph{{Axion Cosmology and the
  Energy Scale of Inflation}},
  \href{https://doi.org/10.1103/PhysRevD.78.083507}{\emph{Phys. Rev.}
  {\bfseries D78} (2008) 083507},
  [\href{https://arxiv.org/abs/0807.1726}{{\ttfamily 0807.1726}}].

\bibitem{Kawasaki:2013iha}
M.~Kawasaki, T.~T. Yanagida and K.~Yoshino, \emph{{Domain wall and isocurvature
  perturbation problems in axion models}},
  \href{https://doi.org/10.1088/1475-7516/2013/11/030}{\emph{JCAP} {\bfseries
  1311} (2013) 030}, [\href{https://arxiv.org/abs/1305.5338}{{\ttfamily
  1305.5338}}].

\bibitem{Arvanitaki:2010sy}
A.~Arvanitaki and S.~Dubovsky, \emph{{Exploring the String Axiverse with
  Precision Black Hole Physics}},
  \href{https://doi.org/10.1103/PhysRevD.83.044026}{\emph{Phys. Rev.}
  {\bfseries D83} (2011) 044026},
  [\href{https://arxiv.org/abs/1004.3558}{{\ttfamily 1004.3558}}].

\bibitem{Daido:2018dmu}
R.~Daido, F.~Takahashi and N.~Yokozaki, \emph{{Enhanced axion–photon coupling
  in GUT with hidden photon}},
  \href{https://doi.org/10.1016/j.physletb.2018.03.039}{\emph{Phys. Lett.}
  {\bfseries B780} (2018) 538--542},
  [\href{https://arxiv.org/abs/1801.10344}{{\ttfamily 1801.10344}}].

\bibitem{Duffy:2009ig}
L.~D. Duffy and K.~van Bibber, \emph{{Axions as Dark Matter Particles}},
  \href{https://doi.org/10.1088/1367-2630/11/10/105008}{\emph{New J. Phys.}
  {\bfseries 11} (2009) 105008},
  [\href{https://arxiv.org/abs/0904.3346}{{\ttfamily 0904.3346}}].

\bibitem{Masso:2002ip}
E.~Masso, \emph{{Axions and axion like particles}},
  \href{https://doi.org/10.1016/S0920-5632(02)01893-5}{\emph{Nucl. Phys. B
  Proc. Suppl.} {\bfseries 114} (2003) 67--73},
  [\href{https://arxiv.org/abs/hep-ph/0209132}{{\ttfamily hep-ph/0209132}}].

\bibitem{Marsh:2015xka}
D.~J.~E. Marsh, \emph{{Axion Cosmology}},
  \href{https://doi.org/10.1016/j.physrep.2016.06.005}{\emph{Phys. Rept.}
  {\bfseries 643} (2016) 1--79},
  [\href{https://arxiv.org/abs/1510.07633}{{\ttfamily 1510.07633}}].

\bibitem{NELSON1984387}
A.~Nelson, \emph{Naturally weak cp violation},
  \href{https://doi.org/https://doi.org/10.1016/0370-2693(84)92025-2}{\emph{Physics
  Letters B} {\bfseries 136} (1984) 387 -- 391}.

\bibitem{PhysRevLett.53.329}
S.~M. Barr, \emph{Solving the strong cp problem without the peccei-quinn
  symmetry}, \href{https://doi.org/10.1103/PhysRevLett.53.329}{\emph{Phys. Rev.
  Lett.} {\bfseries 53} (Jul, 1984) 329--332}.

\bibitem{Carena:2019nnd}
M.~Carena, D.~Liu, J.~Liu, N.~R. Shah, C.~E.~M. Wagner and X.-P. Wang,
  \emph{{$\nu$ solution to the strong CP problem}},
  \href{https://doi.org/10.1103/PhysRevD.100.094018}{\emph{Phys. Rev.}
  {\bfseries D100} (2019) 094018},
  [\href{https://arxiv.org/abs/1904.05360}{{\ttfamily 1904.05360}}].

\bibitem{Choi:2019omm}
G.~Choi and T.~T. Yanagida, \emph{{Solving the strong CP problem with
  horizontal gauge symmetry}},
  \href{https://doi.org/10.1103/PhysRevD.100.095023}{\emph{Phys. Rev.}
  {\bfseries D100} (2019) 095023},
  [\href{https://arxiv.org/abs/1909.04317}{{\ttfamily 1909.04317}}].

\bibitem{Hertzberg:2016tal}
M.~P. Hertzberg, \emph{{Quantum and Classical Behavior in Interacting Bosonic
  Systems}}, \href{https://doi.org/10.1088/1475-7516/2016/11/037}{\emph{JCAP}
  {\bfseries 1611} (2016) 037},
  [\href{https://arxiv.org/abs/1609.01342}{{\ttfamily 1609.01342}}].

\bibitem{diCortona:2015ldu}
G.~Grilli~di Cortona, E.~Hardy, J.~Pardo~Vega and G.~Villadoro, \emph{{The QCD
  axion, precisely}},
  \href{https://doi.org/10.1007/JHEP01(2016)034}{\emph{JHEP} {\bfseries 01}
  (2016) 034}, [\href{https://arxiv.org/abs/1511.02867}{{\ttfamily
  1511.02867}}].

\bibitem{Chavanis:2011zi}
P.-H. Chavanis, \emph{{Mass-radius relation of Newtonian self-gravitating
  Bose-Einstein condensates with short-range interactions: I. Analytical
  results}}, \href{https://doi.org/10.1103/PhysRevD.84.043531}{\emph{Phys. Rev.
  D} {\bfseries 84} (2011) 043531},
  [\href{https://arxiv.org/abs/1103.2050}{{\ttfamily 1103.2050}}].

\bibitem{Chavanis:2011zm}
P.~Chavanis and L.~Delfini, \emph{{Mass-radius relation of Newtonian
  self-gravitating Bose-Einstein condensates with short-range interactions: II.
  Numerical results}},
  \href{https://doi.org/10.1103/PhysRevD.84.043532}{\emph{Phys. Rev. D}
  {\bfseries 84} (2011) 043532},
  [\href{https://arxiv.org/abs/1103.2054}{{\ttfamily 1103.2054}}].

\bibitem{Kolb:1993hw}
E.~W. Kolb and I.~I. Tkachev, \emph{{Nonlinear axion dynamics and formation of
  cosmological pseudosolitons}},
  \href{https://doi.org/10.1103/PhysRevD.49.5040}{\emph{Phys. Rev. D}
  {\bfseries 49} (1994) 5040--5051},
  [\href{https://arxiv.org/abs/astro-ph/9311037}{{\ttfamily
  astro-ph/9311037}}].

\bibitem{Poon:2006}
T.-C. Poon and T.~Kim, \emph{{Engineering optics with Matlab}}, {\emph{World
  Scientific Publishing} {\bfseries Ch. 3} (2006) }.

\bibitem{Zhang:2008}
Q.~Zhang and M.~I. Hayee, \emph{{Symmetrized split-step fourier scheme to
  control global simulation accuracy in fiber-optic communication systems}},
  {\emph{J. Lightwave Technol.} {\bfseries 26} (2008) }.

\bibitem{Guzman:2018evm}
F.~Guzmán and A.~A. Avilez, \emph{{Head-on collision of multistate ultralight
  BEC dark matter configurations}},
  \href{https://doi.org/10.1103/PhysRevD.97.116003}{\emph{Phys. Rev. D}
  {\bfseries 97} (2018) 116003},
  [\href{https://arxiv.org/abs/1804.08670}{{\ttfamily 1804.08670}}].

\bibitem{Schwabe:2016rze}
B.~Schwabe, J.~C. Niemeyer and J.~F. Engels, \emph{{Simulations of solitonic
  core mergers in ultralight axion dark matter cosmologies}},
  \href{https://doi.org/10.1103/PhysRevD.94.043513}{\emph{Phys. Rev.}
  {\bfseries D94} (2016) 043513},
  [\href{https://arxiv.org/abs/1606.05151}{{\ttfamily 1606.05151}}].

\bibitem{Amin:2019ums}
M.~A. Amin and P.~Mocz, \emph{{Formation, gravitational clustering, and
  interactions of nonrelativistic solitons in an expanding universe}},
  \href{https://doi.org/10.1103/PhysRevD.100.063507}{\emph{Phys. Rev. D}
  {\bfseries 100} (2019) 063507},
  [\href{https://arxiv.org/abs/1902.07261}{{\ttfamily 1902.07261}}].

\bibitem{Cotner:2016aaq}
E.~Cotner, \emph{{Collisional interactions between self-interacting
  nonrelativistic boson stars: Effective potential analysis and numerical
  simulations}}, \href{https://doi.org/10.1103/PhysRevD.94.063503}{\emph{Phys.
  Rev. D} {\bfseries 94} (2016) 063503},
  [\href{https://arxiv.org/abs/1608.00547}{{\ttfamily 1608.00547}}].

\bibitem{Levkov:2016rkk}
D.~G. Levkov, A.~G. Panin and I.~I. Tkachev, \emph{{Relativistic axions from
  collapsing Bose stars}},
  \href{https://doi.org/10.1103/PhysRevLett.118.011301}{\emph{Phys. Rev. Lett.}
  {\bfseries 118} (2017) 011301},
  [\href{https://arxiv.org/abs/1609.03611}{{\ttfamily 1609.03611}}].

\bibitem{Paredes:2015wga}
A.~Paredes and H.~Michinel, \emph{{Interference of Dark Matter Solitons and
  Galactic Offsets}},
  \href{https://doi.org/10.1016/j.dark.2016.02.003}{\emph{Phys. Dark Univ.}
  {\bfseries 12} (2016) 50--55},
  [\href{https://arxiv.org/abs/1512.05121}{{\ttfamily 1512.05121}}].

\bibitem{PhysRevD.33.3495}
A.~K. Drukier, K.~Freese and D.~N. Spergel, \emph{Detecting cold dark-matter
  candidates}, \href{https://doi.org/10.1103/PhysRevD.33.3495}{\emph{Phys. Rev.
  D} {\bfseries 33} (Jun, 1986) 3495--3508}.

\bibitem{Evans:2018bqy}
N.~W. Evans, C.~A.~J. O'Hare and C.~McCabe, \emph{{Refinement of the standard
  halo model for dark matter searches in light of the Gaia Sausage}},
  \href{https://doi.org/10.1103/PhysRevD.99.023012}{\emph{Phys. Rev.}
  {\bfseries D99} (2019) 023012},
  [\href{https://arxiv.org/abs/1810.11468}{{\ttfamily 1810.11468}}].

\bibitem{Navarro:1995iw}
J.~F. Navarro, C.~S. Frenk and S.~D.~M. White, \emph{{The Structure of cold
  dark matter halos}}, \href{https://doi.org/10.1086/177173}{\emph{Astrophys.
  J.} {\bfseries 462} (1996) 563--575},
  [\href{https://arxiv.org/abs/astro-ph/9508025}{{\ttfamily
  astro-ph/9508025}}].

\bibitem{1995ApJ...447L..25B}
A.~{Burkert}, \emph{{The Structure of Dark Matter Halos in Dwarf Galaxies}},
  \href{https://doi.org/10.1086/309560}{\emph{apjl} {\bfseries 447} (Jul, 1995)
  L25--L28}, [\href{https://arxiv.org/abs/astro-ph/9504041}{{\ttfamily
  astro-ph/9504041}}].

\bibitem{BAI2018187}
Y.~Bai and Y.~Hamada, \emph{Detecting axion stars with radio telescopes},
  \href{https://doi.org/https://doi.org/10.1016/j.physletb.2018.03.070}{\emph{Physics
  Letters B} {\bfseries 781} (2018) 187 -- 194}.

\bibitem{Eby:2017xaw}
J.~Eby, M.~Leembruggen, J.~Leeney, P.~Suranyi and L.~C.~R. Wijewardhana,
  \emph{{Collisions of Dark Matter Axion Stars with Astrophysical Sources}},
  \href{https://doi.org/10.1007/JHEP04(2017)099}{\emph{JHEP} {\bfseries 04}
  (2017) 099}, [\href{https://arxiv.org/abs/1701.01476}{{\ttfamily
  1701.01476}}].

\bibitem{2011MNRAS.414.2446M}
P.~J. {McMillan}, \emph{{Mass models of the Milky Way}},
  \href{https://doi.org/10.1111/j.1365-2966.2011.18564.x}{\emph{mnras}
  {\bfseries 414} (Jul, 2011) 2446--2457},
  [\href{https://arxiv.org/abs/1102.4340}{{\ttfamily 1102.4340}}].

\bibitem{Nesti:2013uwa}
F.~Nesti and P.~Salucci, \emph{{The Dark Matter halo of the Milky Way, AD
  2013}}, \href{https://doi.org/10.1088/1475-7516/2013/07/016}{\emph{JCAP}
  {\bfseries 1307} (2013) 016},
  [\href{https://arxiv.org/abs/1304.5127}{{\ttfamily 1304.5127}}].

\bibitem{Giovanelli:2005ee}
R.~Giovanelli et~al., \emph{{The Arecibo Legacy Fast ALFA Survey. 1. Science
  goals, survey design and strategy}},
  \href{https://doi.org/10.1086/497431}{\emph{Astron. J.} {\bfseries 130}
  (2005) 2598--2612}, [\href{https://arxiv.org/abs/astro-ph/0508301}{{\ttfamily
  astro-ph/0508301}}].

\bibitem{2011IJMPD..20..989N}
R.~{Nan}, D.~{Li}, C.~{Jin}, Q.~{Wang}, L.~{Zhu}, W.~{Zhu} et~al., \emph{{The
  Five-Hundred Aperture Spherical Radio Telescope (fast) Project}},
  \href{https://doi.org/10.1142/S0218271811019335}{\emph{International Journal
  of Modern Physics D} {\bfseries 20} (Jan., 2011) 989--1024},
  [\href{https://arxiv.org/abs/1105.3794}{{\ttfamily 1105.3794}}].

\bibitem{Gk}
``{G. Karl, Jansky Very Large Array}.''
  \url{https://science.nrao.edu/facilities/vla}.

\bibitem{GBT}
``{Green Bank Observatory}.'' \url{http://greenbankobservatory.org}.

\bibitem{SKA}
``{S.K.A. Baseline, Design document}.'' \url{https://www.skatelescope.org}.

\bibitem{BENTUM2020856}
M.~Bentum, M.~Verma, R.~Rajan, A.~Boonstra, C.~Verhoeven, E.~Gill et~al.,
  \emph{A roadmap towards a space-based radio telescope for ultra-low frequency
  radio astronomy},
  \href{https://doi.org/https://doi.org/10.1016/j.asr.2019.09.007}{\emph{Advances
  in Space Research} {\bfseries 65} (2020) 856 -- 867}.

\bibitem{Condon16}
J.~Condon and A.~Ransom, \emph{Essential radio astronomy}, {\emph{Princeton
  University Press} (2016) }.

\bibitem{Bacon:2018dui}
{\scshape SKA} collaboration, D.~J. Bacon et~al., \emph{{Cosmology with Phase 1
  of the Square Kilometre Array: Red Book 2018: Technical specifications and
  performance forecasts}},
  \href{https://doi.org/10.1017/pasa.2019.51}{\emph{Publ. Astron. Soc.
  Austral.} {\bfseries 37} (2020) e007},
  [\href{https://arxiv.org/abs/1811.02743}{{\ttfamily 1811.02743}}].

\end{thebibliography}\endgroup
\end{document}